%% file: ms.tex
\pdfoutput=1
\documentclass[lettersize,journal]{IEEEtran}
\usepackage{amsmath,amsfonts}
\usepackage{algorithmic}
\usepackage{algorithm}
\usepackage{array}
\usepackage{subfig}
\usepackage{textcomp}
\usepackage{cleveref}
\usepackage{stfloats}
\usepackage{url}
\usepackage{verbatim}
\usepackage{xcolor}
\usepackage{graphicx}
\usepackage{cite}
\usepackage{calligra}
\usepackage{multirow}
\usepackage{tabularx}
\usepackage{booktabs}
\usepackage{multicol}

\begin{document}

\title{Image Denoising Using Green Channel Prior}

\author{Zhaoming Kong,
        Fangxi Deng
        and Xiaowei Yang
\thanks{Z. Kong and X. Yang are with the School of Software Engineering, South China University of Technology, Guangzhou, 510006 China (e-mail: kong.zm@mail.scut.edu.cn; xwyang@scut.edu.cn).}
\thanks{F. Deng is with Tencent Technology (email: fangxideng@outlook.com).}}

\maketitle

\begin{abstract}
Image denoising is an appealing and challenging task, in that noise statistics of real-world observations may vary with local image contents and different image channels. Specifically, the green channel usually has twice the sampling rate in raw data. To handle noise variances and leverage such channel-wise prior information, we propose a simple and effective green channel prior-based image denoising (GCP-ID) method, which integrates GCP into the classic patch-based denoising framework. Briefly, we exploit the green channel to guide the search for similar patches, which aims to improve the patch grouping quality and encourage sparsity in the transform domain. The grouped image patches are then reformulated into RGGB arrays to explicitly characterize the density of green samples. Furthermore, to enhance the adaptivity of GCP-ID to various image contents, we cast the noise estimation problem into a classification task and train an effective estimator based on convolutional neural networks (CNNs). Experiments on real-world datasets demonstrate the competitive performance of the proposed GCP-ID method for image and video denoising applications in both raw and sRGB spaces. Our code is available at https://github.com/ZhaomingKong/GCP-ID.
\end{abstract}


\section{Introduction}
\IEEEPARstart{D}{uring} acquisition, transmission and compression, an image may be inevitably contaminated by noise to varying degrees, which will adversely affect its quality for display or analysis. Therefore, image denoising plays an indispensable role in modern imaging systems\cite{elad2023image}. Meanwhile, the rapid development of camera devices has empowered a wider audience to engage in photography, thus the growth in the number and size of images poses a greater demand on noise removal in terms of both effectiveness and efficiency. \\
\indent The purpose of noise removal is to recover a clean image $\mathcal{X}$ from its noisy observation $\mathcal{Y} = \mathcal{X} + \mathcal{N}$, where $\mathcal{N}$ is usually modeled as additive white Gaussian noise (AWGN) with standard deviation $\sigma$. As an ill-posed inverse problem, image denoising enjoys a long history and draws great attention from both academia and industries. It has a broad application and also serves as a test-bed for assessing new image processing tools. The study of denoising methods may roughly be divided into two categories \cite{elad2023image, kong2023comparison}, namely traditional denoisers and deep neural network (DNN) models. \\
\indent Traditional denoisers normally utilize the internal information of a single input image with different regularization terms and priors. Early works may date back to the $L_2$-based regularization and wavelet filters, leading to the popular patch-based framework \cite{dabov2007image, chatterjee2011patch}, which takes advantage of the nonlocal self-similarity (NLSS) structures of natural images, transform domain techniques \cite{yaroslavsky2001transform, dong2012nonlocal} and sparse modeling \cite{aharon2006k}. Methods falling into this line of paradigm are quite extensive, producing compelling results in various denoising applications. For example, Zoran et al. \cite{zoran2011learning} used the patch-based prior in the likelihood sense and calculated maximum a-posteriori (MAP) estimate. Sunil et al. \cite{jha2010denoising} introduced singular value decomposition (SVD) to handle sensor array data. Chang et al. \cite{chang2017hyper} utilized low-rank tensor assumptions and hyper-laplacian regularization for multi-channel imaging data. Despite the effectiveness of many well-performing classic denoising algorithms, there are several flaws such as complex and recursive optimization schemes, careful manual parameter settings and lack of external information. \\
\indent To overcome these drawbacks, the emergence and development of deep learning techniques in the past decade has brought a fresh wind to the design of novel and highly-effective denoising algorithms \cite{tian2019deep}. To name a few, Zhang et al. \cite{zhang2017beyond} integrated batch normalization and residual learning into CNNs. Yue et al. \cite{yue2019high} and Pl{\"o}tz et al. \cite{plotz2017benchmarking} fused CNNs with the NLSS concept. Chen et al. \cite{chen2022simple} devised an efficient activation free network. Zamir et al. \cite{zamir2022restormer} introduced transformer \cite{vaswani2017attention} to capture long-range image pixel interactions. Li et al. \cite{li2023stimulating} considered the diffusion model from a denoising perspective with adaptive ensembling. Benefiting from the powerful feature extraction capability of deep networks and advanced computing platforms, various DNN models show superb denoising performance and reach the summit of different applications. However, nor is the DNN model flawless. For example, the model training process usually relies heavily on high quality clean-noisy image pairs, which are not always available. Besides, the overparameterized network architectures will drastically increase storage and computation burdens. Moreover, the expensive hardware devices may not be affordable by ordinary users and researchers. \\
\indent Apart from the aforementioned concerns, we notice that the main focus of many methods lies in the modeling of relationship between image pixels and patches. However the correlation among different channels is often understated or ignored. In fact, to reconstruct a full-resolution color image from sensor readings, a digital camera generally goes through different image signal processing (ISP) steps. In particular, demosaicing \cite{menon2011color} usually adopts the Bayer color filter array (CFA) pattern \cite{chung2008lossless}, which measures the green channel at a higher sampling rate than red/blue ones. Such prior knowledge is known as the green channel prior (GCP) \cite{guo2021joint}. It is commonly seen in applications of raw sensor data \cite{liu2020joint, guo2021joint, zhang2022joint}, which are not always accessible. Therefore, it is natural to ask if the GCP may be directly leveraged for both raw and sRGB color spaces. Another interesting question is how the robustness of the DNNs and the simplicity of the classic paradigm can be effectively exploited. In this paper, motivated by the fact that the green channel has a higher signal-to-noise ratio (SNR) than red/blue ones in many natural images, we propose a lightweight GCP-based image denoising (GCP-ID) method to explore the potential of the channel-wise prior. \\
\indent Following the classic patch-based framework, we may conclude that the performance of related denoisers is often subject to three factors: nonlocal similar patch search, image patch representation and local smoothness. The proposed method aims to achieve improvements from three different perspectives with GCP-guided search, RGGB representation and noise estimation, respectively. Briefly, GCP-ID utilizes the green channel to guide the search of similar patches because the green channel is less noisier and preserves better image structure and details. To explicitly model the density of green samples and the importance of the green channel, each image patch is rearranged into RGGB array based on the Bayer pattern, then the spectral correlation among all channels can be exploited with block circulant representation and t-SVD transform \cite{kilmer2011factorization, kilmer2013third}. The similar patches are stacked into a group and the group-level relationship can be captured by performing principal component analysis (PCA) along the grouping dimension \cite{dabov2009bm3d, zhang2010two}. The group-level redundancy is modelled by sparsity in the transform domain, controlled by certain noise level. Instead of using a predefined $\sigma$ value, we reformulate the noise estimation problem as a classification task, and combine a simple CNN framework with the proposed method to enhance local adaptivity and smoothness. \\
\indent Our contributions to this line of research can be summarized in several aspects. First, we integrate the GCP into a unified patch-based denoising framework for both raw and sRGB data. Besides, to get rid of the complicated learning procedures in both training and testing phases, we consider to effectively take advantage of the traditional denoiser and CNNs. Experiments on real-world datasets show the competitive performance of the proposed method for image and video denoising tasks. Furthermore, we also discuss its potential to enable customization of a specific denoiser without ground-truth noise free images, along with extension to other imaging techniques such as the hyperspectral imaging (HSI).\\
\indent This paper is organized as follows. Section II includes background knowledge. Section III introduces related works. Section IV presents our proposed GCP-ID method. Experiments and discussions are provided in Section V. Finally we conclude this paper in Section VI.
\vspace{-3pt}
\section{Background}
\subsection{Symbols and Notations}
\indent In this paper, we mainly adopt the mathematical notations and preliminaries of tensors from \cite{kolda2009tensor} for image representation. Vectors and matrices are denoted by boldface lowercase letters $\mathbf{a}$ and capital letters $\mathbf{A}$, respectively. Tensors are denoted by calligraphic letters, e.g., $\mathcal{A}$. Given an $N$th-order tensor $\mathcal{A} \in \mathbb{R}^{I_1\times I_2\times\cdots\times I_N}$, the Frobenius norm of a tensor $\mathcal{A} \in \mathbb{R}^{I_1\times I_2\times\cdots\times I_N}$ is defined as $\|\mathcal{A}\|_F = \sqrt{\sum_{i_1=1}...\sum_{i_N=1}\mathcal{A}_{i_1...i_N}^2}$. The $n$-mode product of a tensor $\mathcal{A}$ by a matrix $\mathbf{U}\in \mathrm{R}^{P_n\times I_n}$ is denoted by $\mathcal{A}\times _n\mathbf{U}$.
\vspace{-3pt}
\subsection{Block Circulant Matrix}
Circulant matrices and structures occur in many image processing applications \cite{asriani2023real}. Specifically, for a third order tensor $\mathcal{A}\in \mathbb{R}^{n_1\times n_2 \times n_3}$,  we can create an $n_1n_3 \times n_2n_3$ block-circulant matrix denoted by $bcirc(\mathcal{A})$ with the frontal slices of $\mathcal{A}$. It is given by
\begin{equation}\label{Equ_bcirc}
  bcirc(\mathcal{A}) = \begin{bmatrix}
\mathcal{A}^{(1)} & \mathcal{A}^{(n_3)} & \cdots & \mathcal{A}^{(n_2)}\\
\mathcal{A}^{(2)} & \mathcal{A}^{(1)} & \cdots & \mathcal{A}^{(3)}\\
\vdots & \vdots  & \ddots  & \vdots  \\
\mathcal{A}^{(n_3)} & \mathcal{A}^{(n_3 - 1)} & \cdots & \mathcal{A}^{(1)}
\end{bmatrix}
\end{equation}
where the frontal slice $\mathcal{A}(:,:,i)$ is denoted compactly as $\mathcal{A}^{(i)}$.
\vspace{-10pt}
\subsection{T-product and t-SVD}
\vspace{0pt}
The product between matrices can be generalized to the product of two tensors according to t-product \cite{kilmer2013third}. Specifically, the t-product `$*$' between two third-order tensors $\mathcal{A} \in \mathbb{R}^{N_1 \times N_2 \times N_3}$ and $\mathcal{B} \in \mathbb{R}^{N_2 \times N_4 \times N_3}$ is also a third order tensor $\mathcal{C} \in \mathbb{R}^{N_1 \times N_4 \times N_3}$ with $\mathcal{C} = \mathcal{A} * \mathcal{B}$ denoted by
\begin{equation}\label{Equ_t_product}
  bcirc(\mathcal{C}) = bcirc(\mathcal{A}) bcirc(\mathcal{B})
\end{equation}
Directly handling block circulant matrices is time consuming. Eq. (\ref{Equ_t_product}) can be computed efficiently in the Fourier domain
\begin{equation}\label{Equ_t_SVD_Fourier_domain}
  bdiag(\mathcal{C}_{F}) = bdiag(\mathcal{A}_{F}) bdiag(\mathcal{B}_{F})
\end{equation}
where $bdiag$ is the block diagonal representation, $\mathcal{A}_{F}$ is obtained by performing the fast Fourier transform (FFT) along the third mode of $\mathcal{A}$ via $\mathcal{A}_F = \mathcal{A} \times _3\mathbf{W}_{FFT}$, and $\mathbf{W}_{FFT}$ is the fast Fourier transform (FFT) matrix. \\
\indent The tensor-SVD or t-SVD of a third-order tensor $\mathcal{A}$ can then be defined as the t-product of three third-order tensors $\mathcal{U}\in \mathbb{R}^{N_1\times N_1 \times N_3}$, $\mathcal{S}\in \mathbb{R}^{N_1\times N_2 \times N_3}$ and $\mathcal{V}\in \mathbb{R}^{N_2\times N_2 \times N_3}$
\begin{equation}\label{Equ_t-SVD}
  \mathcal{A} = \mathcal{U} * \mathcal{S} * \mathcal{V}^T
\end{equation}
where $\mathcal{U}$ and $\mathcal{V}$ are orthogonal tensors, and $\mathcal{S}$ can be regarded as the corresponding coefficient tensor.
\vspace{3pt}
\section{Related works}
\vspace{3pt}
\subsection{Image Denoising Using Different Priors} To effectively suppress noise, different prior information has been introduced. Specifically, the nonlocal self-similarity (NLSS) prior of natural images is widely adopted, which refers to the fact that a natural image patch often has many patches similar to it across the image\cite{xu2015patch}. The sparsity prior \cite{elad2006image, dabov2007image} shows that an image patch can be sparsely represented by certain proper basis function. The total variation (TV) prior \cite{beck2009fast} assumes that images are locally smooth. The low-rank prior \cite{xu2017multi, chang2017hyper} indicates that a matrix/tensor formed by a group of similar patches has a low-rank property. The consistency prior \cite{ren2021adaptive, chen2013image} argues that relative similarities between pixels should be preserved in the filtered output. The deep image prior \cite{ulyanov2018deep, jo2021rethinking} demonstrates that a randomly initialized CNN acts as an image prior for denoising. In addition to these popular priors, their variants and applications are extensive. For example, Chen et al.\cite{chen2024flex} and Jiang et al.\cite{jiang2022deep} proposed to incorporate flexible and adaptive priors into DNN. Xu et al. \cite{xu2018external} used external patch priors to guide internal prior learning of traditional denoisers. We notice that the success of many approaches resorts to complicated and recursive optimization procedures, which may result in high computation complexity. Therefore, it is interesting to investigate if certain prior information can be exploited for efficient noise removal in practice.

\subsection{Traditional Patch-based Framework} 
The popular patch-based denoising framework \cite{dabov2007image} incorporates the NLSS prior, sparse representation and transform domain techniques \cite{yaroslavsky2001transform, maggioni2012nonlocal} into a subtle paradigm, which mainly follows three consecutive steps: grouping, collaborative filtering and aggregation. Briefly, given a local reference patch $\mathcal{P}_{ref}$, its similar patches are stacked into a group $\mathcal{G}_{noisy}$ with certain patch matching criteria \cite{foi2007pointwise, ehret2017global}. To utilize the nonlocal similarity feature and estimate clean patches $\mathcal{G}_{clean}$, collaborative filtering is then performed on $\mathcal{G}_{noisy}$ with different transforms and regularization terms \cite{beck2009fast, pang2017graph} via
\begin{equation}\label{tensor_collaborative_filtering}
  \hat{\mathcal{G}}_{clean} = \mathop{\arg\min_{\mathcal{G}_{clean}}} \| \mathcal{G}_{noisy} - \mathcal{G}_{clean} \|_{F}^2 + \rho\cdot\Psi(\mathcal{G}_{clean})
\end{equation}
where $\| \mathcal{G}_{noisy} - \mathcal{G}_{clean} \|_{F}^2$  measures the conformity between the clean and noisy groups, and $\Psi(\cdot)$ is a regularization term for certain priors. Finally, the aggregation step adopts an averaging procedure to further smoothout noise. \\
\indent Early works focus on handling single-channel grayscale images \cite{buades2005review, aharon2006k, zoran2011learning}. When the input is an RGB color image, naively filtering each channel separately will lead to unsatisfactory results since the spectral correlation among RGB channels is ignored \cite{xu2017multi}. To improve the channel-by-channel strategy, two solutions are mainly adopted. The first one is to apply a decorrelation transform along the RGB dimension, and then process each channel of the transformed space independently. The representative CBM3D method \cite{dabov2007color} falls into such category and considers the luminance-chrominance (e.g., YCbCr) space as a less correlated color space. An alternative approach is to jointly model the inter-channel correlation. For example, Dai et al. \cite{dai2013multichannel} adopted a multichannel fusion scheme based on a penalty function. Xu et al. \cite{xu2017multi} concatenated the patches of RGB channels as a long vector and introduced weight matrices to characterize the realistic noise property. To avoid explicit vectorization of image patches and preserve more structural information, various tensor-based approaches have recently been introduced \cite{muti2008lower, rajwade2012image, kong2017new} according to different multiway filtering strategies and algebraic expressions. \\
\indent In spite of the reasonably good performance of existing methods, they either treat each image channel as equal or manually assign different weights. Besides, the motivation of many related works stems largely from the relationship among image patches with various regularization terms, which may not exploit the channel-wise correlation and lead to complex optimization problems. In addition, gathering sufficient similar patches is not easy with the presence of noise. Therefore, an efficient and effective scheme is desired for further enhancement of related works.
\subsection{Green Channel Prior} 
The human visual system (HVS) is sensitive to green color because its peak sensitivity lies in the medium wavelengths, corresponding to the green portion. Therefore, the Bayer pattern is widely adopted by many digital cameras \cite{li2008image}, and the greater sampling rate of the green image will result in higher SNR in the green channel \cite{guo2021joint}. Table \ref{Table_SNR_comparison} and Fig. \ref{Fig_SNR_comparison} compare the SNR difference in RGB channels of images from several real-world datasets (CC15 \cite{nam2016holistic}, HighISO \cite{yue2019high}, and IOCI \cite{kong2023comparison}). It can be seen that the green channel of images normally has a higher SNR than the corresponding red/blue channels.  To exploit the GCP, existing methods mainly focus on raw image data processing. For example, GCP-Net \cite{guo2021joint} utilizes the green channel to guide the feature extraction and feature upsampling of the whole image. SGNet \cite{liu2020joint} produces an initial estimate of the green channel and then uses it as guidance to recover all missing values.
\vspace{-2.8pt}
\input{Table_SNR_comparison}

\input{Fig_SNR_comparison}

\indent Interestingly, the importance of the GCP is often understated and rarely introduced as a prior for noise removal in the sRGB color space. In this paper, we intend to fill this gap and propose a simple and effective GCP-guided method for both raw and sRGB denoising tasks.
\section{Method}
In this section, we present the proposed GCP-ID method in detail. Our journey of GCP-ID starts with the sRGB color inputs, and it can be naturally adopted for raw image data. The overall denoising procedure is illustrated in Fig. \ref{Figs_flowchart}. Briefly, GCP-ID consists of four key ingredients: (1) green channel guided patch search, (2) RGGB representation for image patches, (3) nonlocal t-SVD transform with GCP and (4) noise level estimation with GCP-guided CNN.
\input{Figs_flowchart}

\vspace{-3.8pt}
\subsection{Green Channel Guided Patch Search}\label{section_method_patch_search}
Given an image patch $\mathcal{P}_{ref} \in \mathbb{R}^{ps \times ps \times 3}$, where $ps$ refers to patch size, our goal is to search its $K$ most similar patches in a local window by Euclidean distance. Directly comparing image patches will incur a high computational burden. Besides, the presence of complex real-world noise may undermine the patch search process. To efficiently group more closely similar patches, we consider to take advantage of the green channel, which is less noisier and shares structural information with red/blue channels. Specifically, we calculate the distance $d$ between two image patches via
\begin{equation}\label{Equ_distance_calculation}
  d=\left\{
    \begin{aligned}
    &\| \mathcal{P}_{ref}^G -  \mathcal{P}_j^G \|, \, \, \, \|\mathcal{P}_{ref}^G\| \geq \text{max}(\frac{1}{\lambda}\|\mathcal{P}_{ref}^R\|, \frac{1}{\lambda} \|\mathcal{P}_{ref}^B\|) \\
    & \| \mathcal{P}_{ref}^{mean} -  \mathcal{P}_j^{mean} \|, \, \, \, \text{otherwise}
    \end{aligned}
    \right.
 \end{equation}
where $\mathcal{P}^{R}$, $\mathcal{P}^{G}$ and $\mathcal{P}^{B}$ represent R, G and B channels of the image patch, respectively. $\mathcal{P}^{mean}$ is the mean value of RGB channels, which can be regarded as the opponent color space or average pooling operation on RGB channels. $\lambda$ is a weight parameter that controls the search scheme and reflects the importance of the green channel in patch search. In our paper, we empirically set $\lambda = 1.2$. \\
\indent To verify the effectiveness of the GCP-guided strategy, we compare the patch search successful rate, which is measured by the ratio of patches that are also similar in the underlying clean image. Briefly, given a noisy observation, we randomly select 1000 reference patches $\mathcal{P}_{ref}\in \mathbb{R}^{8\times8\times3}$. For each $\mathcal{P}_{ref}$, we gather a group $\mathcal{G}_{noisy} \in \mathbb{R}^{8\times8\times3\times60}$ consisting of 60 similar patches within a local window of size $20\times20$. Table \ref{Table_patch_search_successful_rate} reports the average successful rate of different search schemes on the CC dataset. The advantage of the GCP-guided strategy is able to enhance the conformity between noisy and underlying clean patches, which may help encourage sparsity in the transform domain. Therefore, the correlation among image patches can be better captured by learning group-level transform $\mathbf{U}_{group} \in \mathbb{R}^{K\times K}$ of $\mathcal{G}_{noisy}$ via classic algorithms such as the principal component analysis (PCA) \cite{dabov2009bm3d}.
\input{Table_patch_search_successful_rate}

\vspace{-6.8pt}

\subsection{RGGB Representation for Image Patches}
\input{Fig_Bayer_pattern_RGGB}

As illustrated in Fig. \ref{Fig_Bayer_pattern_RGGB}(a), the popular Bayer CFA pattern measures the green image on a quincunx grid and the red and blue images on rectangular grids\cite{li2008image}. As a result, the density of the green samples of raw data is twice that of the other two. In practice, the raw Bayer pattern data is not always available. Besides, the interpolation algorithm to reconstruct a full-color representation of the image varies according to different camera devices.\\
\indent To explicitly model the density and importance of the green channel in the sRGB space, we reshape each color image patch $\mathcal{P}\in \mathbb{R}^{ps\times ps \times 3}$ into an RGGB array $\hat{\mathcal{P}}\in \mathbb{R}^{ps\times ps \times 4}$ by inserting one extra green channel. Furthermore, to capture the spectral correlation among RGB channels, we use the block circulant matrix $\hat{\mathbf{P}}_{bcirc}\in \mathbb{R}^{4ps\times4ps}$ to represent the RGGB array. Fig. \ref{Fig_Bayer_pattern_RGGB}(b) illustrates the RGGB and block circulant representation (BCR) of an RGB input. It can be seen that the density of green channels remains twice that of red/blue ones. Interestingly, such representation is in consistent with the original Bayer pattern of raw data.
\vspace{-11.8pt}
\subsection{Nonlocal t-SVD Transform with GCP}
The stacked similar patches often share the same feature space, and a nonlocal generalization of SVD \cite{rajwade2012image} can be explored by learning pairwise projection matrices $\mathbf{U}_{row} \in \mathbb{R}^{4ps\times4ps}$ and $\mathbf{U}_{col}\in \mathbb{R}^{4ps\times4ps}$ via
\vspace{-6.9pt}
\begin{equation}\label{Equ_nonlocal_SVD}
\begin{aligned}
  & \text{min} \sum_{i = 1}^K \| \hat{\mathbf{P}}_{bcirc_i} -  \mathbf{U}_{row} \hat{\mathbf{S}}_{bcirc_i} \mathbf{U}_{col}^T\|^2 \\
  & \text{s.t} \quad \mathbf{U}_{row}^T \mathbf{U}_{row} = \mathbf{I}, \, \mathbf{U}_{col}^T \mathbf{U}_{col} = \mathbf{I}
\end{aligned}
\vspace{-1.9pt}
\end{equation}
where $\hat{\mathbf{S}}_{bcirc_i} \in \mathbb{R}^{4ps\times4ps}$ is the coefficient matrix of $\hat{\mathbf{P}}_{bcirc_i}$. From Fig. \ref{Fig_Bayer_pattern_RGGB}, it is noticed that the RGB channels of $\hat{\mathbf{P}}_{bcirc}$ repeat several times, however, directly handling BCR matrices is time consuming. To efficiently leverage patch-level redundancy and exploit the spectral correlation among RGB channels, we can take advantage of the nonlocal t-SVD transform to obtain a pair of projection tensors and simplify Eq. (\ref{Equ_nonlocal_SVD}) via
\begin{equation}\label{Equ_nonlocal_t-SVD}
\begin{aligned}
  & \text{min} \sum_{i = 1}^K \| \hat{\mathcal{P}}_{i} -  \mathcal{U}_{row} * \hat{\mathcal{S}}_{i} * \mathcal{U}_{col}^T\|_F^2 \\
  & \text{s.t} \quad \mathcal{U}_{row}^T * \mathcal{U}_{row} = \mathcal{I}, \, \, \mathcal{U}_{col}^T * \mathcal{U}_{col} = \mathcal{I}
\end{aligned}
\end{equation}
where $\hat{\mathcal{S}}_{i} \in \mathbb{R}^{ps\times ps \times 4}$ is the corresponding coefficient tensor of $\hat{\mathcal{P}}_{i}$, $\mathcal{U}_{row} \in \mathbb{R}^{ps\times ps \times 4}$ and $\mathcal{U}_{col}\in \mathbb{R}^{ps\times ps \times 4}$ are a pair of orthogonal tensors, respectively. Eq. (\ref{Equ_nonlocal_t-SVD}) can be solved by slice-wise SVD in the Fourier domain after performing the fast Fourier transform (FFT) along the third mode of $\hat{\mathcal{P}}$ via
 \begin{equation}\label{Equ_FFT}
 \begin{aligned}
 \hat{\mathcal{P}} \times _3\mathbf{W}_{FFT} & = \begin{pmatrix}
 1 & 1 & 1 & 1\\
 1 & -i & -1 & i \\
 1 & -1 & 1 & -1 \\
 1 & i & -1 & -i \\
 \end{pmatrix}
 \begin{pmatrix}
 R \\
 G \\
 G \\
 B \\
 \end{pmatrix}\\
 & = \begin{pmatrix}
 R + 2G + B \\
 R-G + (B-G)i \\
 R - B \\
 R-G + (G-B)i \\
 \end{pmatrix}
 \end{aligned}
 \end{equation}
where $\mathbf{W}_{FFT}$ is the FFT matrix. Interestingly, from Eq. (\ref{Equ_FFT}) we can observe that in the Fourier domain, the first frontal slice maintains the density of each channel, the second and fourth slices are complex conjugate that exploit the relationship among RGB channels, and the third slice captures the difference between red and blue channels. Therefore, the RGGB representation models the correlation among RGB channels without explicitly assigning weights to each channel. It is worth mentioning that t-SVD has been adopted by previous denoisers \cite{zhang2014novel, kong2019color, shi2021robust}, but they treat each channel equally.\\
\indent Given a group of noisy RGGB patches $\hat{\mathcal{G}}_{noisy} \in \mathbb{R}^{ps\times ps \times 4 \times K}$, the overall learning process of the proposed GCP-ID can be summarized using a one-step extension of the t-SVD transform via
\begin{equation}\label{Equ_summarize_denoising}
  \vspace{-1.8pt}
  \hat{\mathcal{S}}_{noisy} = (\mathcal{U}_{row}^T * \hat{\mathcal{G}}_{noisy} * \mathcal{U}_{col}) \times _4 \mathbf{U}_{group}^T
\end{equation}
According to the sparsity assumption that the energy of the signal resides in a few large coefficients \cite{foi2007pointwise}, the hard-thresholding technique \cite{donoho1994ideal} can be applied to truncate small coefficients of $\hat{\mathcal{S}}_{noisy}$ under a certain threshold $\tau$ via
 \begin{equation}\label{Equ_hard_thresholding}
 \vspace{-1.8pt}
  \hat{\mathcal{S}}_{truncate}=\left\{
    \begin{aligned}
    \hat{\mathcal{S}}_{noisy}, \quad |\hat{\mathcal{S}}_{noisy}| \geq \tau \\
    0, \quad |\hat{\mathcal{S}}_{noisy}| < \tau
    \end{aligned}
    \right.
 \end{equation}
After shrinkage, the estimated clean group $\hat{\mathcal{G}}_{clean}$ can be reconstructed by performing the inverse transform of Eq. (\ref{Equ_summarize_denoising})
\begin{equation}\label{Equ_summarize_inverse_denoising}
\vspace{-1.8pt}
\hat{\mathcal{G}}_{clean} = (\mathcal{U}_{row} * \hat{\mathcal{S}}_{truncate} * \mathcal{U}_{col}^T) \times _4 \mathbf{U}_{group}
\end{equation}
Finally, for an sRGB color image, we fetch the RGB channels of $\hat{\mathcal{G}}_{clean}$, and each denoised image patch $\mathcal{P}_{clean} \in \mathbb{R}^{ps\times ps \times 3}$ is averagely written back to its original location. The implementation of GCP-ID is briefed in Algorithm \ref{Algorithm_GCP-ID}.
\input{Algorithm_GCP-ID}

\vspace{-6pt}
\subsection{Noise Level Estimation with GCP-guided CNNs}
To model local smoothness and sparsity, the noise level $\sigma$ plays a crucial role. Using a predefined $\sigma$ ignores the noise variances and local features of nature images, i.e., some regions are more severely contaminated. In addition, manual parameter tuning for each input is tedious, especially when the number and size of image data grows significantly. Therefore, effective noise estimation is essential for efficient image denoising. There are many noise estimators such as PCA-based methods \cite{liu2013single, pyatykh2012image}, pixel- or block-based algorithms \cite{meer1990fast, ghazal2010homogeneity} and statistic-based approaches\cite{chen2015efficient}. These methods weigh highly on the internal image features, while the importance of subsequent denoisers is often understated. 
\input{Fig_train_CNN_estimator} \\
\indent We propose to combine GCP-ID with DNNs and reformulate the noise level estimation problem as a classification task. As illustrated in Fig. \ref{Fig_train_CNN_estimator}, for each subimage of the noisy observation, we select an optimal $\sigma$ of GCP-ID within a certain range based on a specific evaluation metric such as the peak SNR (PSNR) or structure similarity (SSIM \cite{wang2004image}). In this way, GCP-ID creates labels to train a classifier and the CNN architecture is adopted due to its simplicity and effectiveness. This simple combination scheme enjoys several benefits: (i) it eliminates the need to manually assign a $\sigma$ value for each noisy observation; (ii) the GCP in Eq. (\ref{Equ_distance_calculation}) can be used to guide the model training, which reduces the storage and computation costs by $\frac{2}{3}$; (iii) training with image patches helps capture local features, which will enhance adaptiveness of the denoiser to various image contents with local noise estimation; and (iv) from Fig. \ref{Fig_sensitivity_visual_effects} and Fig. \ref{Fig_sensitivity_PSNR_SSIM_highISO}, we notice that GCP-ID may not be sensitive to subtle changes in $\sigma$, thus `misclassification' may be tolerable, making it easier to train the model.
\input{Fig_sensitivity_visual_effects}

\input{Fig_sensitivity_PSNR_SSIM_highISO}

\subsection{Complexity Analysis}
For each reference patch of noisy observation $\mathcal{Y}$, the computational burden of GCP-ID lies mainly in three parts: (i) the search of $K$ similar patches $O(KW^2ps)$, (ii) the group-level PCA transform $O(Kps^4)$ and (iii) the nonlocal t-SVD transform $O(Kps^3)$. Therefore, the overall complexity of GCP-ID is $O([KW^2ps + Kps^4]n_{ref})$, where $n_{ref}$ is the number of reference patches of $\mathcal{Y}$. The proposed method is efficient, in that it does not involve any complex recursive optimization strategies, thus we are able to utilize parallel computing tools to further reduce its running time.
\vspace{-3.8pt}
\section{Experiments}
In this section, we mainly report results of image and video denoising tasks in both raw and sRGB spaces. Our experiments are performed on a moderate computer equipped with Core(TM) i7-10700F @ 2.9 GHz and 16GB RAM. We employ PSNR and SSIM values for objective evaluations. Normally, the larger these two measures are, the closer the denoised image is to the reference one.
\vspace{-10.8pt}
\subsection{Parameter Settings and Implementation Details}
GCP-ID has several free parameters such as the patch size $ps$, search window size $W$, number of similar patches $K$, hard threshold value $\tau$ and noise level $\sigma$. In our experiments, we set $ps = 8$, $W = 20$, $K = 30$, and $\tau = 1.1\sigma\sqrt{2\text{log}(4Kps^2)}$. To estimate the noise level $\sigma$, we train a simple CNN classifier for 90 epochs according to the official PyTorch implementation\footnote{https://pytorch.org/tutorials/beginner/blitz/cifar10\_tutorial.html}. We use the Adam optimizer with a learning rate of 0.001. The SIDD small dataset is used for both raw and sRGB model training. Each input is divided into subimages of size $128 \times 128$ for training. For raw sensor data, $\sigma$ is chosen from 10 different values \{1.25, 2.5, 5, 10, 15, 20, 25, 30, 40, 50\}. For color inputs, $\sigma$ belongs to $\{1.25, 5, 10, 20, 30, 40, 50, 60, 80, 100, 120, 140\}$. We notice that the SIDD dataset contains a large proportion of severely contaminated images, which may result in relatively high $\sigma$ value and apparent oversmooth effects for different real-world datasets. To alleviate this problem, we also use the clean-clean image pairs and assign small $\sigma$ values (e.g., 1.25, 5 or 10). During inference, the noise level of each subimage is decided by the average $\sigma$ value of 9 neighbouring subimages to attenuate mispredictions and improve consistency among adjacent image patches. 
\vspace{-6.8pt}
\subsection{Datasets and Compared Methods}
\subsubsection{Datasets}
To verify the effectiveness of the proposed method for raw and sRGB denoising, we perform comprehensive experiments on popular real-world datasets listed in Table \ref{Table_dataset}. More details are provided in the supplemental material.
\input{Table_dataset}

\subsubsection{Compared Methods}
We compare GCP-ID with state-of-the-art traditional denoisers and DNN methods. Briefly, traditional denoisers include nonlocal approaches such as BM3D \cite{dabov2007color}, MStSVD \cite{kong2019color} and NLHCC \cite{hou2020nlh}. DNN methods consist of effective supervised (S) and self-supervised (SS) models such as CycleISP\cite{zamir2020cycleisp}, FFDNet \cite{zhang2018ffdnet}, Restormer \cite{zamir2022restormer}, Self2Self \cite{quan2020self2self} and Neigh2Neigh \cite{huang2021neighbor2neighbor}. Parameters and models of all compared methods are carefully chosen to obtain their best possible performance for each dataset.
\vspace{-6.8pt}
\subsection{Performance Evaluations}
\subsubsection{Raw and color image denoising}
Table \ref{Table_Raw_image} and Table \ref{Table_Color_image} report the results of compared methods for raw and color image denoising, respectively. From Table \ref{Table_Raw_image}, it is noticed that the GCP-ID outperforms BM3D by over 2dB on SIDD. Furthermore, with the help of the CNN estimator, the proposed denoiser considerably narrows the gap between classic patch-based denoisers and well-trained DNN models tailored for raw image denoising. Table \ref{Table_Color_image} shows the competitive performance of GCP-ID on a variety of real-world color image datasets. For example, it shares the same PSNR and SSIM values with NAFNet on DND. Besides, compared to the classic nonlocal denoiser CBM3D and DNN methods such as Self2Self and CBDNet, GCP-ID produces PSNR improvements of more than 0.3dB on SIDD and CC. In addition, the advantage of GCP-ID over MStSVD demonstrates the effectiveness of integrating GCP into the nonlocal t-SVD transform.
\input{Table_Raw_image}
\input{Table_Color_image}\\
\indent To better understand and depict the denoising results, we present visual evaluations from Fig. \ref{Fig_nonlocal_denoiser_comparison} to Fig. \ref{Fig_GCP_w_wo_CNN}. In general, the proposed method exhibits competitive denoising performance in terms of both noise removal and detail recovery. By contrast, the pretrained models may not generalize well to unseen noise patterns. From Fig. \ref{Fig_nonlocal_denoiser_comparison}, it is noticed that compared to the state-of-the-art classic denoisers, GCP-ID can effectively suppress noise and reduce color artifacts, which demonstrates the benefit of recursively modeling and utilizing the channel-wise correlations. Fig. \ref{Fig_CC15_visual_evaluation} vividly exhibits and explains the strength of GCP when the green channel is more important. Fig. \ref{Fig_SonyA6500_visual_evaluation} illustrates the effectiveness of GCP-ID as the green channel no longer plays a dominant role. Fig. \ref{Fig_SIDDsRGB_visual_evaluation} shows the robustness of the proposed method to different noise patterns. Furthermore, Fig. \ref{Fig_GCP_w_wo_CNN} indicates that the CNN estimator may empower GCP-ID with better adaptiveness and protect the denoiser from obvious oversmoothness and color shifts.
\input{Fig_nonlocal_denoiser_comparison}

\input{Fig_CC15_visual_evaluation}

\input{Fig_SonyA6500_visual_evaluation}

\input{Fig_SIDDsRGB_visual_evaluation}

\input{Fig_GCP_w_wo_CNN}

\vspace{-3pt}
\indent Apart from the effectiveness of GCP-ID, we are also interested in its efficiency. From the inference time comparison in Table \ref{Table_inference_time}, it can be seen that GCP-ID does not incur a high computation burden. Considering GCP-ID is a one-step denoiser without recursive and complicated optimization procedures, it is also flexible, lightweight and parallelizable. Besides, the introduction of the simple CNN noise estimator can help circumvent manual parameter tuning and model selection, making the proposed scheme a practical alternative to very deep networks for denoising.
\input{Table_inference_time}

\subsubsection{Raw and color video denoising}
GCP-ID can be effortlessly extended to handle video sequences by performing patch search along both spatial and temporal dimensions. Specifically, we set $ps = 8$, $K = 30$, $\tau = 1.1\sigma\sqrt{2\text{log}(4ps^2N_fK)}$, where $N_f$ is the number of frames. Besides, the CNN noise estimator of image data can be directly applied without retraining. Table \ref{Table_Raw_Color_video} lists the results of different methods for real-world video denoising tasks. It is noted that for the CRVD raw dataset, compared DNN models used almost $60\%$ of data for training. Nevertheless, the proposed method achieves a comparable performance, which manifests the robustness of GCP-ID and the corresponding CNN estimator. For color video denoising, GCP-ID produces a steady improvement in PSNR over compared methods on both CRVD and IOCV datasets. Specifically, our method obtains a performance boost of 0.81dB over the recent algorithm ASwin \cite{lindner2023lightweight} and 0.38dB over the previous best method RVRT \cite{paliwal2021multi}. \\
\indent Visual quality evaluations of different real-world scenes are provided in Fig. \ref{Fig_CRVDsRGB_indoor_new} and Fig. \ref{Fig_CRVD_outdoor}. Similar to our observations of image denoising, the proposed method handles textures and structures of video sequences well, as it leverages spatiotemporal patch and group level redundancy. In particular, Fig. \ref{Fig_CRVDsRGB_indoor_new} shows the advantage of the proposed method in terms of adaptiveness to noise variances and different image contents. In comparison, the pretrained/predefined models start to oversmooth textures at low ISO value and leave more obvious color artifacts as the noise level and ISO value increases. Fig. \ref{Fig_CRVD_outdoor} demonstrates the capability of the proposed method in exploiting local structures and suppressing severe noise in flat part, thus enhancing temporal coherence and reducing annoying flickering effects.
\input{Table_Raw_Color_video}

\input{Fig_CRVDsRGB_indoor_new}

\input{Fig_CVRD_outdoor}

\subsection{Parameter Analysis and Ablation Study}
\subsubsection{Parameter analysis}
In addition to the noise level $\sigma$, the proposed GCP-ID denoiser has several important free parameters. For example, the GCP-guided search weight $\lambda$ decides the nonlocal similar patch search strategy, and the number of similar patches $K$ helps to model group-level redundancy. Fig. \ref{Fig_parameter_analysis} illustrates the sensitivity of GCP-ID to different parameters based on the SIDD (sRGB) validation dataset. In particular, Fig. \ref{Fig_parameter_analysis}\subref{Fig_guided_search_weight_analysis} shows that imposing a high weight on the green channel may degrade the denoising performance since the green channel does not always dominate the patch search step. Similarly, Fig. \ref{Fig_parameter_analysis}\subref{Fig_number_of_patches_analysis} indicates that simply increasing the number of grouped patches $K$ does not guarantee better results, in that it is challenging to find sufficient similar patches in the presence of severe and complex noise.
\input{Fig_parameter_analysis}

\subsubsection{Ablation study}
The effectiveness of the proposed GCP-ID for image denoising lies mainly in two key steps, namely the GCP-guided patch search and RGGB representation. It is interesting to investigate the difference among each step, thus we use the SIDD validation dataset and compare various implementations of GCP-ID in Table \ref{Table_ablation_study}. From the reported results, we notice that the RGGB representation has a higher PSNR value, while the GCP-guided patch search step is superior in terms of the SSIM metric.
\input{Table_ablation_study}

\indent To better understand the observed results, we visualize the denoising performance in Fig. \ref{Fig_Ablation_study_GCP_RGGB}. It can be seen that the GCP-guided search step filters out noise at the cost of more obvious over-smoothness, while the RGGB representation preserves detail and textures but also produces unpleasant artifacts. This interesting observation indicates that the group-level redundancy from the GCP-guided similar patch search weighs highly on the sparsity in the transform domain. Furthermore, the patch level redundancy introduced by the recursive use of RGGB retains structural information, meanwhile, the noise patterns also repeat, leaving obvious color artifacts when the image is severely contaminated. As a result, the combination of these two steps by GCP-ID tends to balance noise removal and detail restoration for better denoising effects.
\input{Fig_Ablation_study_GCP_RGGB}

\vspace{-9pt}
\subsection{Discussion}
\subsubsection{Effectiveness of the CNN estimator}
Our experiments show that the GCP-based CNN noise estimator contributes substantially to the improvements of GCP-ID in many cases. To study the performance of the CNN estimator, we report its prediction accuracy on real-world datasets. Besides, from Fig. \ref{Fig_sensitivity_PSNR_SSIM_highISO}, it is noticed that the proposed GCP-ID denoiser is robust to subtle changes of noise levels, thus deviation within a certain range may be acceptable in practice. We say the estimator is \textit{approximate} if the absolute difference between the predicted label and the true label is less than or equal to 1. From Table \ref{Table_CNN_approximate}, it is observed that the approximate results are much better and achieve an accuracy of over $98\%$ and $83\%$ on raw and color image datasets, respectively. The accuracy for noise prediction of raw data is considerably higher since the nonlinear steps on the camera’s ISP often result in a more complex noise distribution in the sRGB domain \cite{kousha2022modeling}. Such observations are in consistent with results in Tables \ref{Table_Raw_image} and \ref{Table_Color_image}, which also help explain the performance of the proposed GCP-ID and the corresponding CNN estimator.
\input{Table_CNN_approximate}

\indent To further evaluate the proposed CNN estimator, we compare it with a popular efficient noise estimation algorithm proposed by Chen et al. \cite{chen2015efficient}. Table \ref{Table_compare_CNN_and_Chen} shows that under the CNN estimator, the proposed method outperforms the compared algorithm significantly. As illustrated in Fig. \ref{Fig_CNN_compare_Chen}, the CNN estimator is more effective for GCP-ID in terms of both noise removal and detail preservation. This demonstrates the importance and effectiveness of learning a noise estimator with the subsequent denoiser.
\input{Table_compare_CNN_and_Chen}

\input{Fig_CNN_compare_Chen}

\subsubsection{Customization of a denoiser}
Despite the impressive achievements of existing approaches, the lack of high quality clean-noisy image pairs still haunts the effectiveness of both traditional and DNN methods in terms of parameter tuning and model training, resulting in unsatisfactory generalization ability. Therefore, it is interesting to investigate how to modify and customize the proposed method for different cameras without clean reference data. From Fig. \ref{Fig_train_CNN_estimator}, it is noticed that the training of CNN estimator does not rely on noise-free images, thus for a noisy input, we can select from the outputs of GCP-ID the most visually pleasant one to create labels for the CNN. Each labeled noisy input will generate a number of training samples, as it can be divided into many image patches of smaller size. This strategy is less expensive and much more efficient than repeatedly sampling numerous noisy observations for a mean image. Furthermore, it provides a higher degree of flexibility, in that every user may hold a different view on noise removal.\\
\indent To verify the effectiveness of the proposed modification strategy, we use the DND (sRGB) dataset and label 15 images based on the visual effects of GCP-ID outputs with $\sigma \in \{10, 20, 30, 40, 50, 60, 80, 100, 120, 140\}$. The new data are included to retrain the CNN estimator. Table \ref{Table_customize_GCP} and Fig. \ref{Fig_GCP_ID_customize} compare different implementations of GCP-ID. By reducing color artifacts and preserving more structural information, the simple modification strategy brings about improvements in both objective evaluations and visual quality. 
\input{Table_customize_GCP}

\input{Fig_GCP_ID_customize}

\subsubsection{Extension to HSI}
 The effectiveness of GCP on raw and sRGB images motivates us to explore its extension to other imaging techniques, \textit{e.g.}, HSI data. We notice that the GCP is based on the sensitivity of human eyes to medium range of spectrum, resulting in the fact that green light at this wavelength produces the impression of highest `brightness' \cite{Gigahertz}. Therefore, such spectrum-wise prior may be borrowed to guide denoising for HSI. We first evenly divide an HSI input into several groups along the spectral dimension. Spectral bands within the medium spectrum range are used for patch search and nonlocal transform learning according to Eq. (\ref{Equ_summarize_inverse_denoising}). The similar patch indices and learned transforms are propagated to other groups of spectral bands.\\
\indent For performance evaluations, we use the Real-HSI dataset \cite{zhang2021hyperspectral}, including 59 real-world clean-noisy HSI pairs of size $696 \times 520 \times 34$. The overall quantitative results and average running time by competing approaches are reported in Table \ref{Table_Real-HSI_denoising_results}. Interestingly, the proposed method is also able to achieve competitive performance for real HSI noise removal, which demonstrates the importance and effectiveness of leveraging spectrum-wise prior information to guide HSI denoising. Besides, by getting rid of a large number of similar patches and the repetitive calculation of local transforms, GCP-ID is at least 12 times faster than the state-of-the-art low rank tensor algorithms LTDL and OLRT. Fig. \ref{Fig_Real-HSI_visual_evaluation} shows that GCP-ID retains the details of small shapes and effectively removes noise in flat areas. In comparison, BM4D and MAN show unwanted artifacts, as the predefined transforms/pretrained models may be less adaptive to spectrum variations. Furthermore, since the optimal tensor rank is hard to obtain, the representative low rank method OLRT tends to blur textures and suffers from oversmoothness, while LTDL fails to remove stripe noise.
\input{Table_Real-HSI_denoising_results}

\input{Fig_Real-HSI_visual_evaluation}
\vspace{19.8pt}
\section{Conclusion}
In this paper, we present a novel denoising method termed GCP-ID, which leverages channel-wise prior information of image data to guide noise removal. Following the traditional patch-based denoising paradigm, the proposed GCP-ID incorporates the green channel guided search strategy for nonlocal similarity and the RGGB representation for local patches into a unified framework. The GCP-based CNN noise estimator is integrated to enhance adaptiveness as well as boost the denoising performance. We investigate the feasibility of a simple one-step algorithm for different denoising tasks. The qualitative and quantitative experimental results show the effectiveness and efficiency of the proposed method to handle raw and sRGB color image data. Moreover, the flexibility and potential of GCP-ID is demonstrated by a simple modification strategy along with its extension to HSI noise removal. It is interesting to delve deeper into the spectrum prior, which may be better exploited by advanced DNN models for a further enhancement and more applications.

\bibliographystyle{IEEEtran}
\bibliography{IEEEabrv, Reference}

\end{document}

%% file: Table_SNR_comparison.tex
\begin{table}[htbp]
\small
  \centering
  \caption{Average SNR in RGB channels of images from real-world datasets. `C: Canon', `N: Nikon', `F: Fujifilm', `H: Huawei', `S: Sony', `X: Xiaomi'.}
  \vspace{-1.8pt}
  \scalebox{0.588}{
    \begin{tabular}{cccccccccc}
    \toprule
    \multirow{2}[3]{*}{Channel} & \multirow{2}[3]{*}{CC15} & \multirow{2}[3]{*}{HighISO} & \multicolumn{7}{c}{IOCI} \\
\cmidrule{4-10}          &       &       & C-600D & F-X100T & H-honor6X & N-D5300 & S-A6500 & X-8   & C-5DMark4 \\
    \midrule
    R     & 24.5  & 29.3  & 30.5  & 29.8  & 30.2  & 28.3  & 31.0  & 26.9  & 33.7  \\
    \midrule
    G     & \textbf{27.3} & \textbf{29.4} & \textbf{32.9} & \textbf{29.9} & \textbf{31.2} & \textbf{29.4} & \textbf{33.0} & \textbf{27.5} & \textbf{33.9} \\
    \midrule
    B     & 21.3  & 27.2  & 29.6  & 28.9  & 30.3  & 25.5  & 32.8  & 26.3  & 32.6  \\
    \bottomrule
    \end{tabular}%
  \label{Table_SNR_comparison}}%
  \vspace{-13.8pt}
\end{table}%

%% file: Fig_SNR_comparison.tex
\begin{figure}[htbp]
\centering
\graphicspath{{Figs/others/}}
  \subfloat[CC15]{\includegraphics[width= 1.669in, height = 1.399in]{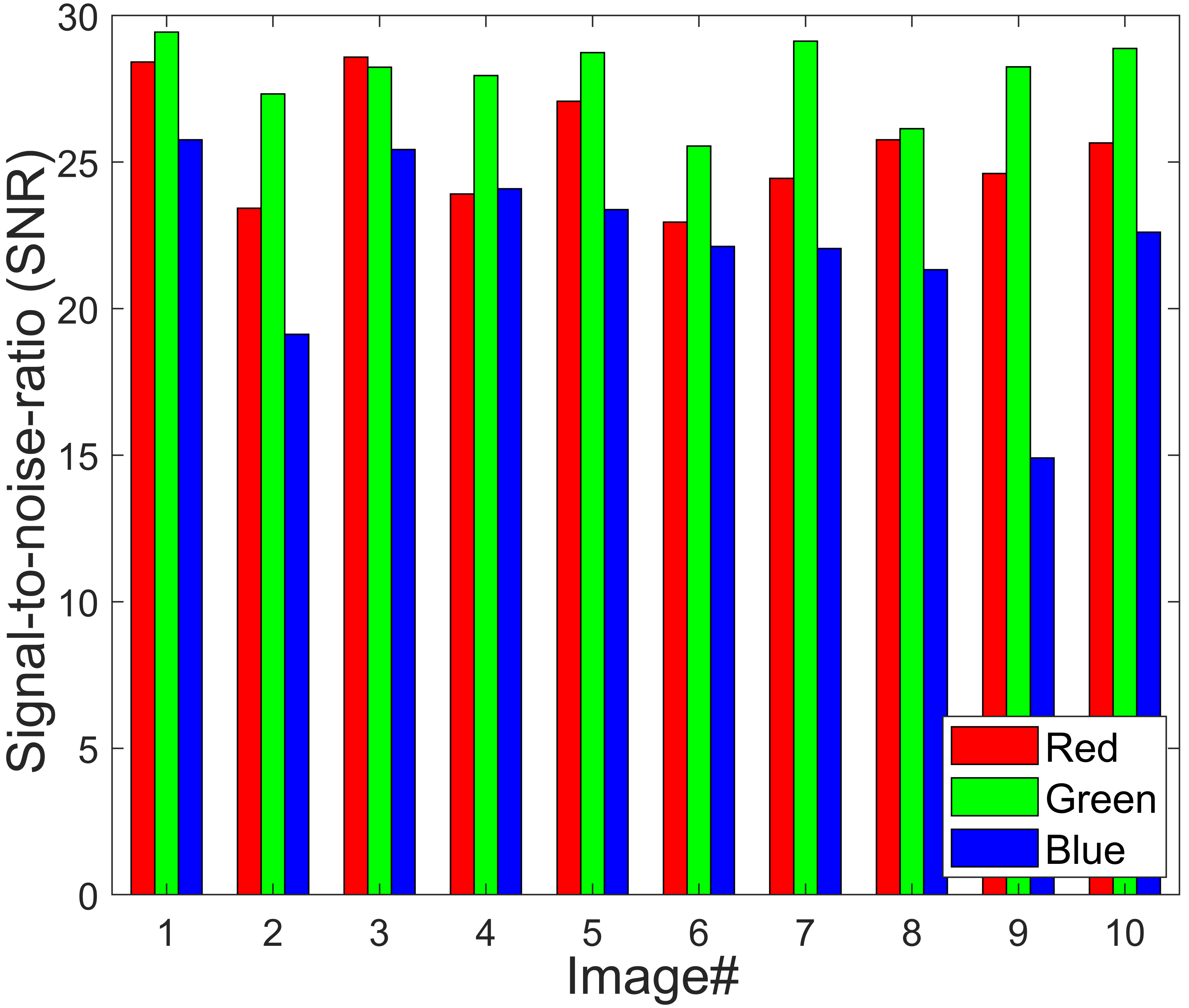}} \, \,
  \subfloat[HighISO]{\includegraphics[width= 1.669in, height = 1.399in]{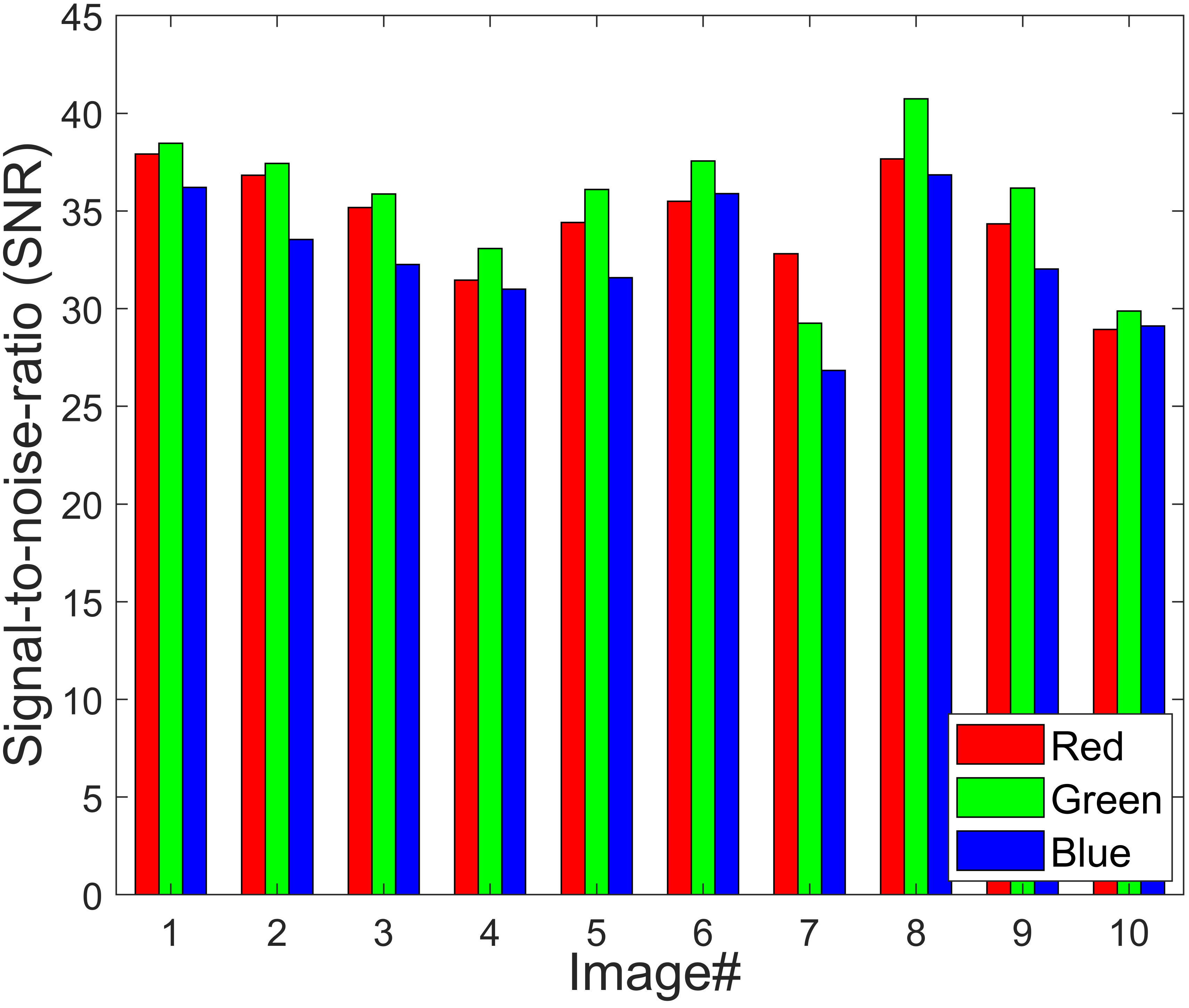}}  
  \caption{SNR in RGB channels of real-world images.}
  \label{Fig_SNR_comparison}
  \vspace{-9pt}
\end{figure}

%% file: Figs_flowchart.tex
\begin{figure*}[!htb]
  \vspace{-12pt}
  \centering
  \graphicspath{{Figs/}}
  \includegraphics[width=6.999in, height=1.759in]{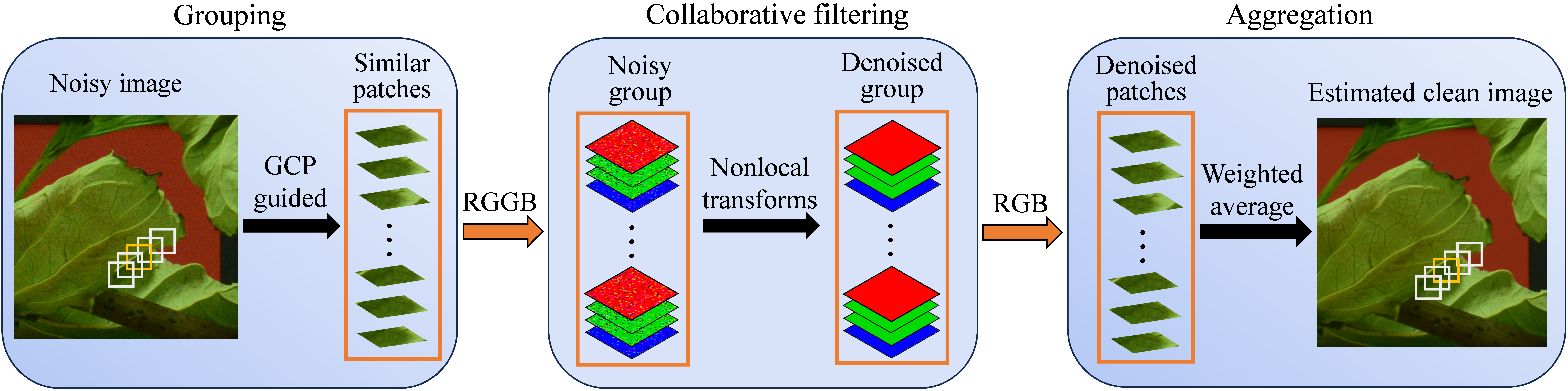}\\
  \vspace{-1.6pt}
  \caption{Overview of the proposed GCP-ID method for image denoising.}
  \label{Figs_flowchart}
  \vspace{-10.8pt}
\end{figure*}

%% file: Table_patch_search_successful_rate.tex
\begin{table}[!htb]
\vspace{-1.8pt}
\small
  \centering
  \caption{Average patch search successful rate of different schemes on the CC15 dataset.}
  \vspace{-3.8pt}
  \renewcommand{\arraystretch}{0.6168}
  \scalebox{0.8518}{
    \begin{tabular}{ccccc}
    \toprule
    Search scheme & Green channel \, & \, YCbCr & \, Opponent  & \, GCP-guided \\
    \midrule
    Successful rate & 0.488 & 0.499 & 0.493 & \textbf{0.504} \\
    \bottomrule
    \end{tabular}}%
  \label{Table_patch_search_successful_rate}%
  \vspace{-3.6pt}
\end{table}%

%% file: Fig_Bayer_pattern_RGGB.tex

\begin{figure}[htbp]
\centering
\graphicspath{{Figs/Demo/}}
  \subfloat[Bayer CFA pattern of raw data]{\includegraphics[width=2.969in, height=0.719in]{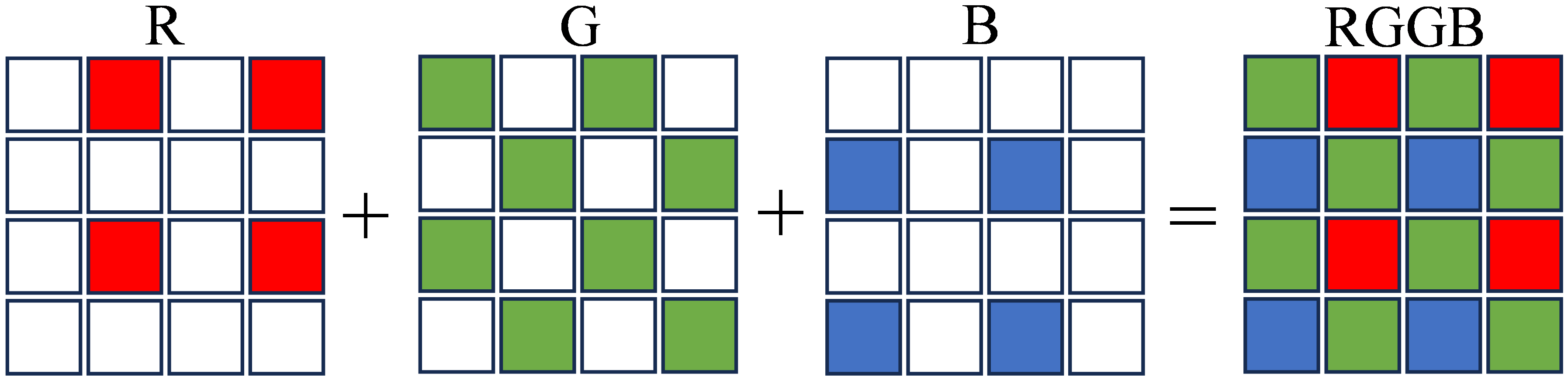}} \\
  \vspace{-5.19pt}
  \subfloat[RGGB and block circulant representation of an RGB input]{\includegraphics[width= 3.169in, height=0.816in]{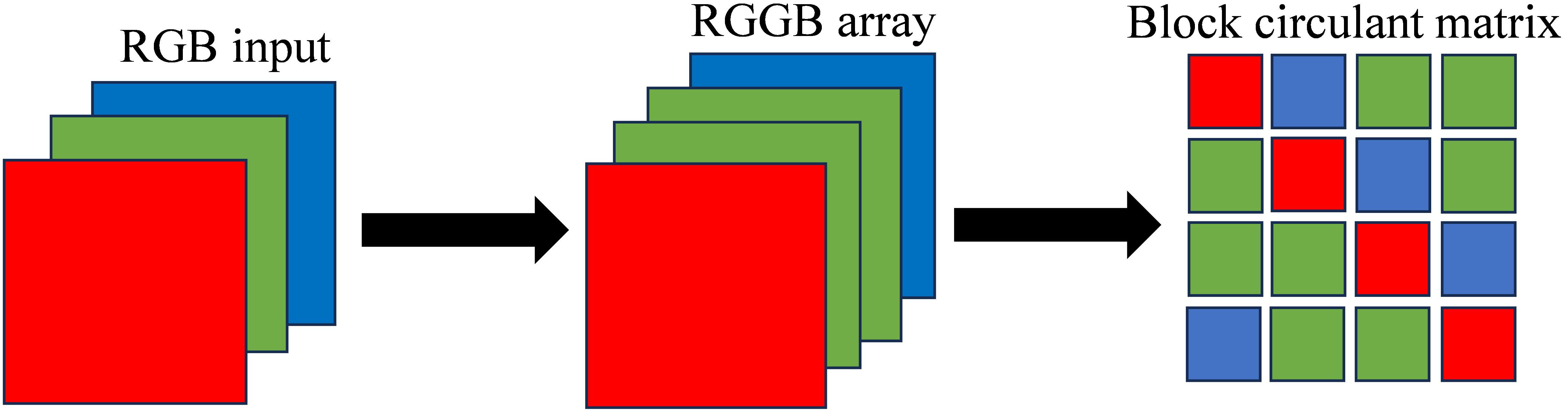}}  
  \vspace{-2.8pt}
  \caption{The Bayer CFA pattern and RGGB representation.}
  \label{Fig_Bayer_pattern_RGGB}
  \vspace{-19.8pt}
\end{figure}

%% file: Algorithm_GCP-ID.tex
\setlength{\textfloatsep}{1pt}
\begin{algorithm}[!htbp]
    \caption{GCP-ID}
    {\bf Input:} Noisy raw or color image $\mathcal{Y}$, patch size $ps$, number of similar patches $K$, search window size $W$ and noise level $\sigma$.\\
    {\bf Output:} Estimated clean image $\mathcal{X}$.\\
    {\bf Step 1} (Grouping): For each reference patch, stack its $K$ similar patches within $W$ according to Eq. (\ref{Equ_distance_calculation}) in a group $\hat{\mathcal{G}}_{noisy} \in \mathbb{R}^{ps\times ps \times 4 \times K}$ based on the RGGB representation.\\
    {\bf Step 2} (Collaborative filtering):\\
     \hspace*{0.18in}(1) Perform the forward nonlocal t-SVD based on Eq. (\ref{Equ_summarize_denoising}) to obtain the coefficient $\hat{\mathcal{S}}_{noisy}$, projection tensors $\mathcal{U}_{row}$, $\mathcal{V}_{row}$ and group-level transform $\mathbf{U}_{group}$.\\
     \hspace*{0.18in}(2) Truncate $\hat{\mathcal{S}}_{noisy}$ via hard-thresholding in Eq. (\ref{Equ_hard_thresholding}).\\
     \hspace*{0.18in}(3) Apply the inverse nonlocal t-SVD transform in Eq. (\ref{Equ_summarize_inverse_denoising}) to obtain estimated clean group $\hat{\mathcal{G}}_{clean}$.\\
    {\bf Step 3} (Aggregation): Fetch and averagely write all denoised image patches of $\hat{\mathcal{G}}_{clean}$ to their original locations.
    \label{Algorithm_GCP-ID}
\end{algorithm}

%% file: Fig_train_CNN_estimator.tex
\begin{figure}[!htb]
  \vspace{-3.9pt}
  \centering
  \graphicspath{{Figs/others/}}
  \includegraphics[width=3.518in, height=0.9816in]{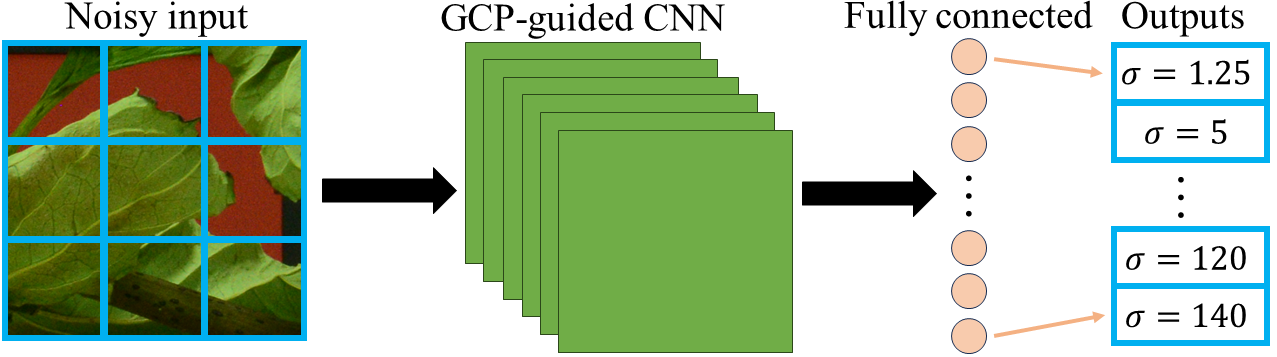}\\
  \vspace{-3pt}
  \caption{Flowchart of the GCP-based CNN noise estimator.}
  \label{Fig_train_CNN_estimator}
  \vspace{-2.8pt}
\end{figure} 

%% file: Fig_sensitivity_visual_effects.tex
\begin{figure}[htbp]
\vspace{-8.6pt}
\centering
\graphicspath{{Figs/others/}}
  \subfloat[Noisy]{\includegraphics[width=0.812in, height=0.812in]{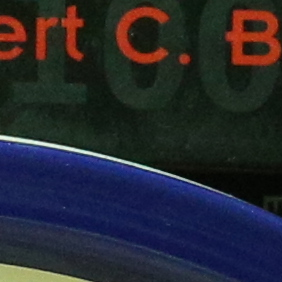}} \,
  \subfloat[$\sigma = 15$]{\includegraphics[width=0.812in]{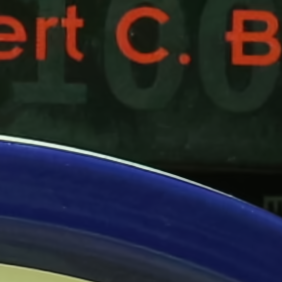}} \, 
  \subfloat[$\sigma = 20$]{\includegraphics[width=0.812in]{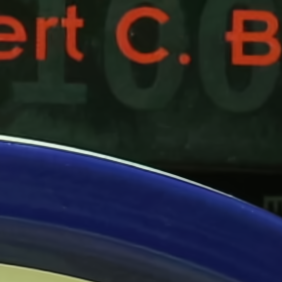}} \,
  \subfloat[$\sigma = 25$]{\includegraphics[width=0.812in]{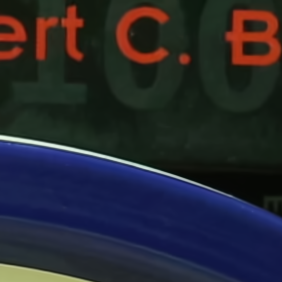}} \,
  \vspace{-2.8pt}
  \caption{Illustration of GCP-ID with different $\sigma$.}
  \vspace{-6.8pt}
  \label{Fig_sensitivity_visual_effects}
  \vspace{-10.8pt}
\end{figure}

%% file: Fig_sensitivity_PSNR_SSIM_highISO.tex
\begin{figure}[htbp]
\centering
\graphicspath{{Figs/others/}}
  \subfloat[PSNR]{\includegraphics[width=1.699in, height=1.398in]{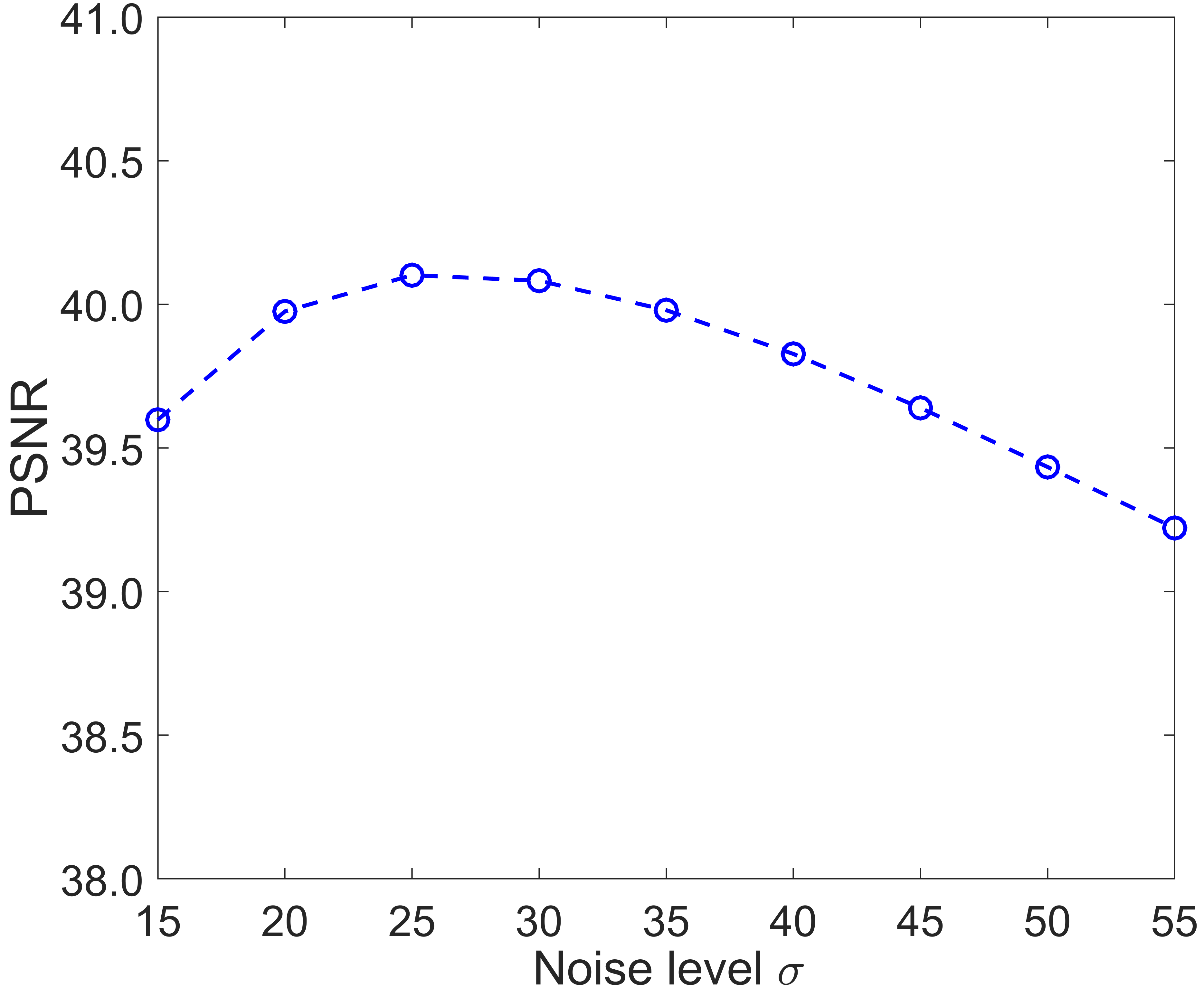}}
  \vspace{-2.6pt}
  \subfloat[SSIM]{\includegraphics[width=1.699in, height=1.398in]{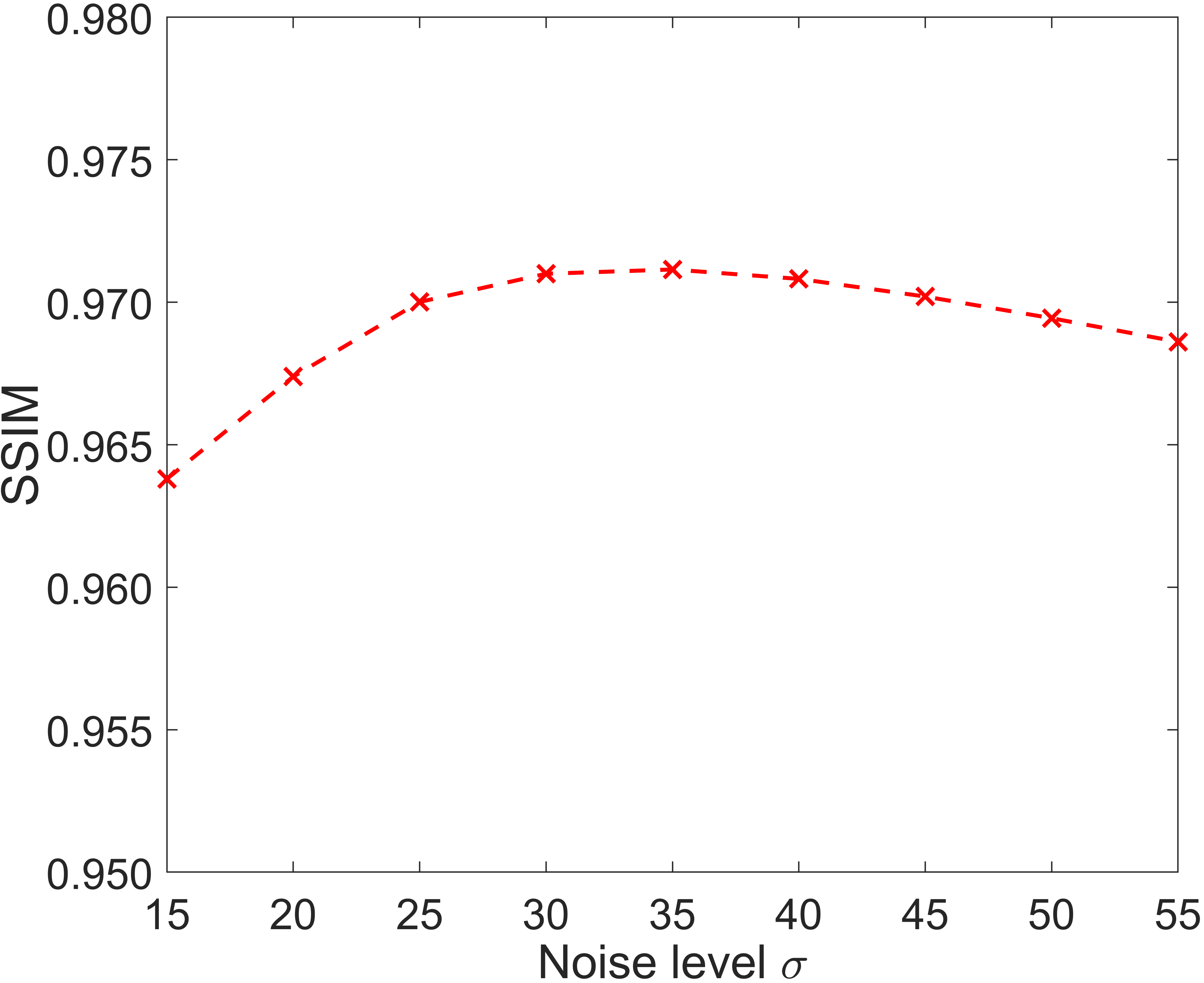}}  
  \vspace{-1.6pt}
  \caption{Impact of $\sigma$ on GCP-ID with the HighISO dataset.}
  \label{Fig_sensitivity_PSNR_SSIM_highISO}
\end{figure}

%% file: Table_dataset.tex
\begin{table}[htbp]
\vspace{-2.9pt}
  \centering
  \renewcommand{\arraystretch}{1.5}
  \caption{Real-world datasets for different denoising tasks.}
  \vspace{-1.6pt}
  \scalebox{0.908}{
    \begin{tabular}{|c|c|}
    \hline
    Applications & Real-world datasets \\
    \hline
    Raw image & SIDD \cite{abdelhamed2018high}, DND \cite{plotz2017benchmarking} \\
    \hline
    Raw video & CRVD indoor \cite{yue2020supervised}, CRVD outdoor \cite{yue2020supervised} \\
    \hline
    Color image & SIDD, DND, CC15 \cite{nam2016holistic}, PolyU \cite{xu2018real}, HighISO \cite{yue2019high}, IOCI \cite{kong2023comparison} \\
    \hline
    Color video & CRVD \cite{yue2020supervised}, IOCV \cite{kong2023comparison} \\
    \hline
    \end{tabular}}%
  \label{Table_dataset}%
  \vspace{-5.16pt}
\end{table}%

%% file: Table_Raw_image.tex
\begin{table}[htbp]
\vspace{-2pt}
  \centering
  \caption{Raw image denoising results on SIDD and DND. `*': the results are from the authors' papers.}
  \renewcommand{\arraystretch}{0.818}
  \scalebox{0.9018}{
    \begin{tabular}{ccccc}
    \toprule
    \multicolumn{2}{c}{Methods} & SIDD val & SIDD test & DND \\
    \midrule
    \multicolumn{2}{c}{\# Images} & 1280  & 1280  & 1000 \\
    \midrule
    \multirow{6}[11]{*}{\shortstack[c]{Traditional \\ denoisers}}& BM3D \cite{dabov2007image}  & 44.4/0.964 & 45.5/0.980 & 46.6/0.972 \\
\cmidrule{2-5}          & Bitonic \cite{treece2022real} & 44.2/0.978 & - & 46.6/0.971 \\
\cmidrule{2-5}          & KSVD \cite{elad2006image} & -     & 43.3/0.969 & 45.5/0.968 \\
\cmidrule{2-5}          & WNNM \cite{gu2014weighted} & 44.3/0.969 & 44.9/0.975 & 46.3/0.971 \\
\cmidrule{2-5}          & GCP-ID & 47.9/0.981 & 47.5/0.980 & 47.7/0.977 \\
\cmidrule{2-5}          & GCP-ID + CNN & \textcolor{blue}{\textbf{51.2/0.991}} & \textcolor{blue}{\textbf{50.5/0.991}} & \textcolor{blue}{\textbf{48.1/0.978}} \\
\midrule
\midrule
    \multirow{6}[10]{*}{\shortstack[c]{DNN \\ methods}} & Blind2Unblind \cite{wang2022blind2unblind} & 51.4/0.991 & 50.8/0.991 & 48.2/0.980 \\
\cmidrule{2-5}          & CycleISP \cite{zamir2020cycleisp} & \textbf{52.5/0.993} & \textbf{52.4*/0.993*} & \textbf{49.1/0.983} \\
\cmidrule{2-5}          & FBI \cite{byun2021fbi} & 51.1/0.991 & 50.6/0.991 & 48.1/0.979 \\
\cmidrule{2-5}          & MCU-Net \cite{bao2020real} & -     & 48.8*/0.990* & - \\
\cmidrule{2-5}          & TNRD \cite{chen2016trainable} & -     & 42.8/0.945 & 44.9/0.962 \\
\cmidrule{2-5}          & UPI \cite{brooks2019unprocessing} & -     & -     & 48.9*/0.982* \\
\toprule
    \end{tabular}}%
  \label{Table_Raw_image}%
  \vspace{-1.8pt}
\end{table}%

%% file: Table_Color_image.tex
\begin{table*}[htbp]
\footnotesize
  \centering
  \caption{Color image denoising results of compared methods on real-world datasets. The best results are bolded.}
  \renewcommand{\arraystretch}{0.568}
  \scalebox{0.628}{
    \begin{tabular}{ccccccccccccccc}
    \toprule
    \multicolumn{2}{c}{\multirow{2}[2]{*}{Methods/Models}} & \multirow{2}[2]{*}{DnD} & \multirow{2}[2]{*}{SIDD} & \multirow{2}[2]{*}{CC15} & \multirow{2}[2]{*}{PolyU} & \multirow{2}[2]{*}{HighISO} & \multicolumn{8}{c}{IOCI} \\
\cmidrule{8-15}    \multicolumn{2}{c}{} &       &       &       &       &       & CANON 600D & FUJIFilm X100T & HUAWEI honor6X & IPHONE13 & NIKON D5300 & OPPO R11s & SONY A6500 & XIAOMI8 \\
    \midrule
    \multicolumn{2}{c}{\# Images} & 1000  & 1280  & 15    & 100   & 100   & 25    & 71    & 30    & 174   & 56    & 39    & 36    & 50 \\
    \midrule
    \multirow{20}[39]{*}{\shortstack[c]{Traditional \\ denoisers}} & \multirow{2}[2]{*}{NLHCC\cite{hou2020nlh}} & 38.85  & -     & \textbf{38.49} & 38.36  & 40.29  & 42.72  & 42.80  & 38.84  & 41.13  & 43.25  & -     & \textcolor[rgb]{0,  0,  1}{\textbf{46.02}} & 35.73  \\
\cmidrule{3-15}          &       & 0.953  & -     & \textbf{0.965} & 0.965  & 0.971  & 0.984  & 0.978  & 0.959  & 0.976  & 0.981  & -     & \textcolor[rgb]{0,  0,  1}{\textbf{0.991}} & 0.955  \\
\cmidrule{2-15}          & \multirow{2}[2]{*}{Bitonic \cite{treece2022real}} & 37.85  & 36.67  & 35.22  & 36.64  & 37.37  & 39.63  & 41.05  & 37.71  & 39.09  & 39.22  & 38.87  & 43.25  & 34.92  \\
\cmidrule{3-15}          &       & 0.936  & 0.933  & 0.925  & 0.940  & 0.943  & 0.952  & 0.964  & 0.940  & 0.952  & 0.954  & 0.959  & 0.979  & 0.941  \\
\cmidrule{2-15}          & \multirow{2}[2]{*}{LLRT \cite{chang2017hyper}} & 35.45  & 30.74  & 37.77  & 38.28  & 39.59  & 42.24  & 42.22  & 37.91  & 42.02  & 41.76  & 38.69  & 45.17  & 35.71  \\
\cmidrule{3-15}          &       & 0.897  & 0.766  & 0.957  & 0.970  & 0.972  & 0.983  & 0.975  & 0.969  & 0.984  & 0.979  & 0.972  & 0.989  & 0.962  \\
\cmidrule{2-15}          & \multirow{2}[2]{*}{MCWNNM \cite{xu2017multi}} & 37.38  & 29.54  & 37.02  & 38.26  & 39.89  & 42.07  & 42.48  & 39.46  & 41.33  & 41.74  & 40.71  & 45.38  & 35.84  \\
\cmidrule{3-15}          &       & 0.929  & 0.888  & 0.950  & 0.965  & 0.970  & 0.979  & 0.976  & 0.961  & 0.976  & 0.975  & 0.973  & 0.990  & 0.952  \\
\cmidrule{2-15}          & \multirow{2}[2]{*}{TWSC \cite{xu2018trilateral}} & 37.96  & -     & 37.90  & 38.62  & \textbf{40.62} & 42.52  & 42.26  & 38.71  & 41.71  & 42.23  & 40.65  & 45.49  & 35.40  \\
\cmidrule{3-15}          &       & 0.942  & -     & 0.959  & 0.967  & \textbf{0.975} & 0.982  & 0.973  & 0.945  & 0.980  & 0.975  & 0.972  & 0.990  & 0.939  \\
\cmidrule{2-15}          & \multirow{2}[2]{*}{CBM3D \cite{dabov2007image}} & 37.73  & 34.74  & 37.70  & 38.69  & 40.35  & 42.54  & 42.65  & 39.97  & 42.03  & 42.20  & 40.75  & 45.72  & 36.38  \\
\cmidrule{3-15}          &       & 0.934  & 0.922  & 0.957  & 0.970  & 0.974  & 0.984  & 0.977  & 0.967  & 0.983  & 0.979  & 0.974  & 0.990  & 0.961  \\
\cmidrule{2-15}          & \multirow{2}[2]{*}{4DHOSVD\cite{rajwade2012image}} & 37.58  & 34.49  & 37.52  & 38.54  & 40.27  & 42.19  & 42.60  & 39.82  & 41.75  & 41.82  & 40.71  & 45.58  & 36.27  \\
\cmidrule{3-15}          &       & 0.929  & 0.911  & 0.956  & 0.968  & 0.973  & 0.982  & 0.976  & 0.966  & 0.980  & 0.977  & 0.974  & 0.990  & 0.961  \\
\cmidrule{2-15}          & \multirow{2}[2]{*}{MStSVD \cite{kong2019color}} & 38.01  & 34.38  & 37.95  & 38.85  & 40.49  & 42.75  & \textcolor[rgb]{0,  0,  1}{\textbf{42.68}} & 40.08  & \textcolor[rgb]{0,  0,  1}{\textbf{42.05}} & 42.72  & 40.87  & 45.91  & 36.40  \\
\cmidrule{3-15}          &       & 0.938  & 0.901  & 0.959  & 0.971  & 0.974  & 0.984  & \textcolor[rgb]{0,  0,  1}{\textbf{0.977}} & 0.967  & \textcolor[rgb]{0,  0,  1}{\textbf{0.982}} & 0.980  & 0.974  & 0.991  & 0.962  \\
\cmidrule{2-15}          & \multirow{2}[2]{*}{GCP-ID} & 38.15  & 35.03  & 38.30  & \textbf{38.90} & \textcolor[rgb]{0,  0,  1}{\textbf{40.56}} & \textcolor[rgb]{0,  0,  1}{\textbf{43.02}} & \textbf{42.70} & \textcolor[rgb]{0,  0,  1}{\textbf{40.18}} & \textbf{42.05} & \textbf{42.98} & \textbf{40.96} & \textbf{46.06} & \textcolor[rgb]{0,  0,  1}{\textbf{36.48}} \\
\cmidrule{3-15}          &       & 0.939  & 0.915  & 0.962  & \textbf{0.971} & \textcolor[rgb]{0,  0,  1}{\textbf{0.974}} & \textcolor[rgb]{0,  0,  1}{\textbf{0.985}} & \textbf{0.977} & \textcolor[rgb]{0,  0,  1}{\textbf{0.968}} & \textbf{0.982} & \textbf{0.981} & \textbf{0.975} & \textbf{0.991} & \textcolor[rgb]{0,  0,  1}{\textbf{0.962}} \\
\cmidrule{2-15}          & \multirow{2}[2]{*}{GCP-ID + CNN} & 38.36  & 35.56  & 38.25  & \textcolor[rgb]{0,  0,  1}{\textbf{38.86}} & 40.45  & 42.84  & 42.58  & \textbf{40.23} & 42.00  & \textcolor[rgb]{0,  0,  1}{\textbf{42.88}} & \textcolor[rgb]{0,  0,  1}{\textbf{40.92}} & 45.59  & \textbf{36.51} \\
\cmidrule{3-15}          &       & 0.943  & 0.926  & 0.962  & \textcolor[rgb]{0,  0,  1}{\textbf{0.971}} & 0.972  & 0.985  & 0.977  & \textbf{0.969} & 0.982  & \textcolor[rgb]{0,  0,  1}{\textbf{0.981}} & \textcolor[rgb]{0,  0,  1}{\textbf{0.974}} & 0.989  & \textbf{0.963} \\
    \midrule
    \midrule
    \multirow{12}[26]{*}{\shortstack[c]{DNN methods \\ (self-supervised)}} & \multirow{2}[2]{*}{AP-BSN \cite{lee2022ap}} & 37.29  & 35.97  & 35.44  & 36.99  & 38.26  & 41.29  & 41.40  & 37.87  & 40.83  & 40.86  & 39.55  & 43.73  & 34.37  \\
\cmidrule{3-15}          &       & 0.932  & 0.925  & 0.936  & 0.956  & 0.965  & 0.979  & 0.970  & 0.949  & 0.977  & 0.973  & 0.969  & 0.983  & 0.937  \\
\cmidrule{2-15}          & \multirow{2}[2]{*}{Blind2Unblind \cite{wang2022blind2unblind}} & -     & -     & 36.51  & 38.25  & 39.03  & 41.72  & 41.17  & 39.35  & 40.88  & 40.90  & 39.76  & 43.78  & 35.75  \\
\cmidrule{3-15}          &       & -     & -     & 0.935  & 0.968  & 0.962  & 0.981  & 0.970  & 0.967  & 0.976  & 0.970  & 0.971  & 0.986  & 0.956  \\
\cmidrule{2-15}          & \multirow{2}[2]{*}{C2N \cite{jang2021c2n}}  & 37.28  & -     & 37.02  & 37.69  & 38.86  & 41.95  & 41.33  & 38.73  & 40.50  & 40.78  & 40.05  & 44.64  & 35.53  \\
\cmidrule{3-15}          &       & 0.924  & -     & 0.945  & 0.958  & 0.960  & 0.980  & 0.966  & 0.954  & 0.977  & 0.964  & 0.970  & 0.987  & 0.952  \\
\cmidrule{2-15}          & \multirow{2}[2]{*}{Neigh2Neigh \cite{huang2021neighbor2neighbor}} & -     & -     & 34.47  & 37.10  & 36.59  & 40.41  & 41.24  & 38.13  & 39.63  & 38.05  & 39.17  & 44.49  & 35.30  \\
\cmidrule{3-15}          &       & -     & -     & 0.883  & 0.942  & 0.908  & 0.960  & 0.958  & 0.940  & 0.952  & 0.919  & 0.958  & 0.984  & 0.944  \\
\cmidrule{2-15}          & \multirow{2}[2]{*}{SASL \cite{li2023spatially}} & 38.00  & -     & 34.93  & 37.13  & 38.24  & 41.86  & 40.29  & 37.22  & 40.85  & 41.64  & 39.10  & 42.89  & 34.25  \\
\cmidrule{3-15}          &       & 0.9364  & -     & 0.9356  & 0.9540  & 0.9641  & 0.9815  & 0.9632  & 0.9473  & 0.9768  & 0.9762  & 0.9650  & 0.9796  & 0.9298  \\
\cmidrule{2-15}          & \multirow{2}[2]{*}{Self2Self \cite{quan2020self2self}} & -     & -     & 36.26  & -     & 39.49  & -     & -     & -     & -     & -     & -     & -     & - \\
\cmidrule{3-15}          &       & -     & -     & 0.947  & -     & 0.963  & -     & -     & -     & -     & -     & -     & -     & - \\
    \midrule
    \midrule
    \multirow{30}[58]{*}{\shortstack[c]{DNN methods \\ (supervised)}} & \multirow{2}[2]{*}{AINDNet \cite{Kim_2020_CVPR}} & 39.77  & 39.08  & 36.14  & 37.33  & 38.00  & 39.33  & 38.50  & 36.53  & 37.27  & 38.11  & 37.44  & 40.17  & 34.65  \\
\cmidrule{3-15}          &       & 0.959  & 0.953  & 0.935  & 0.954  & 0.946  & 0.976  & 0.966  & 0.954  & 0.964  & 0.961  & 0.968  & 0.981  & 0.953  \\
\cmidrule{2-15}          & \multirow{2}[2]{*}{CBDNet \cite{guo2019toward}} & 38.06  & 33.26  & 36.20  & 37.81  & 38.18  & 42.41  & 41.88  & 38.35  & 40.63  & 40.92  & 39.54  & 44.38  & 35.54  \\
\cmidrule{3-15}          &       & 0.942  & 0.869  & 0.919  & 0.956  & 0.942  & 0.981  & 0.971  & 0.946  & 0.968  & 0.963  & 0.965  & 0.981  & 0.950  \\
\cmidrule{2-15}          & \multirow{2}[2]{*}{CycleISP \cite{zamir2020cycleisp}} & 39.57  & 39.42  & 35.40  & 37.61  & 37.70  & 41.84  & 41.59  & 38.47  & 40.61  & 40.03  & 39.86  & 44.75  & 35.54  \\
\cmidrule{3-15}          &       & 0.955  & 0.956  & 0.916  & 0.955  & 0.936  & 0.977  & 0.969  & 0.952  & 0.970  & 0.953  & 0.968  & 0.987  & 0.949  \\
\cmidrule{2-15}          & \multirow{2}[2]{*}{DeamNet \cite{ren2021adaptive}} & 39.63  & 39.35  & 36.63  & 37.70  & 36.93  & 40.86  & 39.72  & 33.67  & 40.25  & 38.92  & 39.56  & 40.52  & 34.61  \\
\cmidrule{3-15}          &       & 0.953  & 0.955  & 0.936  & 0.958  & 0.944  & 0.978  & 0.967  & 0.953  & 0.966  & 0.957  & 0.969  & 0.980  & 0.952  \\
\cmidrule{2-15}          & \multirow{2}[2]{*}{DIDN \cite{yu2019deep}} & 39.64  & 39.78  & 36.06  & 37.36  & 38.24  & 41.68  & 41.17  & 38.15  & 40.03  & 40.32  & 39.73  & 44.07  & 35.28  \\
\cmidrule{3-15}          &       & 0.953  & 0.958  & 0.946  & 0.953  & 0.950  & 0.977  & 0.966  & 0.951  & 0.963  & 0.959  & 0.966  & 0.985  & 0.949  \\
\cmidrule{2-15}          & \multirow{2}[2]{*}{DnCNN \cite{zhang2017beyond}} & 37.90  & 37.73  & 37.47  & 38.51  & 40.01  & 41.91  & 41.57  & 39.92  & 42.03  & 42.12  & 40.26  & 43.97  & 36.30  \\
\cmidrule{3-15}          &       & 0.943  & 0.941  & 0.954  & 0.966  & 0.971  & 0.979  & 0.969  & 0.962  & 0.983  & 0.977  & 0.970  & 0.984  & 0.960  \\
\cmidrule{2-15}          & \multirow{2}[2]{*}{FFDNet \cite{zhang2018ffdnet}} & 37.61  & 38.27  & 37.67  & 38.76  & 40.28  & 42.55  & 42.44  & 40.05  & 42.43  & 42.44  & 40.75  & 45.71  & 36.47  \\
\cmidrule{3-15}          &       & 0.942  & 0.948  & 0.956  & 0.970  & 0.973  & 0.982  & 0.976  & 0.967  & 0.984  & 0.979  & 0.973  & 0.990  & 0.961  \\
\cmidrule{2-15}          & \multirow{2}[2]{*}{InvDN \cite{liu2021invertible}} & 39.57  & 39.28  & 34.55  & 35.94  & 38.01  & 40.09  & 40.02  & 33.32  & 40.06  & 40.98  & 39.10  & 40.74  & 34.15  \\
\cmidrule{3-15}          &       & 0.952  & 0.955  & 0.937  & 0.947  & 0.952  & 0.973  & 0.960  & 0.930  & 0.966  & 0.971  & 0.964  & 0.969  & 0.935  \\
\cmidrule{2-15}          & \multirow{2}[2]{*}{NAFNet \cite{chen2022simple}} & 38.36  & \textbf{40.15} & 34.39  & 36.38  & 37.88  & 40.23  & 40.01  & 36.13  & 36.53  & 39.93  & 39.32  & 40.45  & 34.82  \\
\cmidrule{3-15}          &       & 0.943  & \textbf{0.960} & 0.923  & 0.947  & 0.954  & 0.943  & 0.962  & 0.919  & 0.884  & 0.951  & 0.966  & 0.924  & 0.948  \\
\cmidrule{2-15}          & \multirow{2}[2]{*}{PNGAN \cite{cai2021learning}} & 39.38  & \textcolor[rgb]{0,  0,  1}{\textbf{39.81}} & \textcolor[rgb]{0,  0,  1}{\textbf{38.44}} & -     & 40.51  & \textbf{43.13} & 42.55  & 39.84  & 41.85  & 43.15  & 40.88  & 45.32  & 36.40  \\
\cmidrule{3-15}          &       & 0.953  & \textcolor[rgb]{0,  0,  1}{\textbf{0.959}} & \textcolor[rgb]{0,  0,  1}{\textbf{0.963}} & -     & 0.974  & \textbf{0.985} & 0.977  & 0.964  & 0.981  & 0.982  & 0.974  & 0.988  & 0.962  \\
\cmidrule{2-15}          & \multirow{2}[2]{*}{Restormer \cite{zamir2022restormer}} & \textbf{40.03} & 39.79  & 36.33  & 37.66  & 38.29  & 41.84  & 41.47  & 38.42  & 40.13  & 40.53  & 39.56  & 44.19  & 35.65  \\
\cmidrule{3-15}          &       & \textbf{0.956} & 0.959  & 0.941  & 0.956  & 0.948  & 0.979  & 0.968  & 0.952  & 0.963  & 0.962  & 0.963  & 0.986  & 0.953  \\
\cmidrule{2-15}          & \multirow{2}[2]{*}{SCUNet \cite{zhang2023practical}} & -     & -     & 37.44  & 38.08  & 39.47  & 42.24  & 41.93  & 40.74  & 41.79  & 42.61  & 40.64  & 44.24  & 36.13  \\
\cmidrule{3-15}          &       & -     & -     & 0.958  & 0.962  & 0.971  & 0.980  & 0.966  & 0.974  & 0.970  & 0.978  & 0.973  & 0.981  & 0.956  \\
\cmidrule{2-15}          & \multirow{2}[2]{*}{VDIR \cite{soh2022variational}} & -     & 39.26  & 34.90  & 37.01  & 36.84  & 41.40  & 40.81  & 37.65  & 39.20  & 39.41  & 38.98  & 44.40  & 35.04  \\
\cmidrule{3-15}          &       & -     & 0.955  & 0.894  & 0.940  & 0.911  & 0.970  & 0.953  & 0.936  & 0.946  & 0.938  & 0.956  & 0.985  & 0.939  \\
    \bottomrule
    \end{tabular}}%
  \label{Table_Color_image}%
  \vspace{-16pt}
\end{table*}%

%% file: Fig_nonlocal_denoiser_comparison.tex

\begin{figure}[htbp]
\vspace{2pt}
\centering
\graphicspath{{Figs/Compare_nonlocal_denoisers/new/Sample1/}}
  \subfloat{\includegraphics[width=0.8in, height=0.8in]{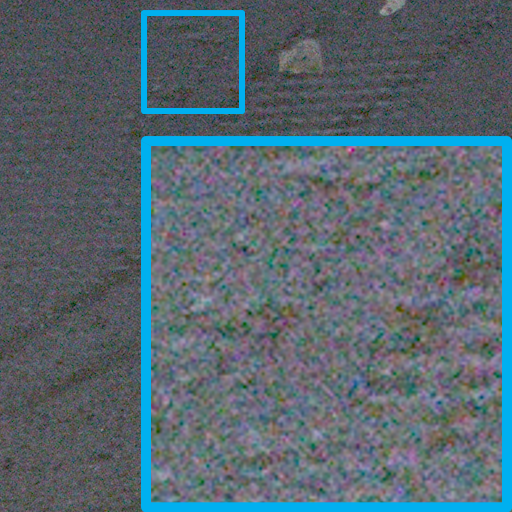}} \,
  \subfloat{\includegraphics[width=0.8in, height=0.8in]{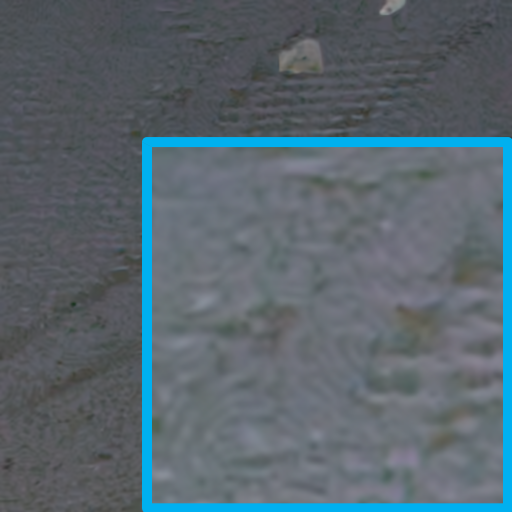}} \, 
  \subfloat{\includegraphics[width=0.8in, height=0.8in]{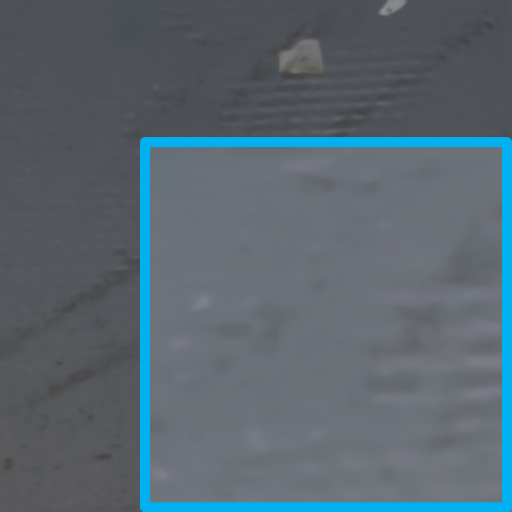}} \,
  \subfloat{\includegraphics[width=0.8in, height=0.8in]{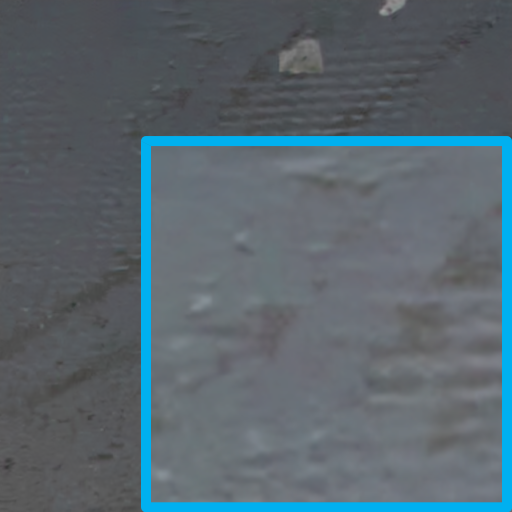}} \\
  \vspace{-6pt}
  \setcounter{subfigure}{0}
\graphicspath{{Figs/Compare_nonlocal_denoisers/new/Sample2/}}
  \subfloat[Noisy]{\includegraphics[width=0.8in, height=0.8in]{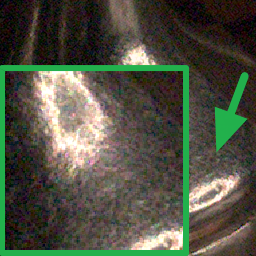}} \,
  \subfloat[CBM3D]{\includegraphics[width=0.8in, height=0.8in]{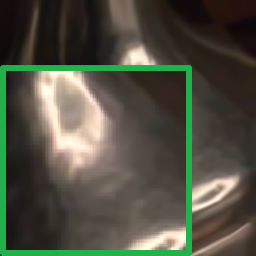}} \, 
  \subfloat[NLHCC]{\includegraphics[width=0.8in, height=0.8in]{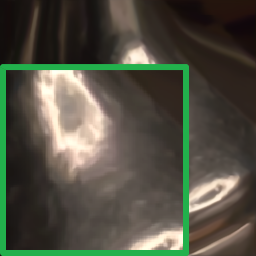}} \,
  \subfloat[GCP-ID]{\includegraphics[width=0.8in, height=0.8in]{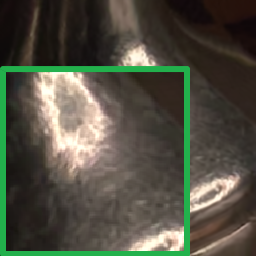}}
  \vspace{-2.9pt}
  \caption{Comparison of representative classic denoisers.}
  \label{Fig_nonlocal_denoiser_comparison}
  \vspace{-1.8pt}
\end{figure}

%% file: Fig_CC15_visual_evaluation.tex
\begin{figure*}[tb]
\vspace{-10pt}
\centering
\graphicspath{{Figs/CC15/combined_new/}}
  \subfloat[Mean]{\includegraphics[width=1.109in, height = 1.109in]{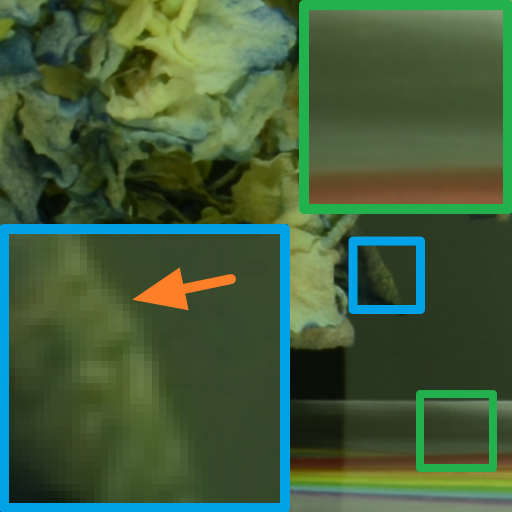}} \,
  \subfloat[Noisy]{\includegraphics[width=1.109in, height = 1.109in]{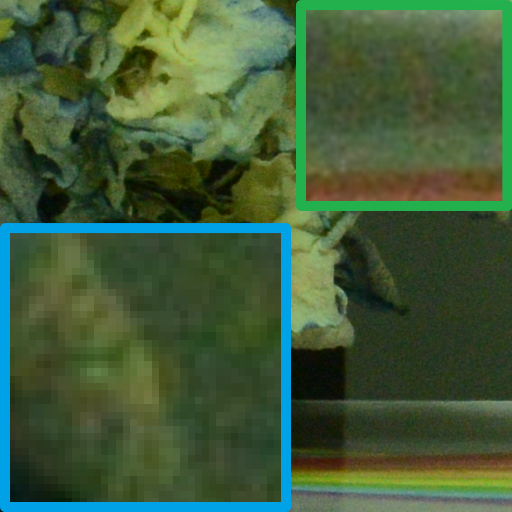}} \, 
  \subfloat[SASL]{\includegraphics[width=1.109in, height = 1.109in]{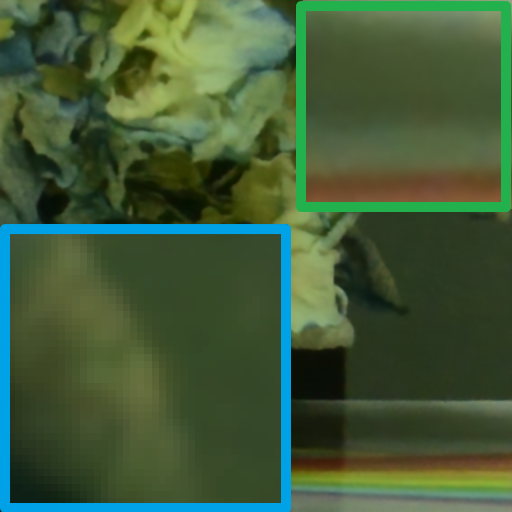}} \,
  \subfloat[FFDNet]{\includegraphics[width=1.109in, height = 1.109in]{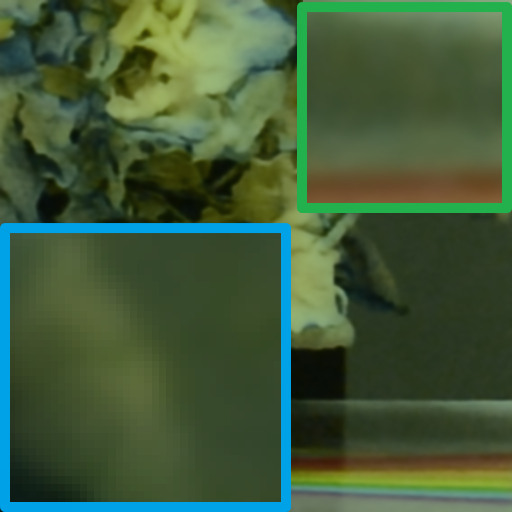}} \,
  \subfloat[Restormer]{\includegraphics[width=1.109in, height = 1.109in]{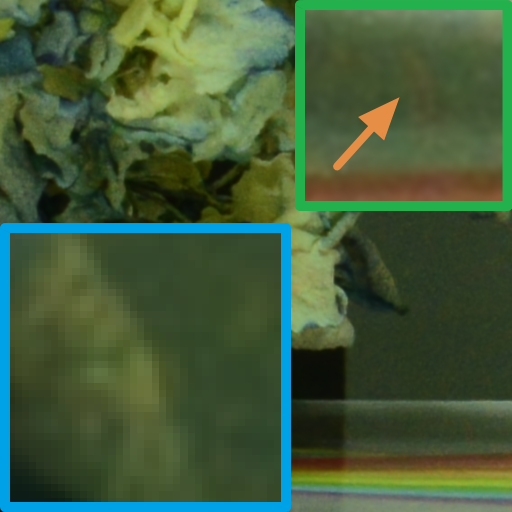}}\,
  \subfloat[PNGAN]{\includegraphics[width=1.109in, height = 1.109in]{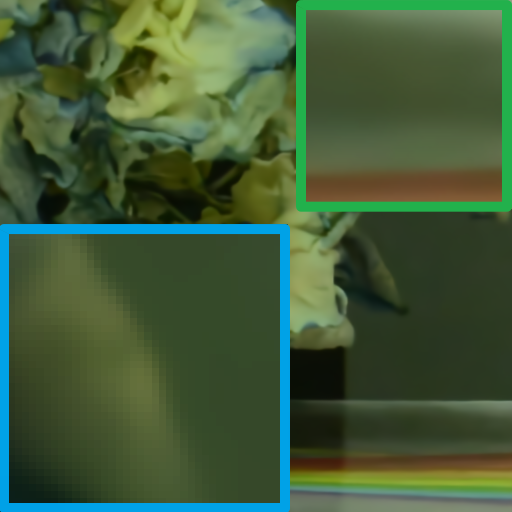}} \\
  \vspace{-8.68pt}
  \subfloat[MCWNNM]{\includegraphics[width=1.109in, height = 1.109in]{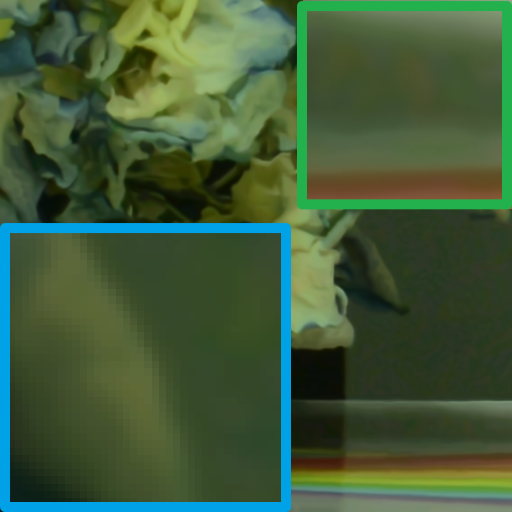}} \,
  \subfloat[Bitonic]{\includegraphics[width=1.109in, height = 1.109in]{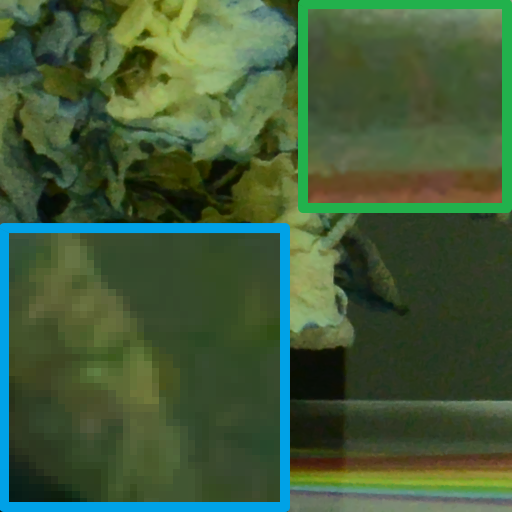}} \,
  \subfloat[NLHCC]{\includegraphics[width=1.109in, height = 1.109in]{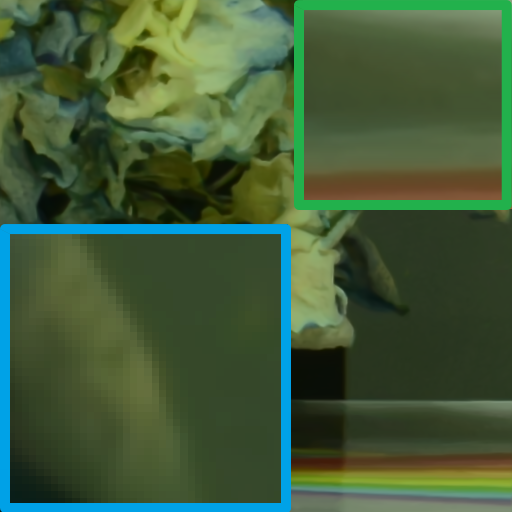}} \,
  \subfloat[CBM3D]{\includegraphics[width=1.109in, height = 1.109in]{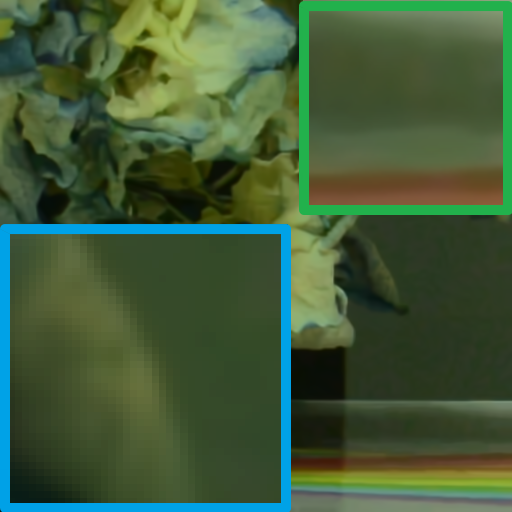}} \,
  \subfloat[MStSVD]{\includegraphics[width=1.109in, height = 1.109in]{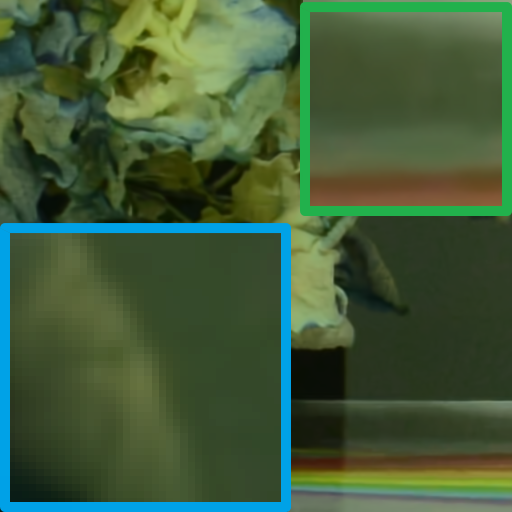}} \,
  \subfloat[GCP-ID]{\includegraphics[width=1.109in, height = 1.109in]{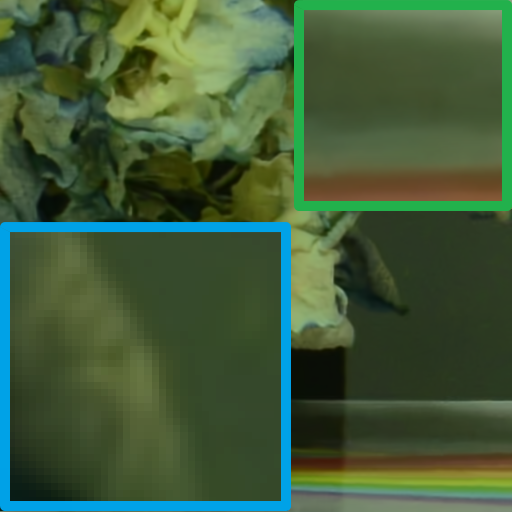}}
  \vspace{-2.8pt}
  \caption{Denoising results of compared methods on the CC dataset.}
  \label{Fig_CC15_visual_evaluation}
  \vspace{-13.8pt}
\end{figure*}

%% file: Fig_SonyA6500_visual_evaluation.tex
\begin{figure}[tb]
\vspace{-3.8pt}
\centering
\graphicspath{{Figs/Sony_A6500/combined_new/}}
  \subfloat[Mean]{\includegraphics[width=0.8in, height=0.8in]{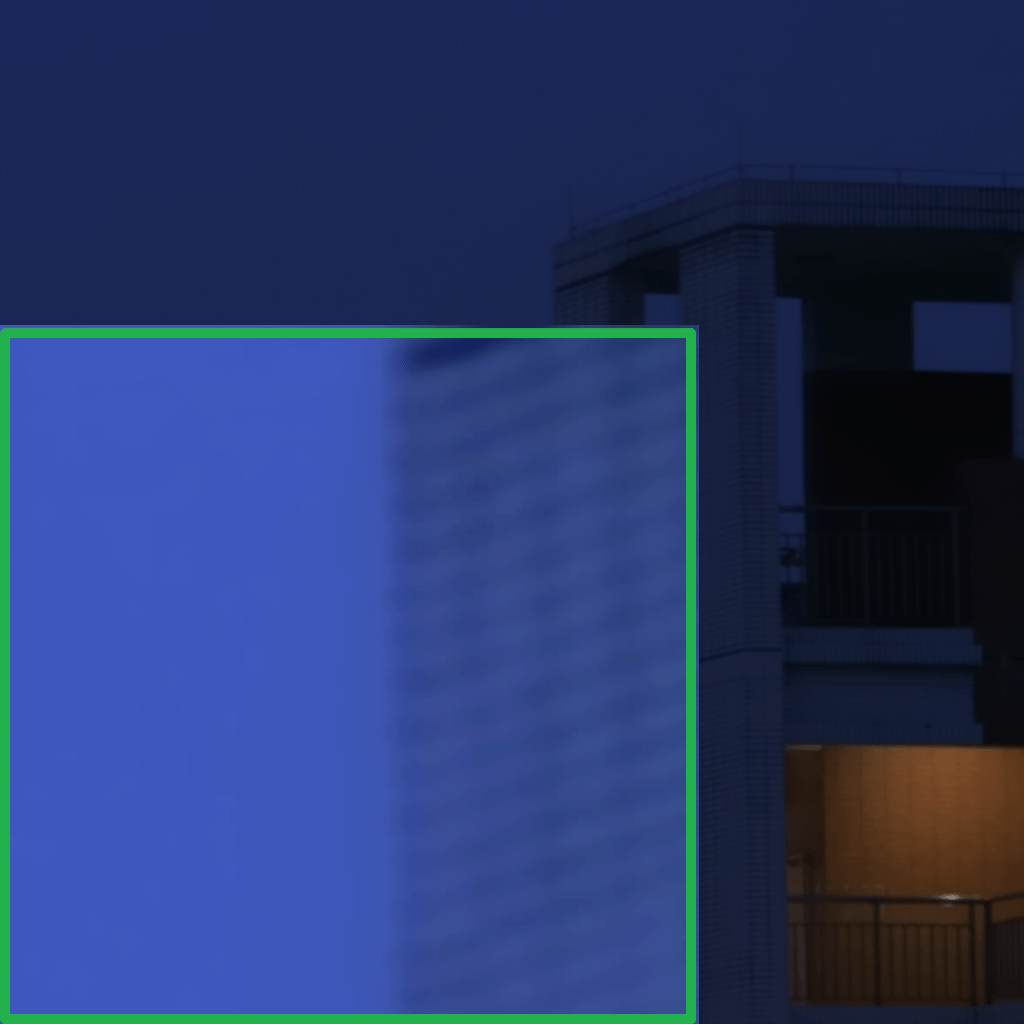}} \,
  \subfloat[Noisy]{\includegraphics[width=0.8in, height=0.8in]{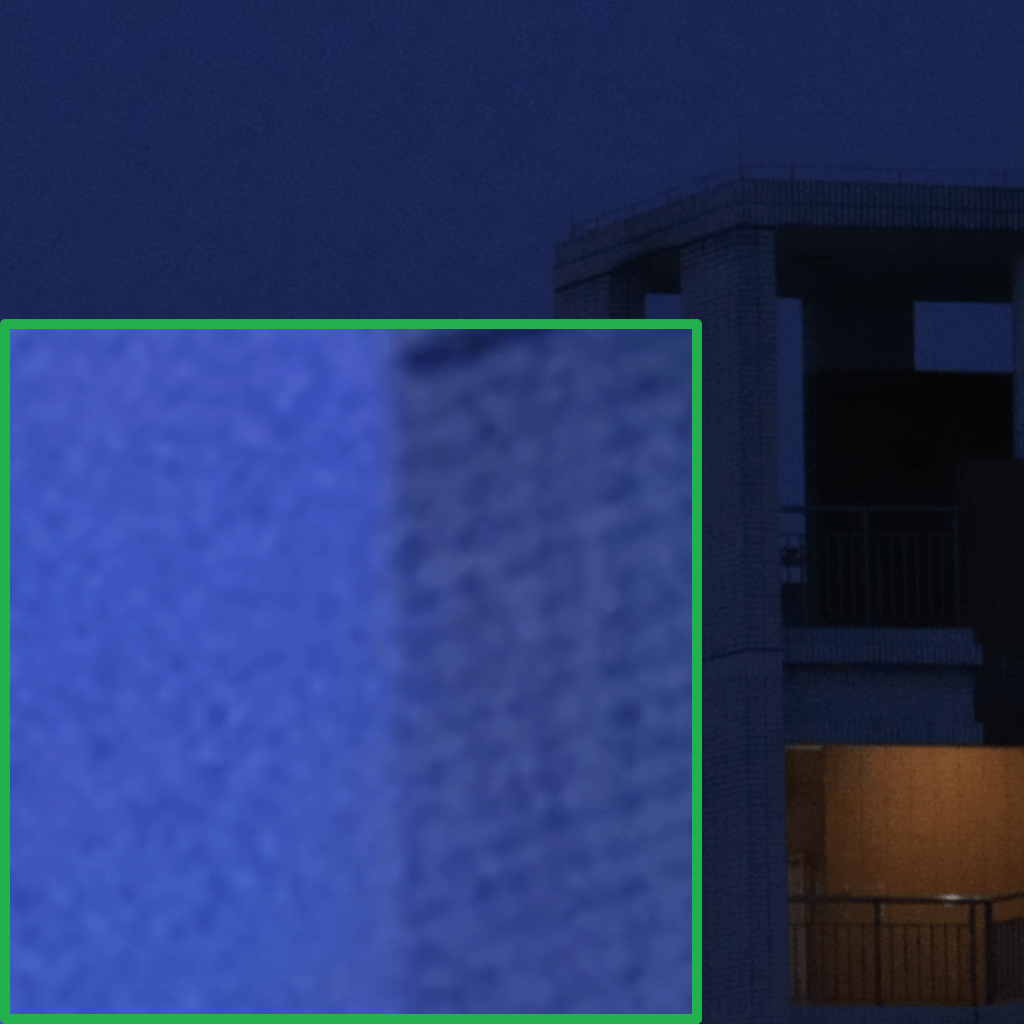}} \, 
  \subfloat[AINDNet]{\includegraphics[width=0.8in, height=0.8in]{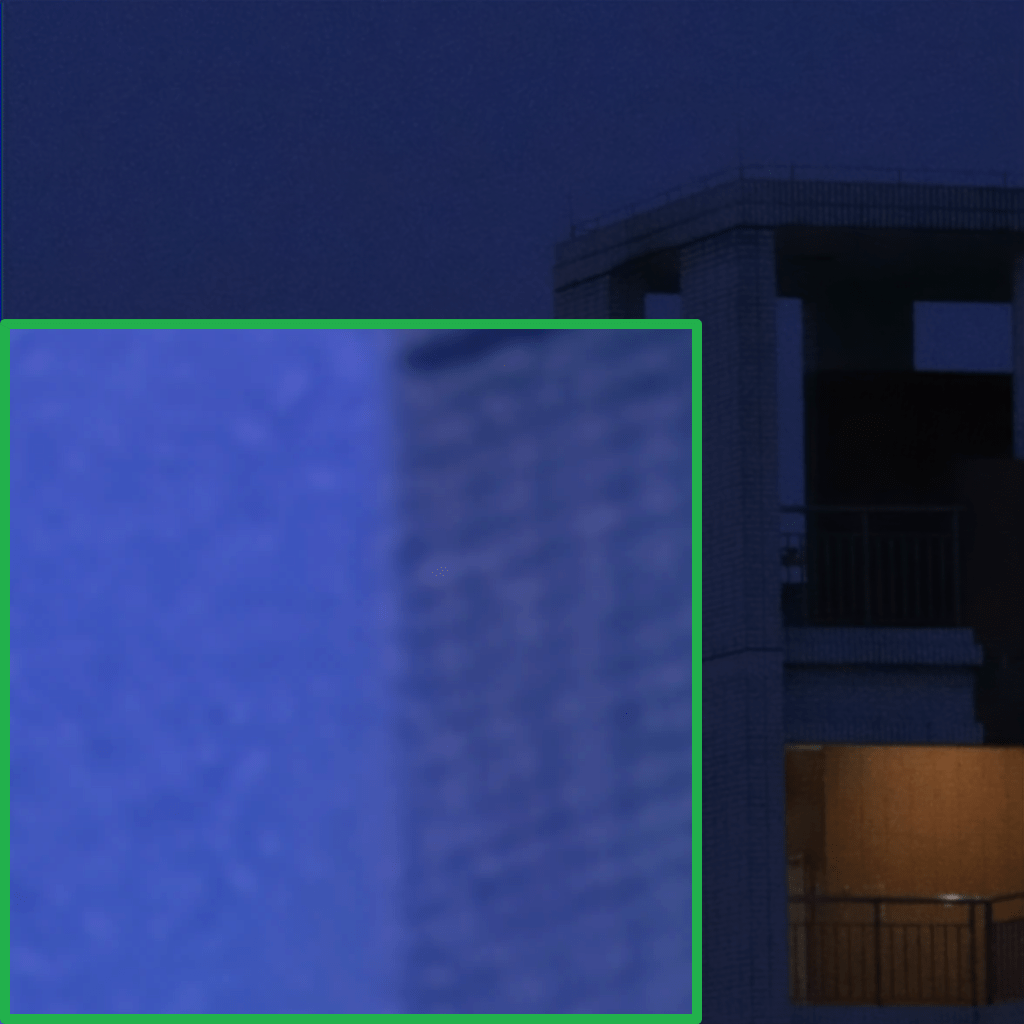}} \, 
  \subfloat[FFDNet]{\includegraphics[width=0.8in, height=0.8in]{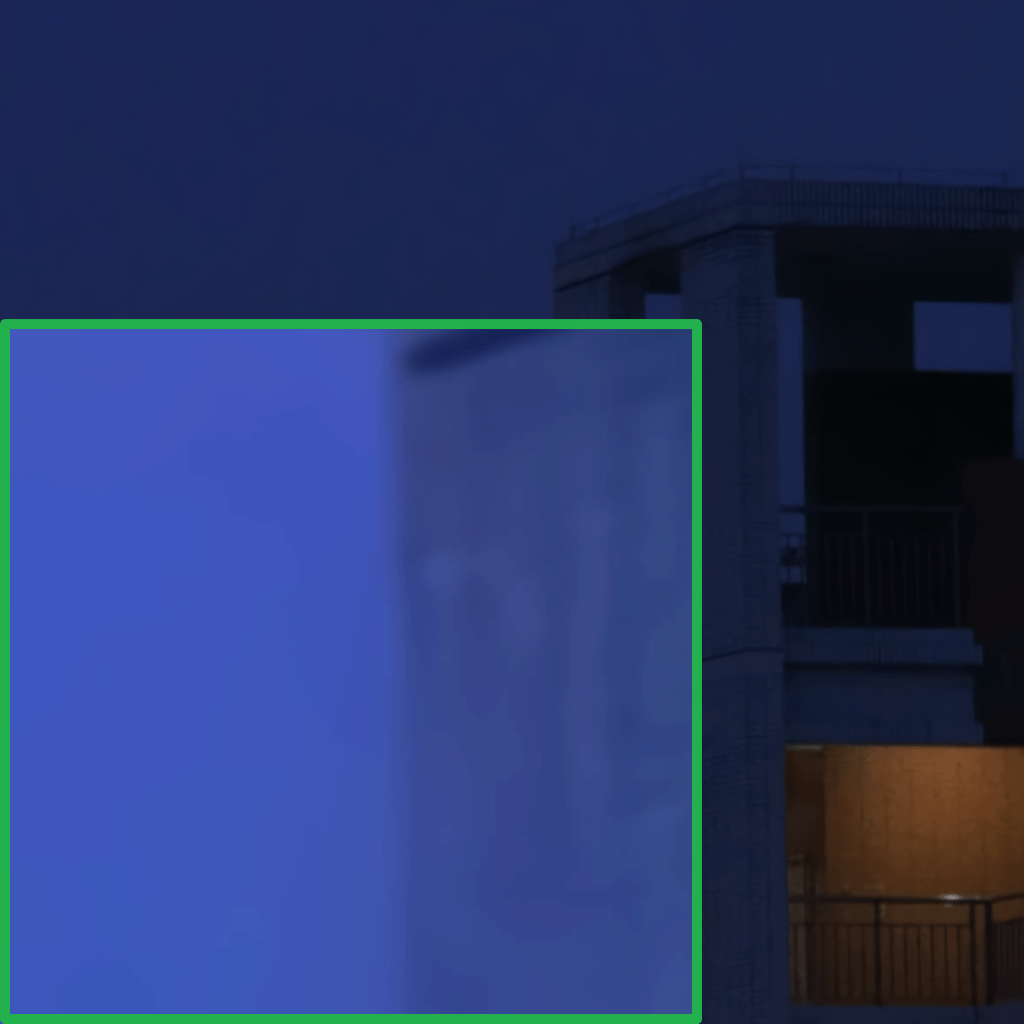}} \\
  \vspace{-8.68pt}
  \subfloat[PNGAN]{\includegraphics[width=0.8in, height=0.8in]{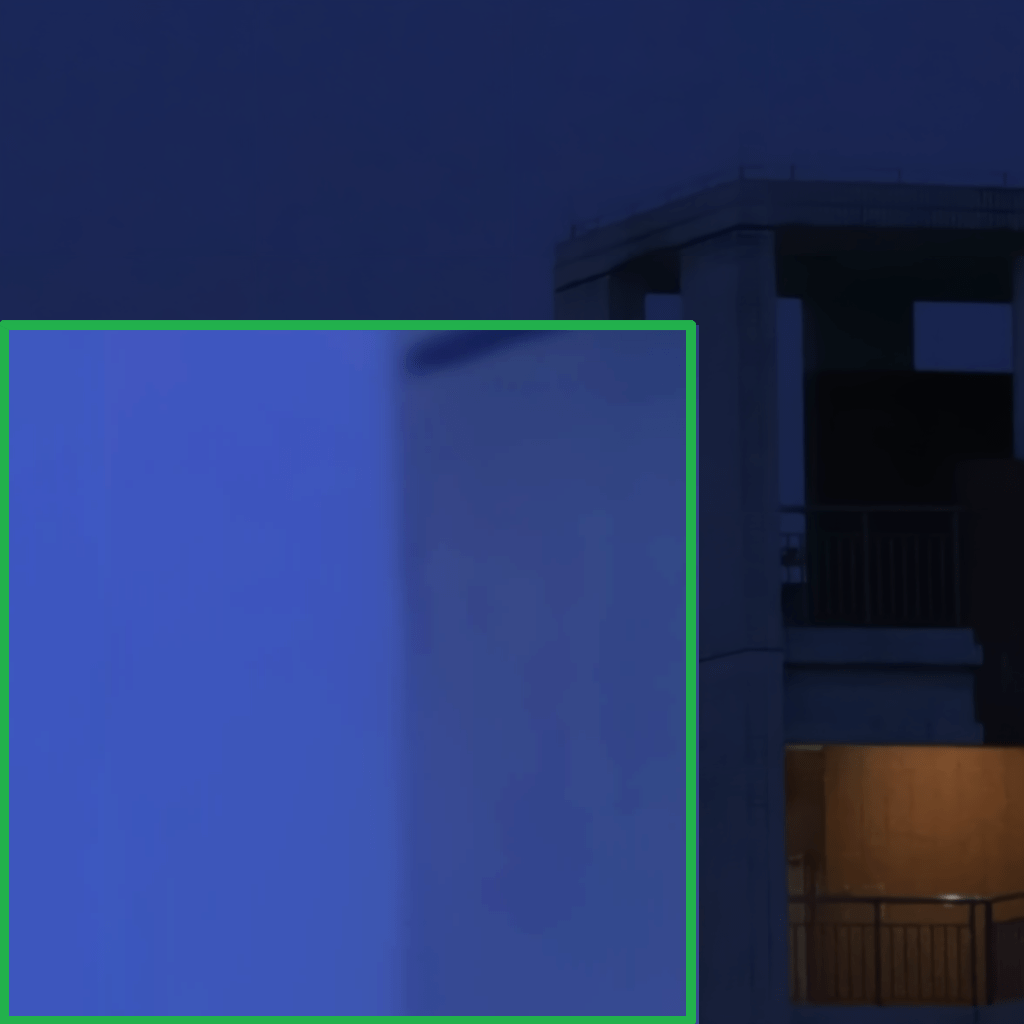}} \,
  \subfloat[Restormer]{\includegraphics[width=0.8in, height=0.8in]{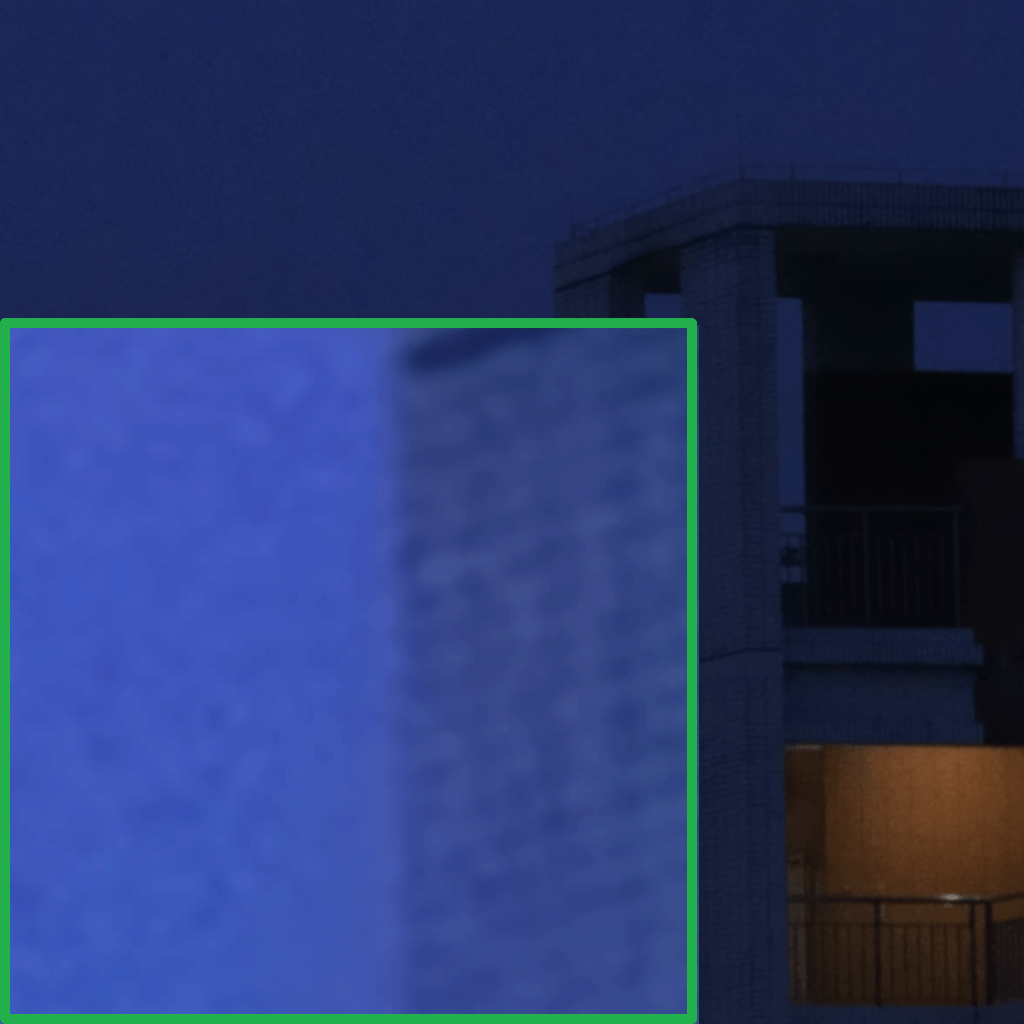}} \,
  \subfloat[VDIR]{\includegraphics[width=0.8in, height=0.8in]{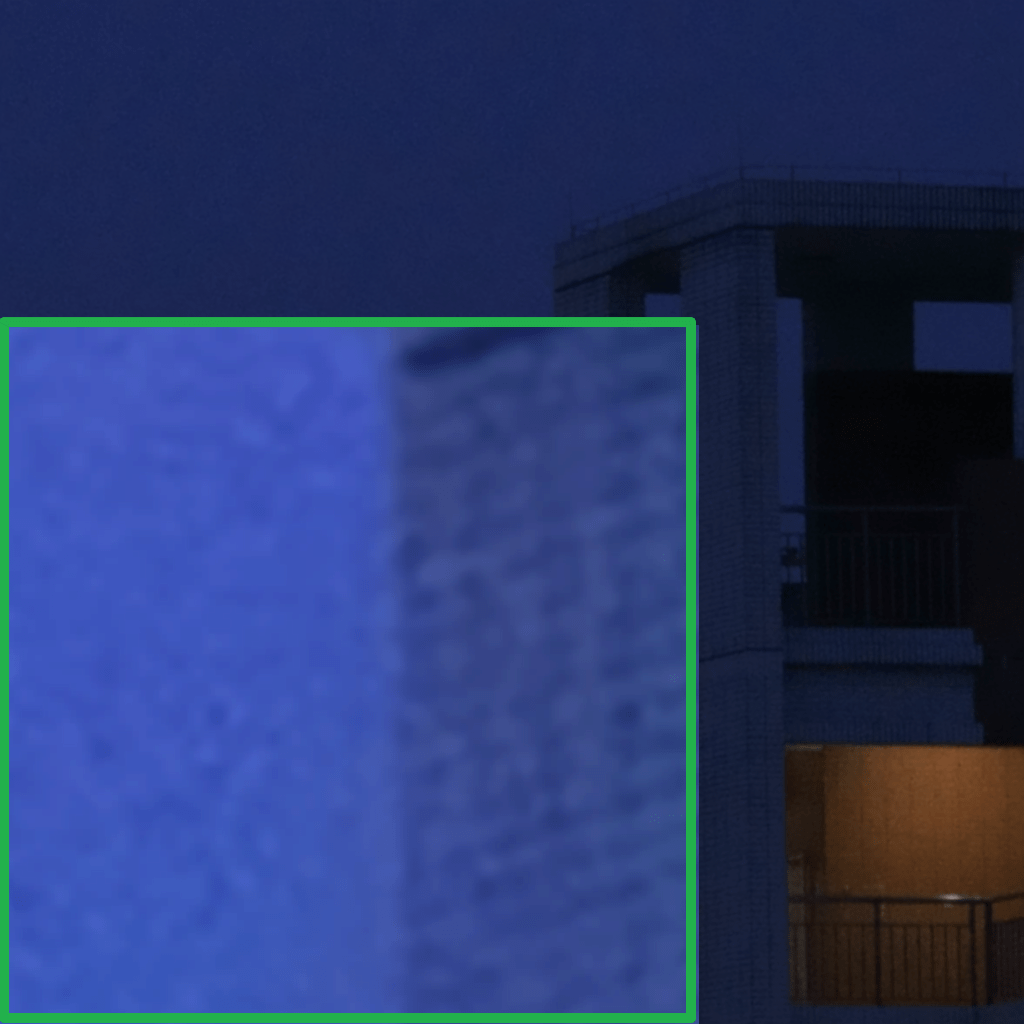}} \, 
  \subfloat[GCP-ID]{\includegraphics[width=0.8in, height=0.8in]{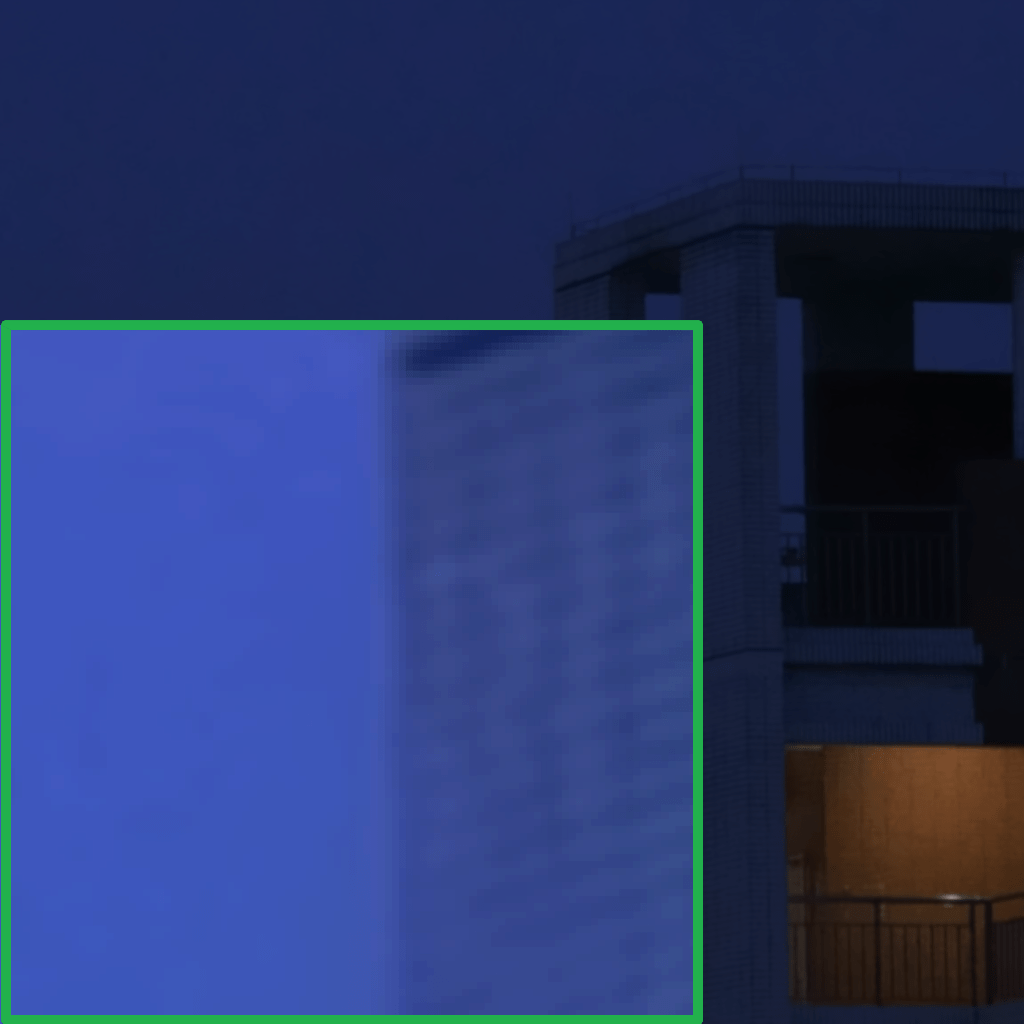}}  
  \vspace{-3pt}
  \caption{Denoising results comparison on the IOCI dataset.}
  \label{Fig_SonyA6500_visual_evaluation}
  \vspace{-8.8pt}
\end{figure}

%% file: Fig_SIDDsRGB_visual_evaluation.tex

\begin{figure}[tb]
\vspace{-1.8pt}
\centering
\graphicspath{{Figs/SIDD_sRGB/Sample2/combined/}}
  \subfloat[Mean]{\includegraphics[width=0.8in, height = 0.8in]{noisy_img10_1_combined}} \,
  \subfloat[AINDNet]{\includegraphics[width=0.8in, height = 0.8in]{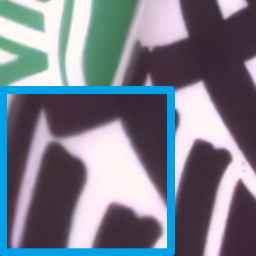}} \, 
  \subfloat[CBDNet]{\includegraphics[width=0.8in, height = 0.8in]{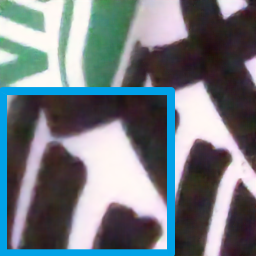}}\,
  \subfloat[CycleISP]{\includegraphics[width=0.8in, height = 0.8in]{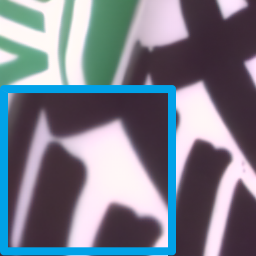}}\\
  \vspace{-8.68pt}
  \subfloat[Restormer]{\includegraphics[width=0.8in, height = 0.8in]{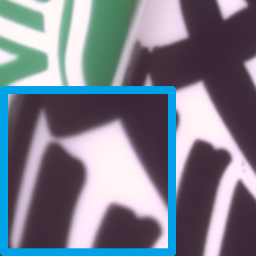}} \,
  \subfloat[VDIR]{\includegraphics[width=0.8in, height = 0.8in]{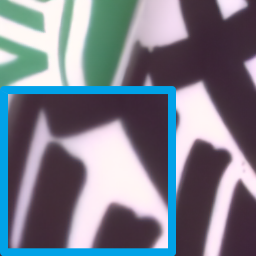}} \, 
  \subfloat[NLHCC]{\includegraphics[width=0.8in, height = 0.8in]{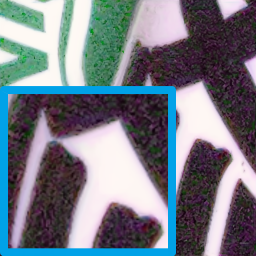}}\,
  \subfloat[GCP-ID]{\includegraphics[width=0.8in, height = 0.8in]{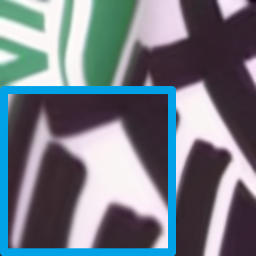}}  
  \vspace{-1.8pt}
  \caption{Denoising results comparison on the SIDD dataset.}
  \label{Fig_SIDDsRGB_visual_evaluation}
  \vspace{-0.8pt}
\end{figure}

%% file: Fig_GCP_w_wo_CNN.tex
\begin{figure}[htbp]
\vspace{-0.8pt}
\centering
\graphicspath{{Figs/GCD_w_wo_CNN/combined/}}
  \subfloat{\includegraphics[width=1.068in, height=1.068in]{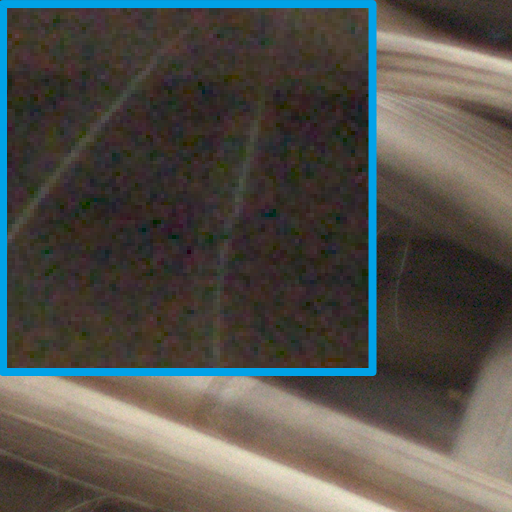}} \, 
  \subfloat{\includegraphics[width=1.068in, height=1.068in]{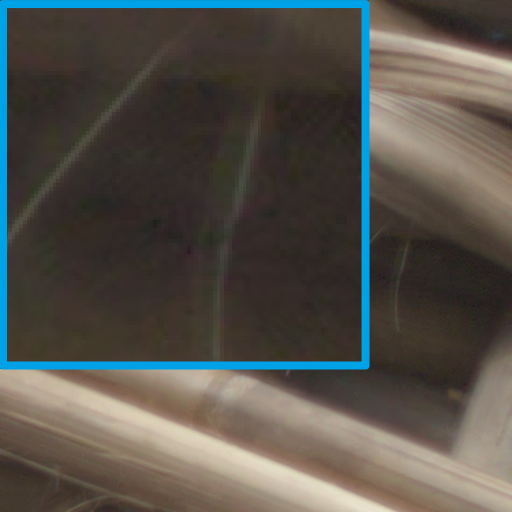}} \, 
  \subfloat{\includegraphics[width=1.068in, height=1.068in]{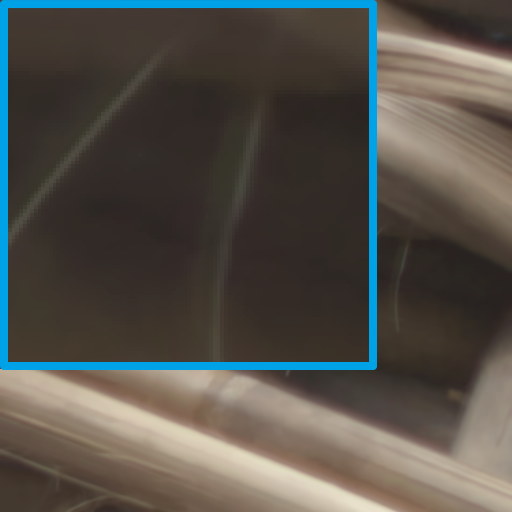}} \\
  \vspace{-6.6pt}
  \setcounter{subfigure}{0}
  \subfloat[Noisy]{\includegraphics[width=1.068in, height=1.068in]{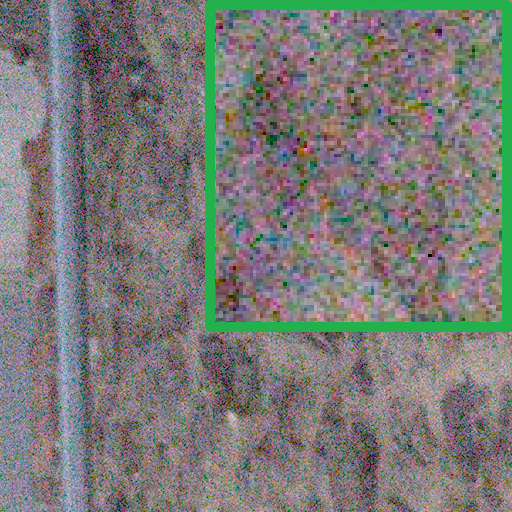}} \,
  \subfloat[GCP-ID]{\includegraphics[width=1.068in, height=1.068in]{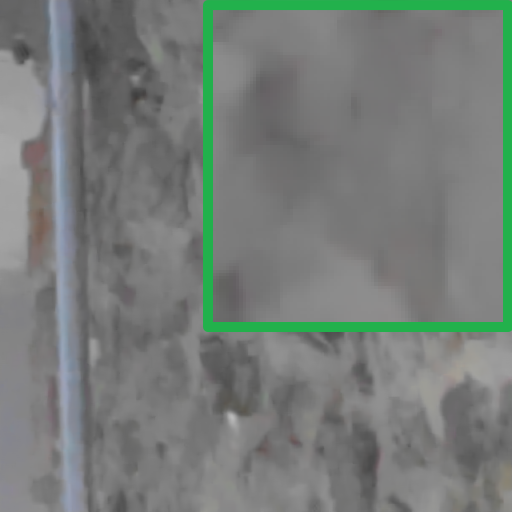}} \, 
  \subfloat[GCP-ID + CNN]{\includegraphics[width=1.068in, height=1.068in]{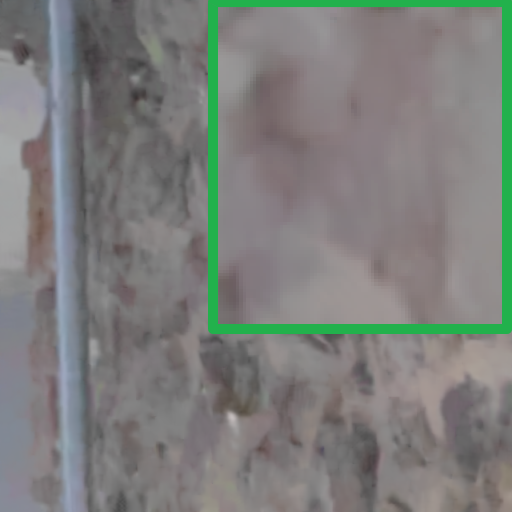}}
  \vspace{-2.6pt}
  \caption{The effectiveness of the CNN estimator on DND.}
  \label{Fig_GCP_w_wo_CNN}
  \vspace{0pt}
\end{figure}

%% file: Table_inference_time.tex
%

\begin{table}[!htbp]
\vspace{-1.8pt}
\scriptsize
  \centering
  \caption{Average inference time on the SIDD raw dataset.}
  \scalebox{0.958}{
    \begin{tabular}{cccccc}
    \toprule
    Method & GCP-ID & Blind2Unblind  & Blind2Unblind & CycleISP & FBI  \\
    \midrule
    Platform & CPU   & CPU   & GPU (T4)   & GPU (T4)   & GPU (T4) \\
    \midrule
    Time (s) & 0.54  & 3.62  & 0.11  & 0.51  & 0.13  \\
    \bottomrule
    \end{tabular}}%
  \label{Table_inference_time}%
\vspace{-1.8pt}  
\end{table}%

%% file: Table_Raw_Color_video.tex
\begin{table*}[htbp]
\scriptsize
  \centering
  \caption{Real-world Raw and color video denoising results. `*': the results are from the authors' papers.}
  \scalebox{0.8516}{
    \begin{tabular}{ccccccccccccccc}
    \toprule
    \multirow{3}[4]{*}{Dataset} & \multicolumn{4}{c}{Traditional denoisers} & \multicolumn{10}{c}{DNN methods} \\
\cmidrule{2-15}          & VBM4D & VIDOSAT & GCP-ID & GCP-ID  + CNN & ASwin & DIDN  & DVDNet & FastDVDNet & FloRNN & MaskDnGAN & RViDeNet & RVRT  & UDVD  & VNLNet \\
          & \cite{maggioni2012video} &  \cite{wen2018vidosat} & (ours) & (ours) & \cite{lindner2023lightweight}& \cite{yu2019deep} & \cite{tassano2019dvdnet} & \cite{Tassano_2020_CVPR} & \cite{li2022unidirectional} & \cite{paliwal2021multi} & \cite{yue2020supervised} & \cite{liang2022recurrent} & \cite{sheth2021unsupervised} & \cite{davy2019non} \\
    \midrule
    \multirow{2}[2]{*}{CRVD (Raw)} & -     & -     & 42.63  & \textbf{43.22}  & -     & 43.25*  & - &  -  & -     & \textcolor{blue}{\textbf{43.08}}  & 43.37* & -  & 44.69* & - \\
\cmidrule{2-15}          & -     & -     & 0.981  & \textcolor{blue}{\textbf{0.984}}  & -     & 0.984*  & -     &   -   & -     & \textbf{0.985}  & 0.985*  & -  & -  & - \\
    \midrule
    \multirow{2}[2]{*}{CRVD (sRGB)} & 34.14  & 34.16  & 36.79  & \textbf{37.32} & 36.51  & -     & 34.50  & 35.84  & 36.66  & -     & -     & \textcolor{blue}{\textbf{36.94}}  & -     & 36.11  \\
\cmidrule{2-15}          & 0.908  & 0.938  & 0.951  & {\textbf{0.965}}  & 0.958  & -     & 0.949  & 0.931  & \textcolor{blue}{\textbf{0.961}} & -     & -     & 0.956  & -     & 0.945  \\
    \midrule
    \multirow{2}[2]{*}{IOCV (sRGB)} & 38.76  & -     & \textbf{39.08} & \textcolor{blue}{\textbf{38.86}}  & 38.81  & -     & 38.53  & 37.57  & 38.64  & -     & -     & 38.50  & 35.02  & 38.76  \\
\cmidrule{2-15}          & 0.977  & -     & \textbf{0.977} & \textcolor{blue}{\textbf{0.977}}  & 0.977  & -     & 0.975  & 0.970  & 0.974  & -     & -     & 0.967  & 0.966  & 0.977  \\
    \bottomrule
    \end{tabular}}%
  \label{Table_Raw_Color_video}%
  \vspace{-2.8pt}
\end{table*}%

%% file: Fig_CRVDsRGB_indoor_new.tex
\begin{figure*}[tb]
\vspace{-2.8pt}
\centering
\graphicspath{{Figs/CRVD/CRVD_indoor_sRGB_comparison_new/ISO3200/combined/}}
  \subfloat{\includegraphics[width=1.1299in, height = 0.946in]{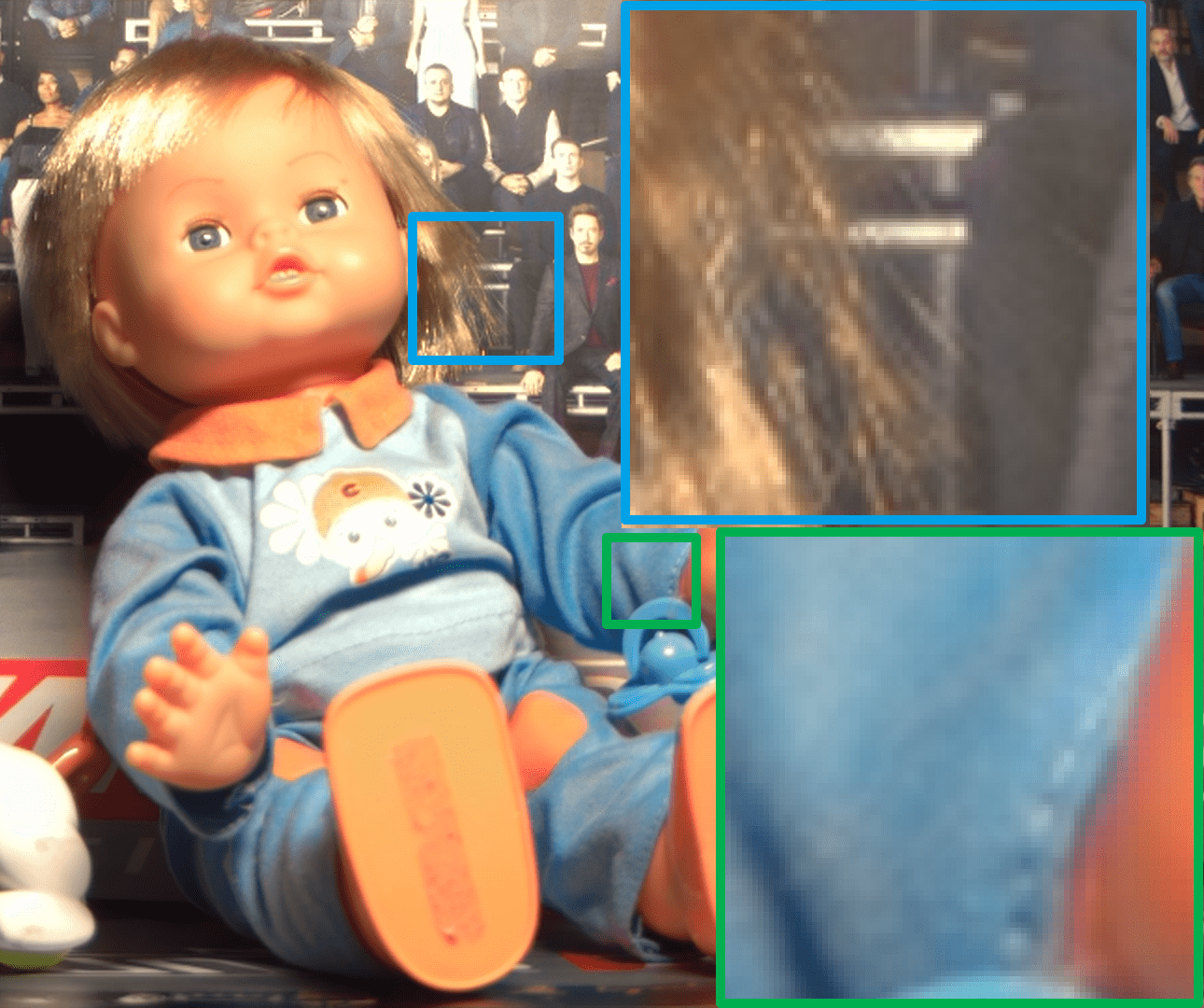}} \,
  \subfloat{\includegraphics[width=1.1299in, height = 0.946in]{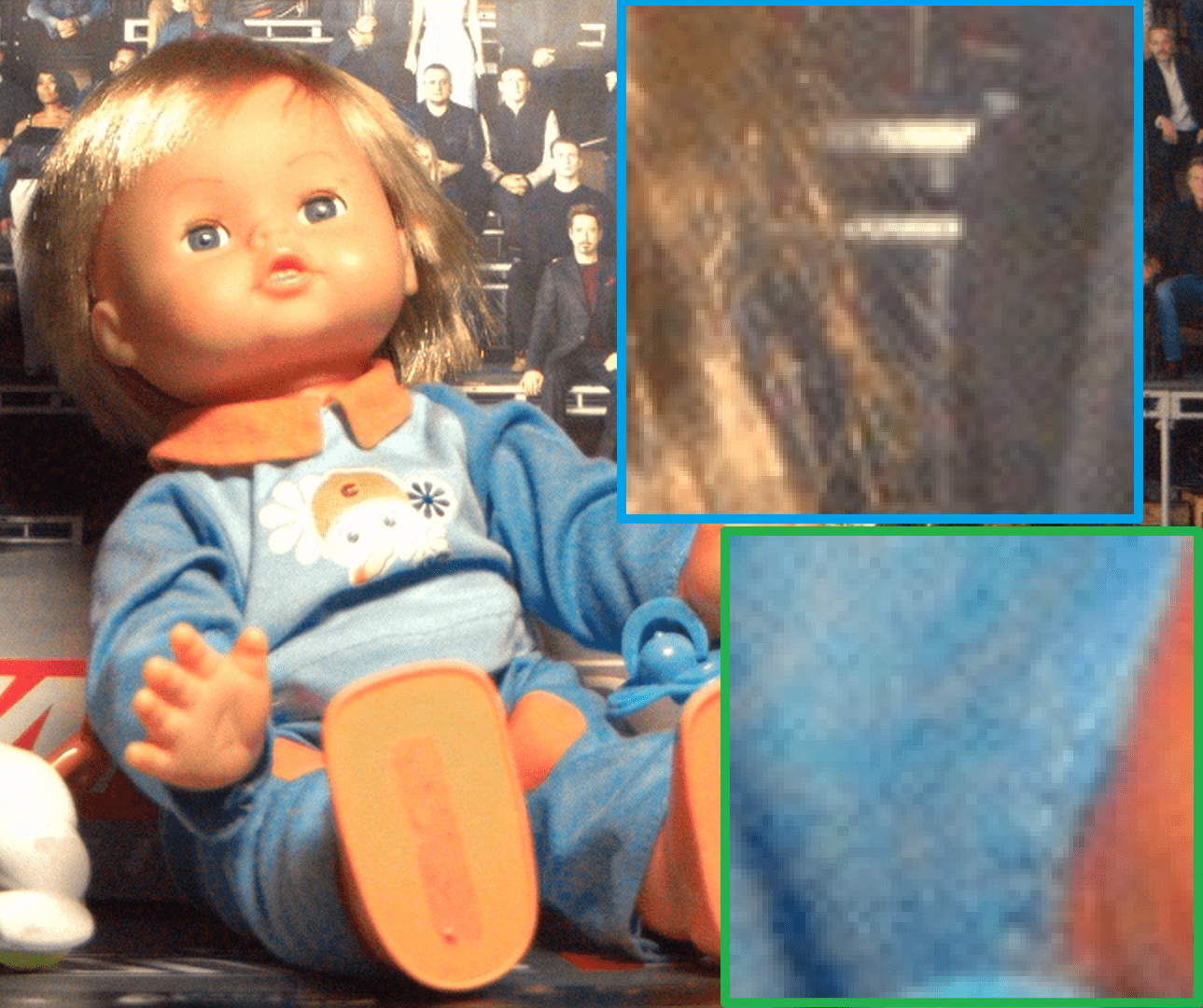}} \, 
  \subfloat{\includegraphics[width=1.1299in, height = 0.946in]{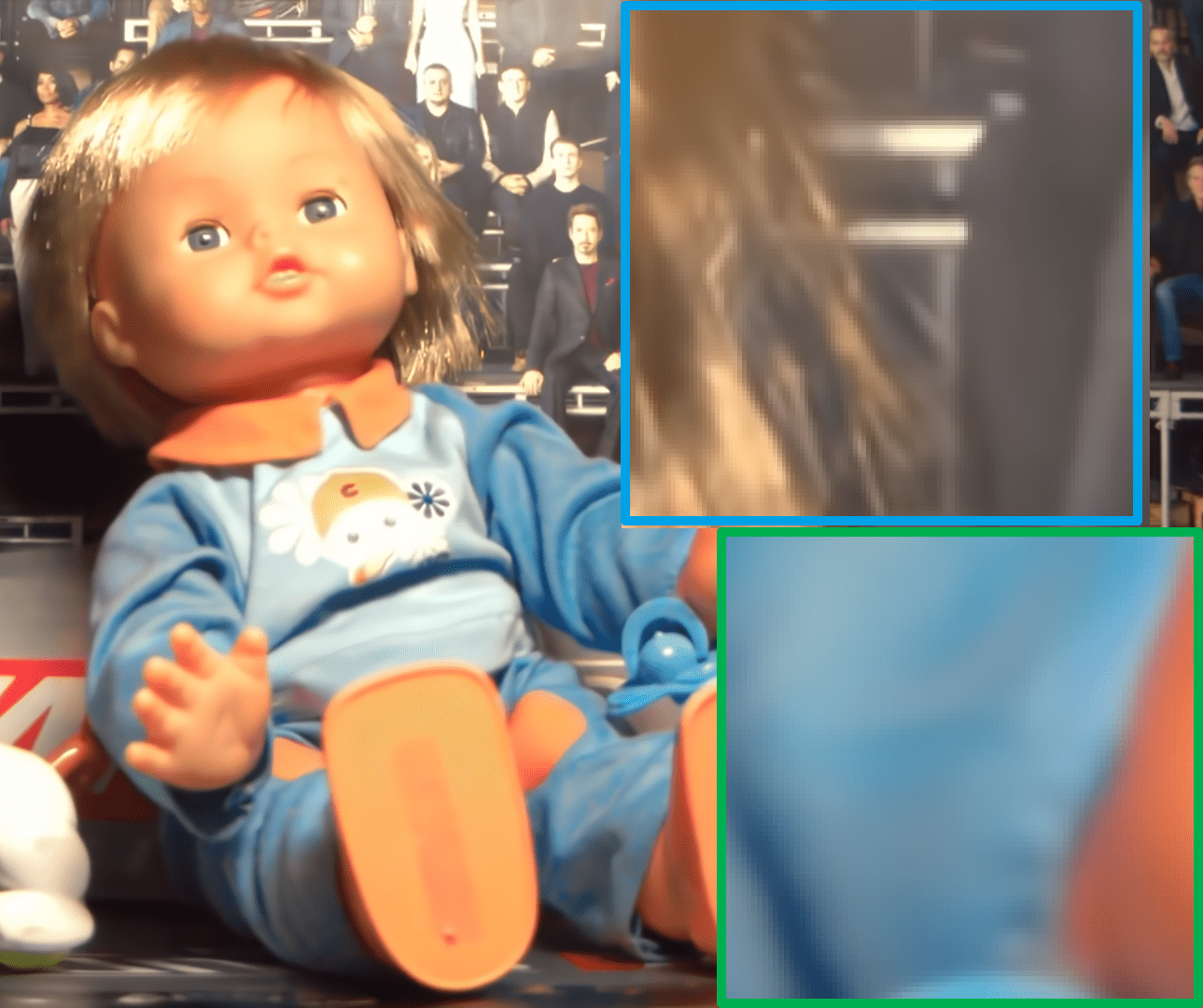}} \,
  \subfloat{\includegraphics[width=1.1299in, height = 0.946in]{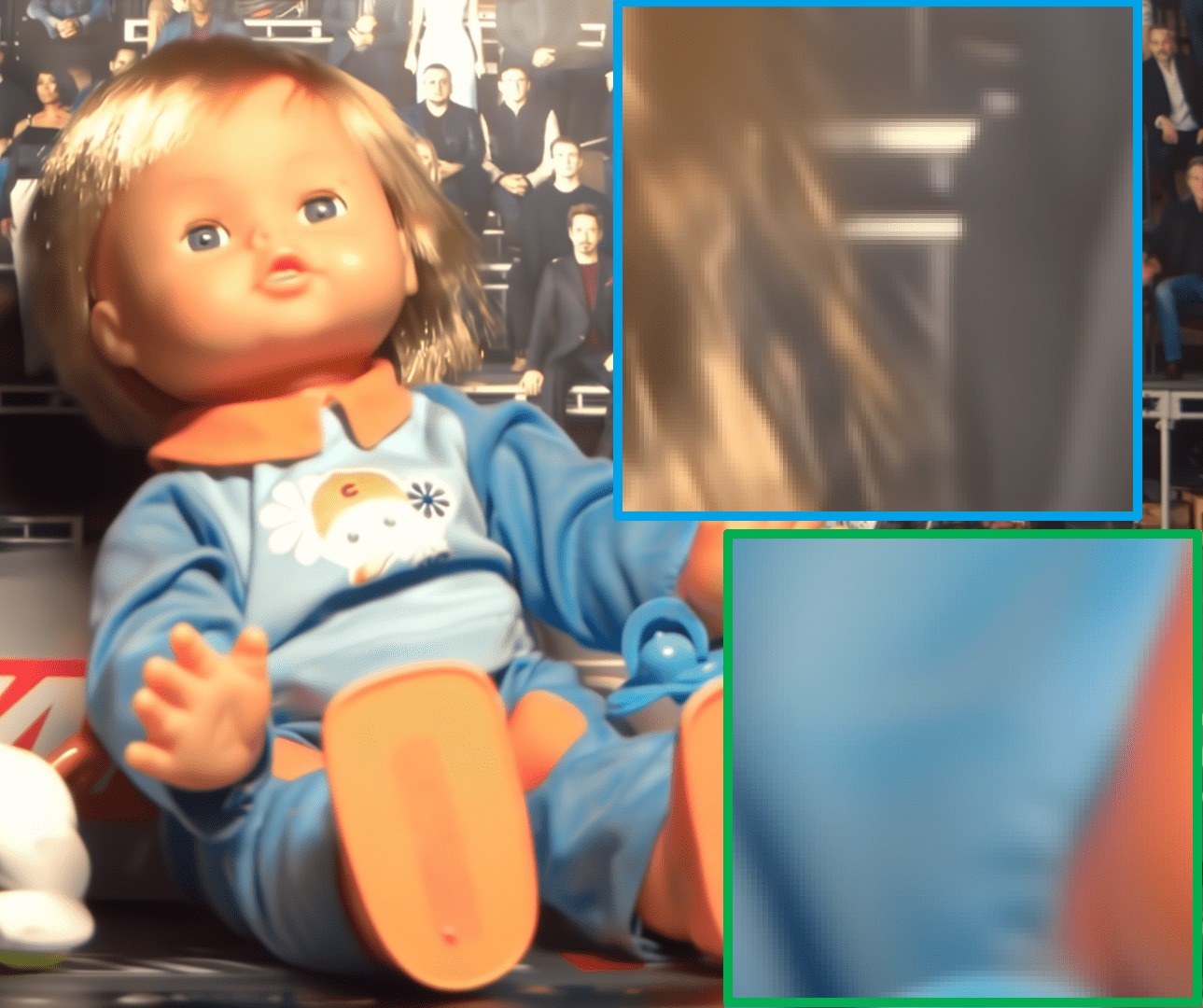}} \,
  \subfloat{\includegraphics[width=1.1299in, height = 0.946in]{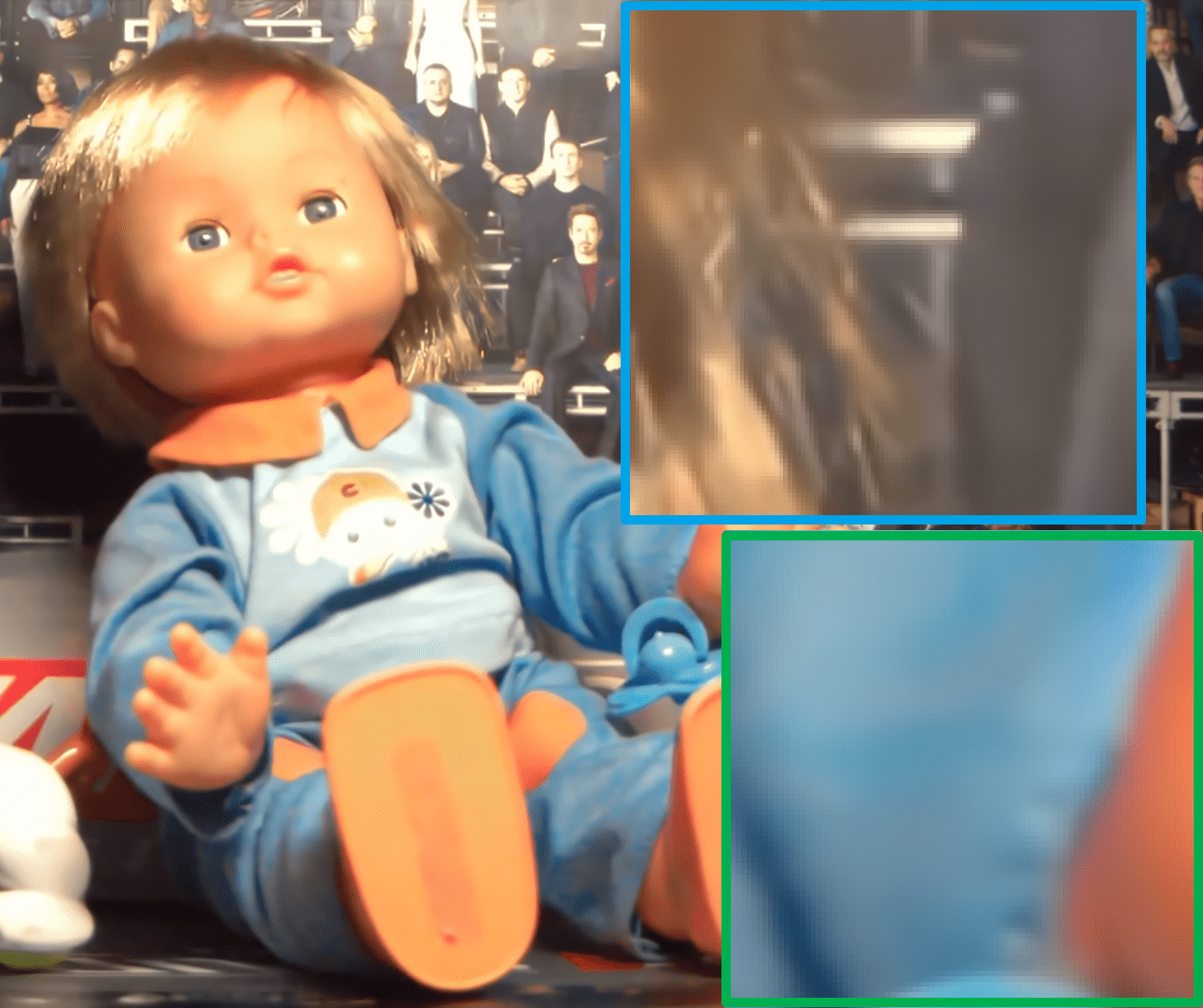}} \,
  \subfloat{\includegraphics[width=1.1299in, height = 0.946in]{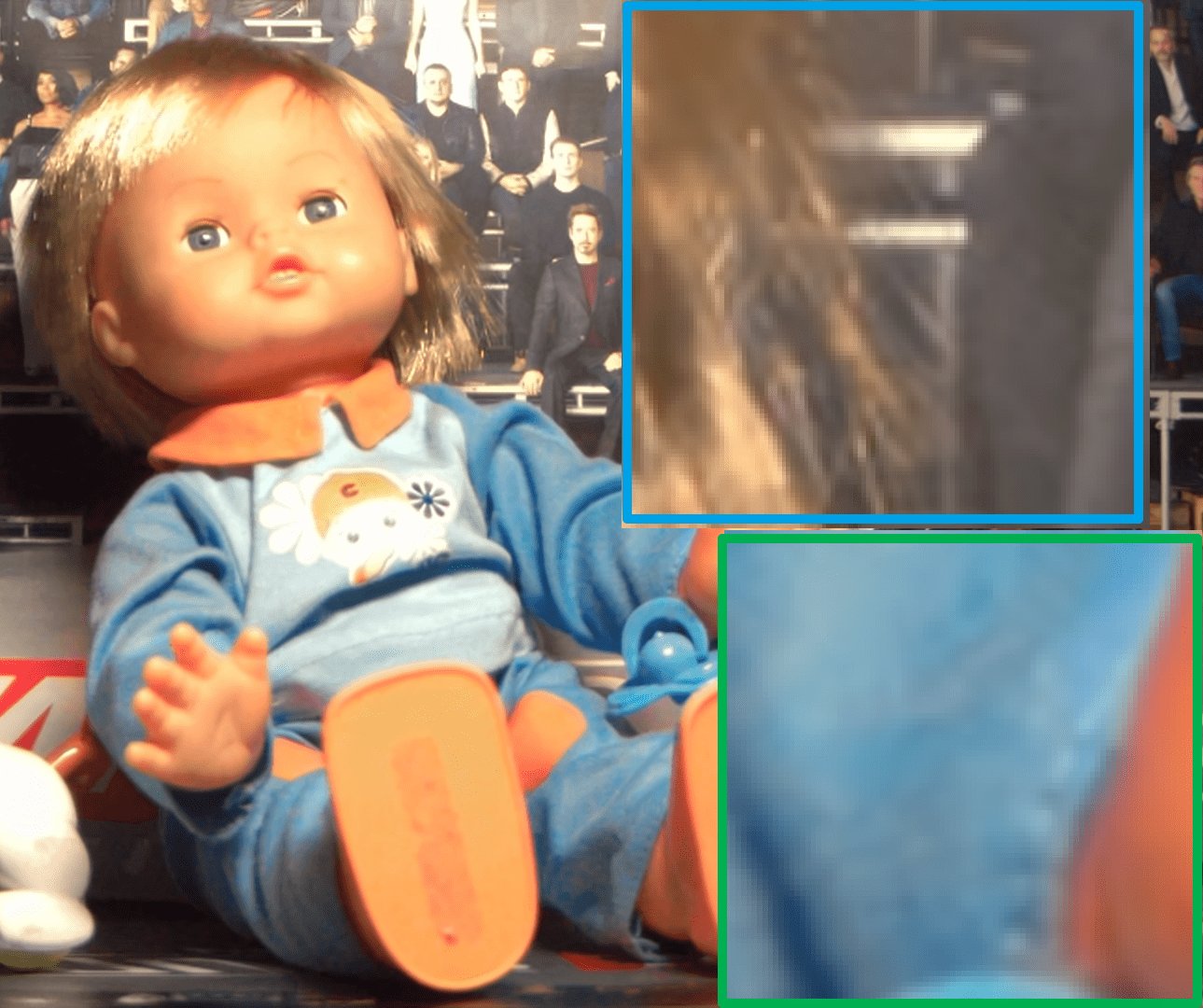}}
  \vspace{1.98pt}
\graphicspath{{Figs/CRVD/CRVD_indoor_sRGB_comparison_new/ISO12800_example2/combined_new/}}
  \subfloat{\includegraphics[width=1.1299in, height = 0.946in]{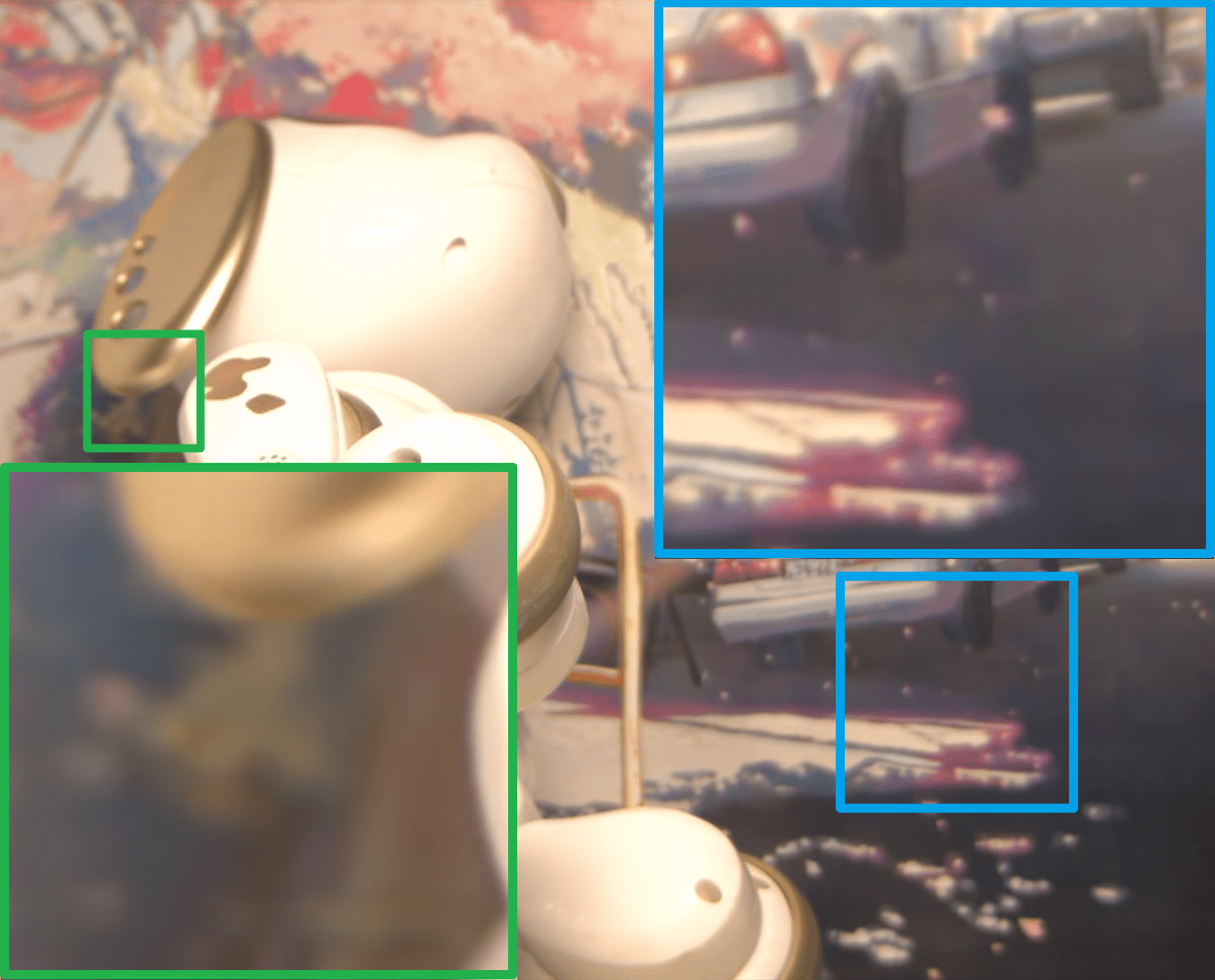}} \,
  \subfloat{\includegraphics[width=1.1299in, height = 0.946in]{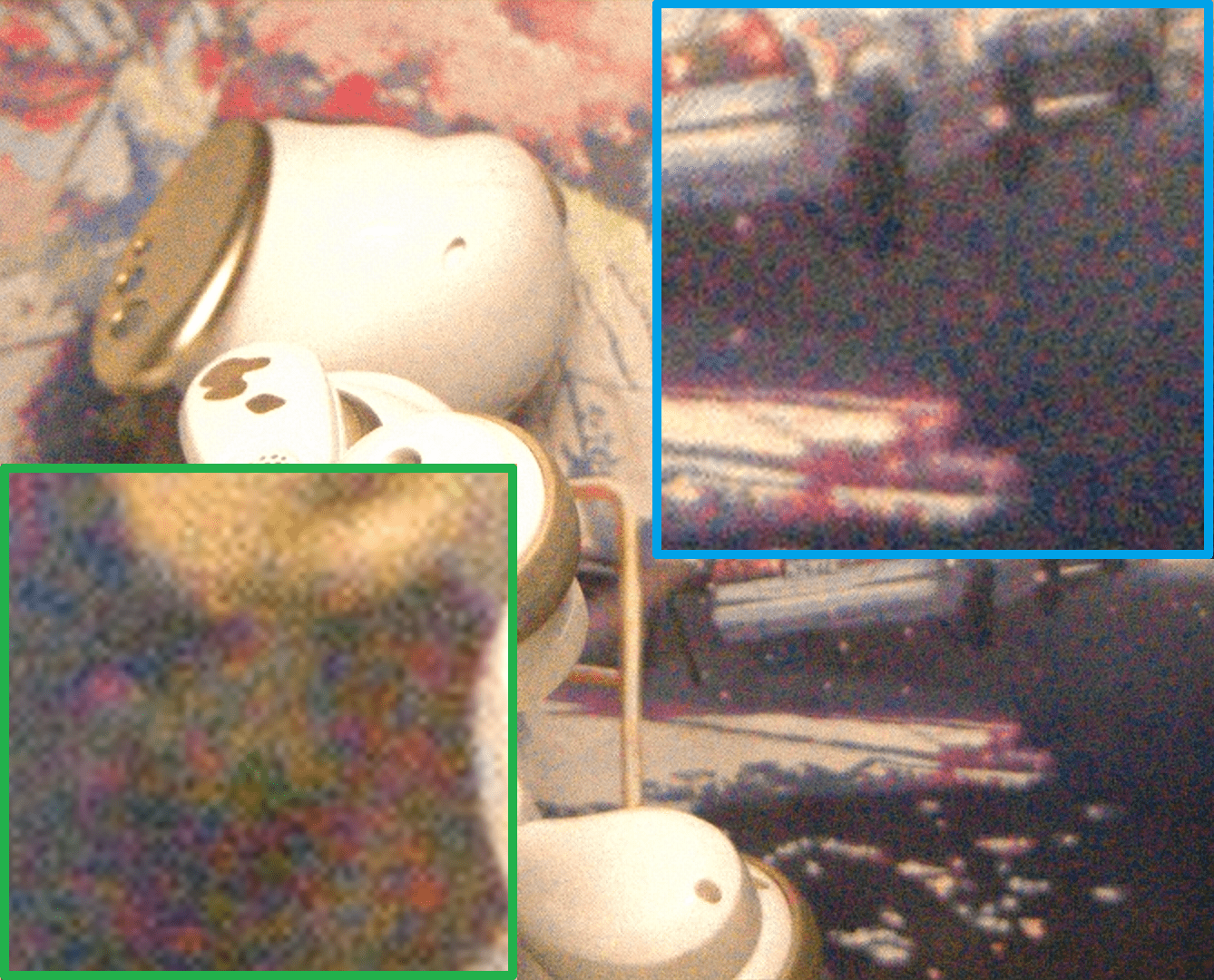}} \, 
  \subfloat{\includegraphics[width=1.1299in, height = 0.946in]{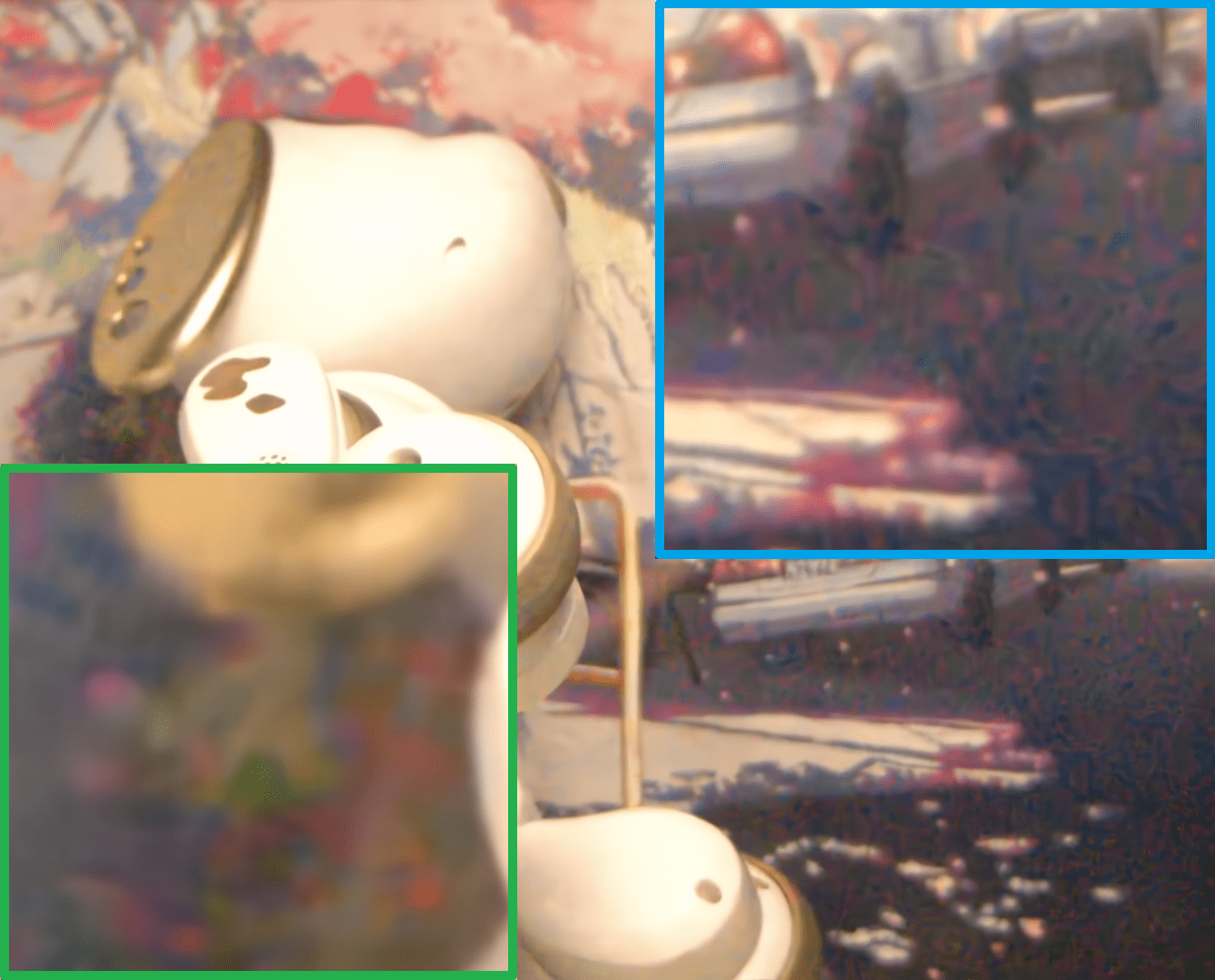}} \,
  \subfloat{\includegraphics[width=1.1299in, height = 0.946in]{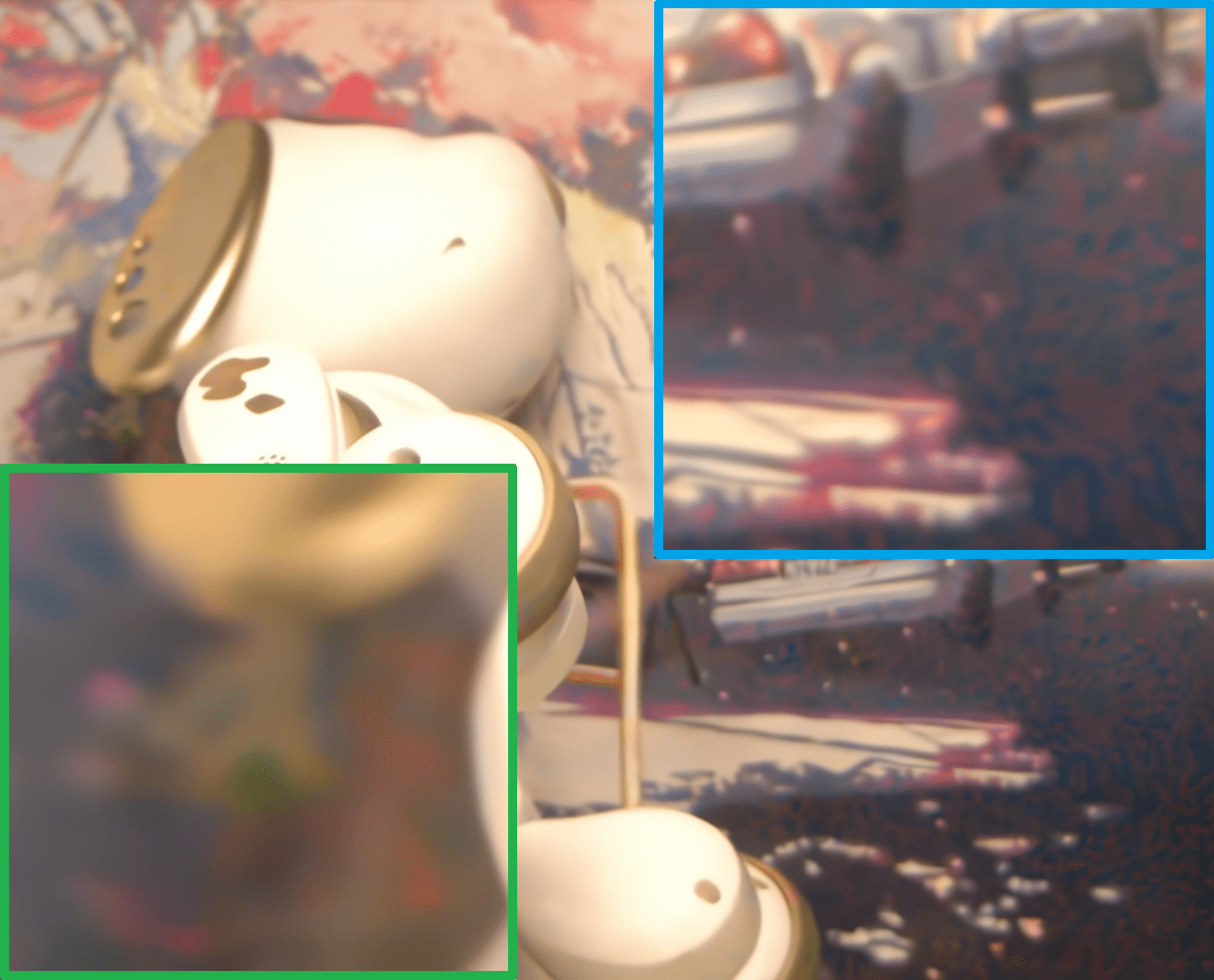}} \,
  \subfloat{\includegraphics[width=1.1299in, height = 0.946in]{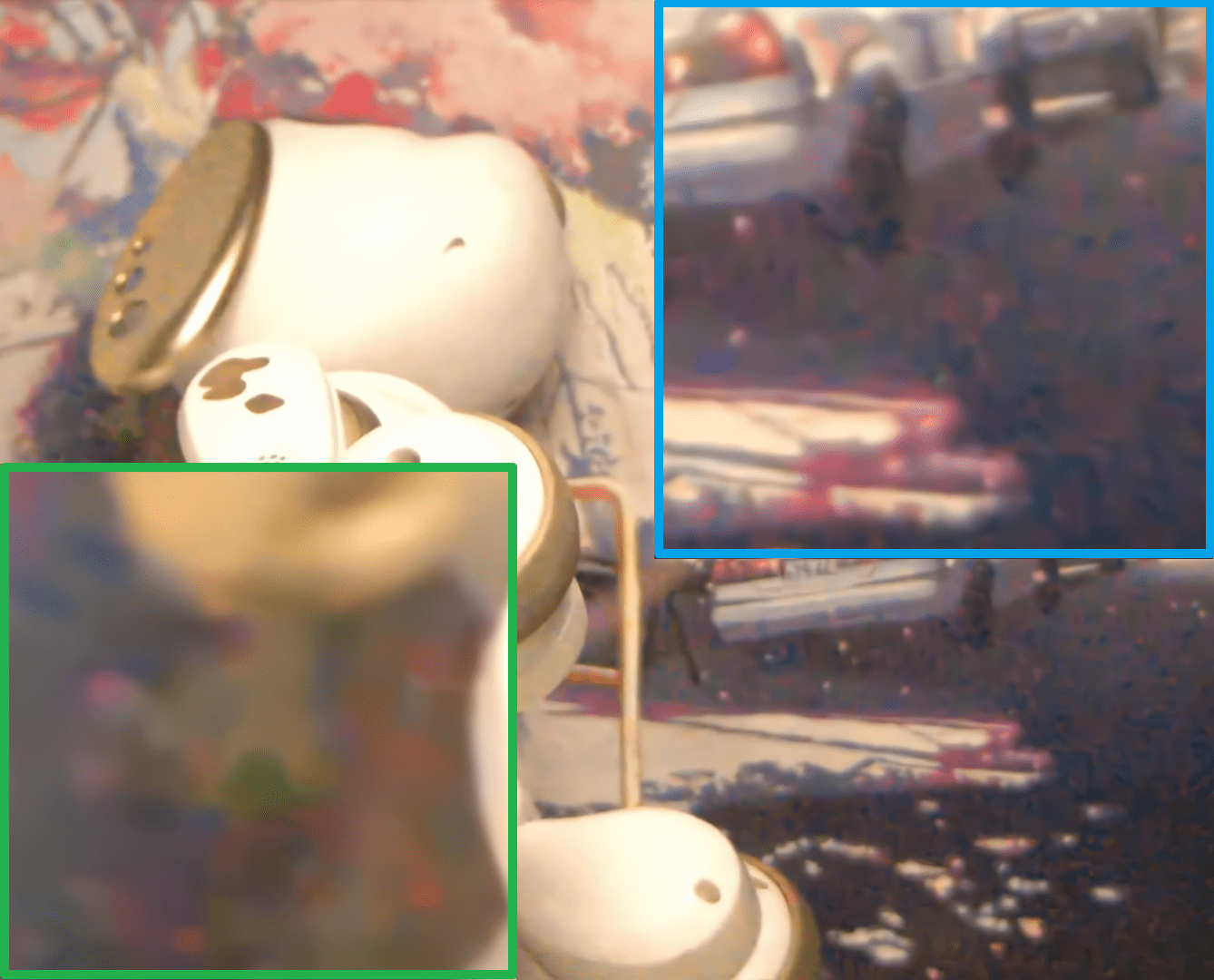}} \,
  \subfloat{\includegraphics[width=1.1299in, height = 0.946in]{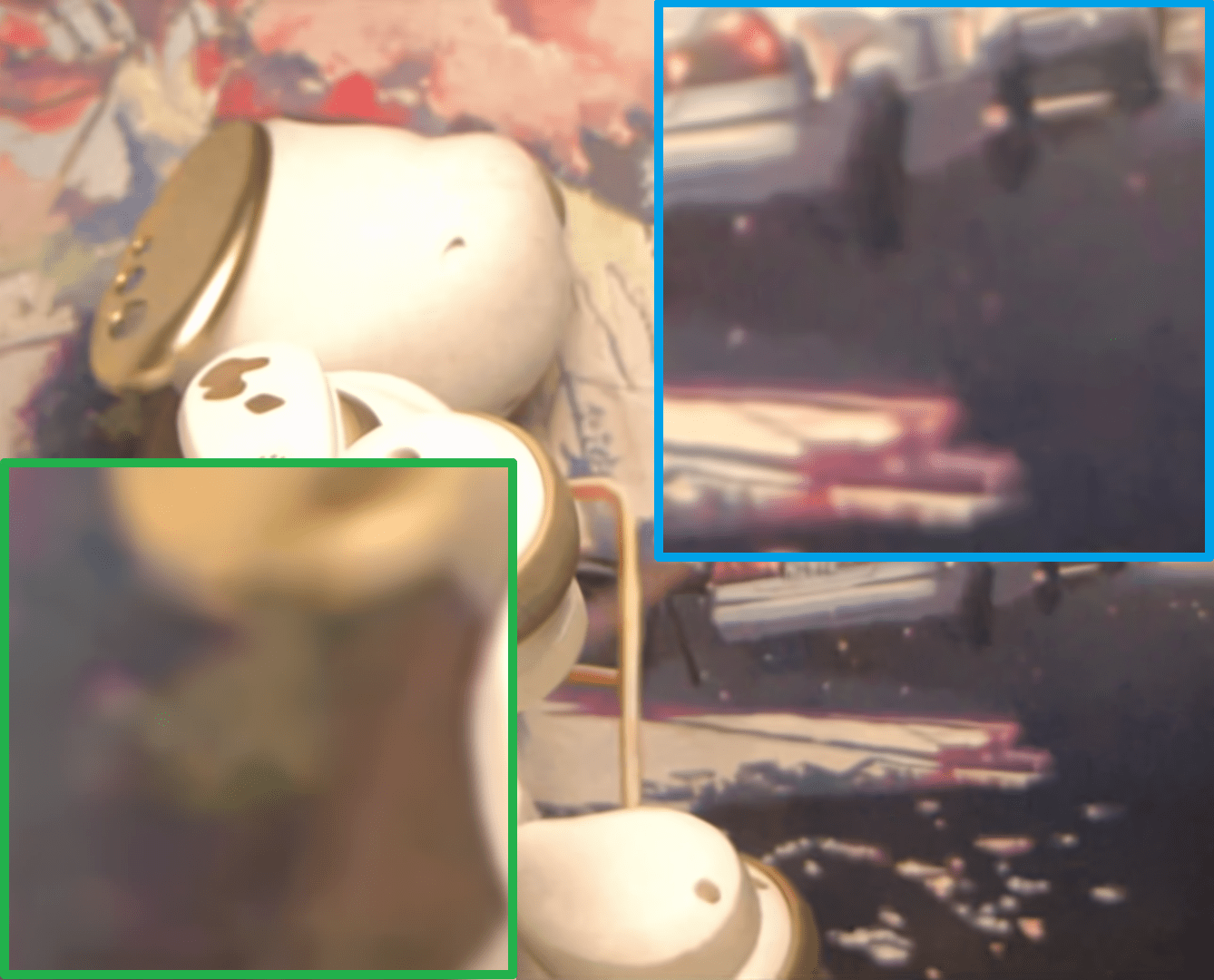}}
  \vspace{1.98pt}
  \graphicspath{{Figs/CRVD/CRVD_indoor_sRGB_comparison_new/ISO25600/combined_new/}}
  \setcounter{subfigure}{0}
  \subfloat[Mean]{\includegraphics[width=1.1299in, height = 0.6356in]{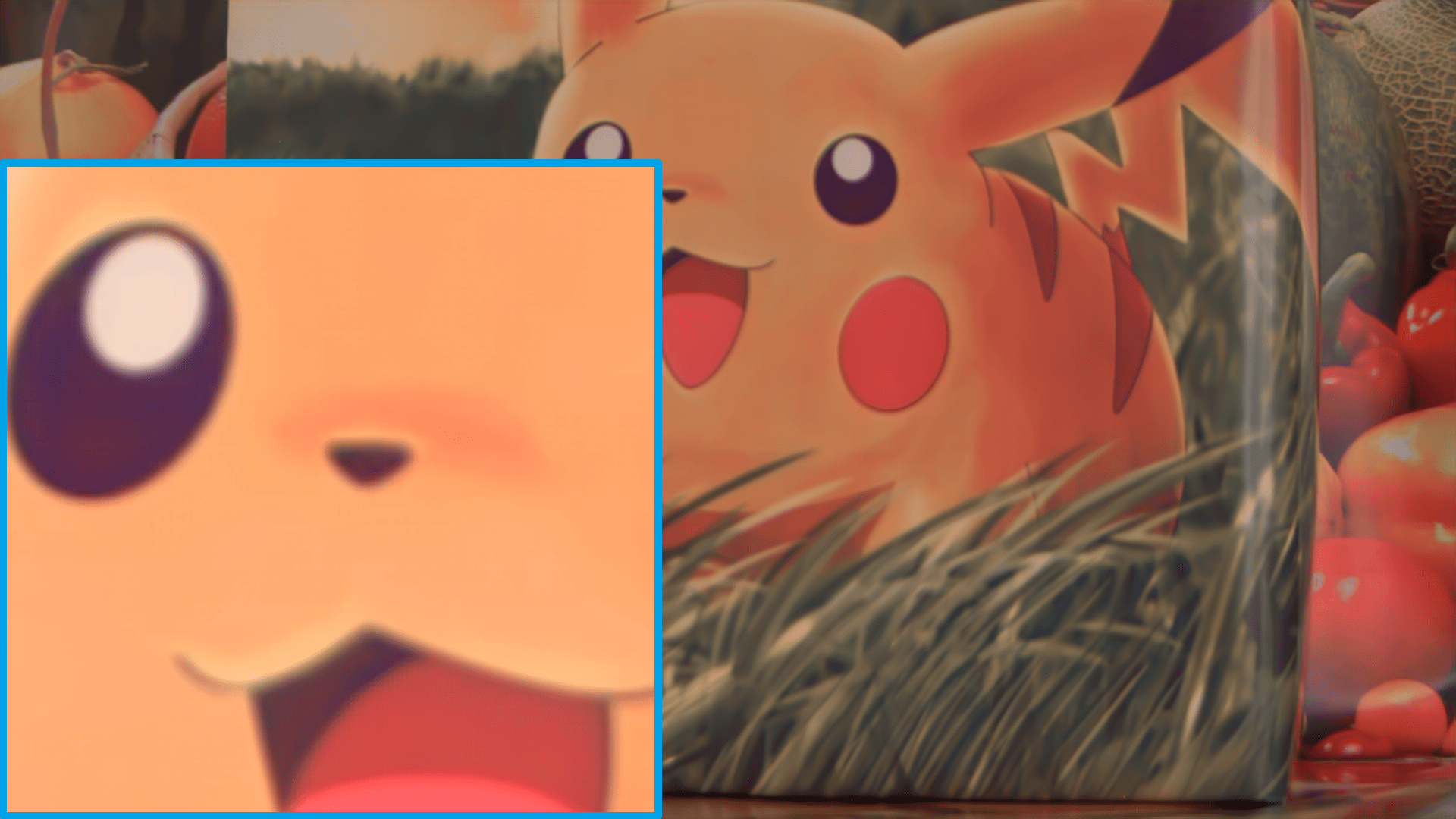}} \,
  \subfloat[Noisy]{\includegraphics[width=1.1299in, height = 0.6356in]{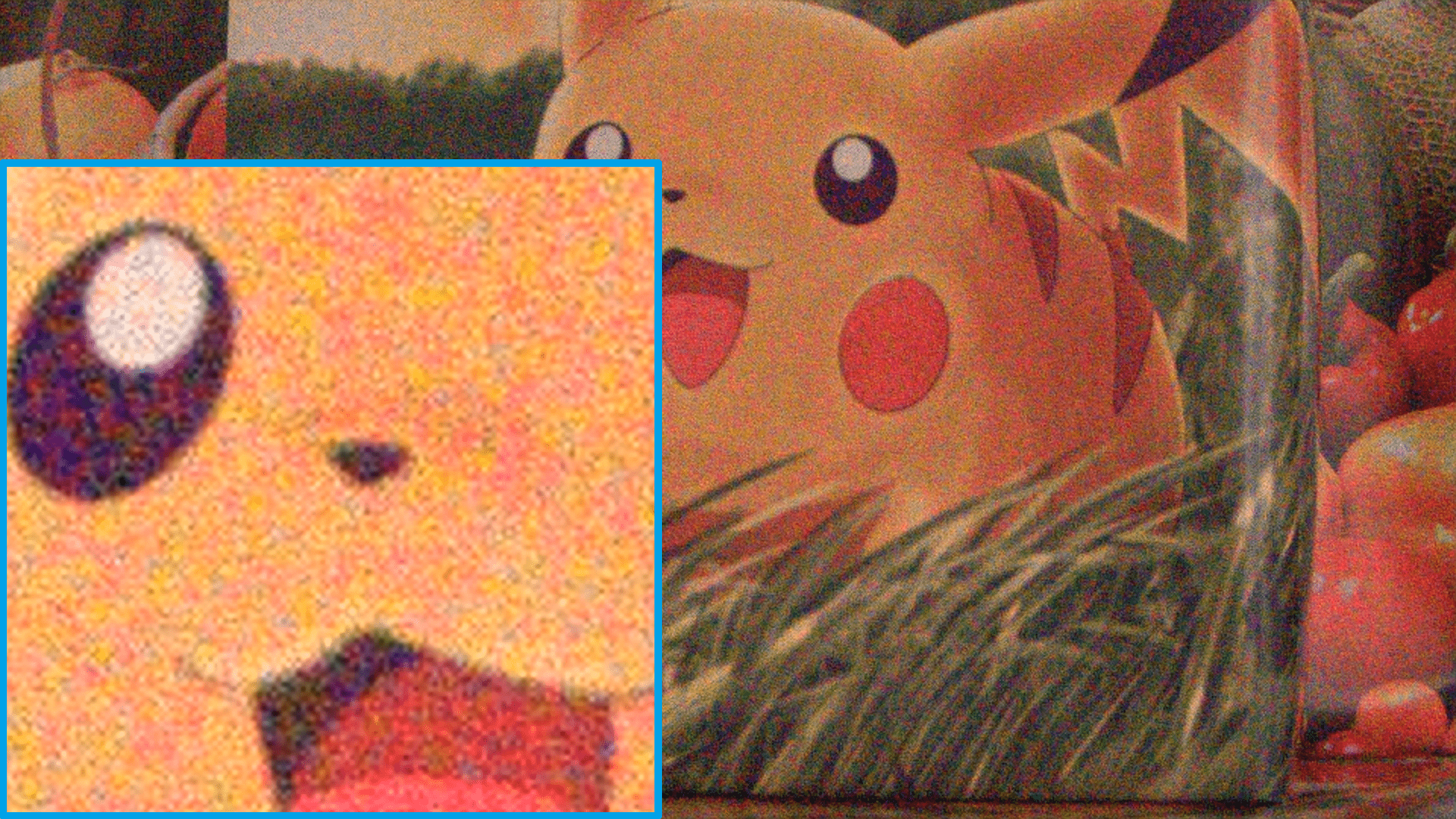}} \, 
  \subfloat[FastDVDnet]{\includegraphics[width=1.1299in, height = 0.6356in]{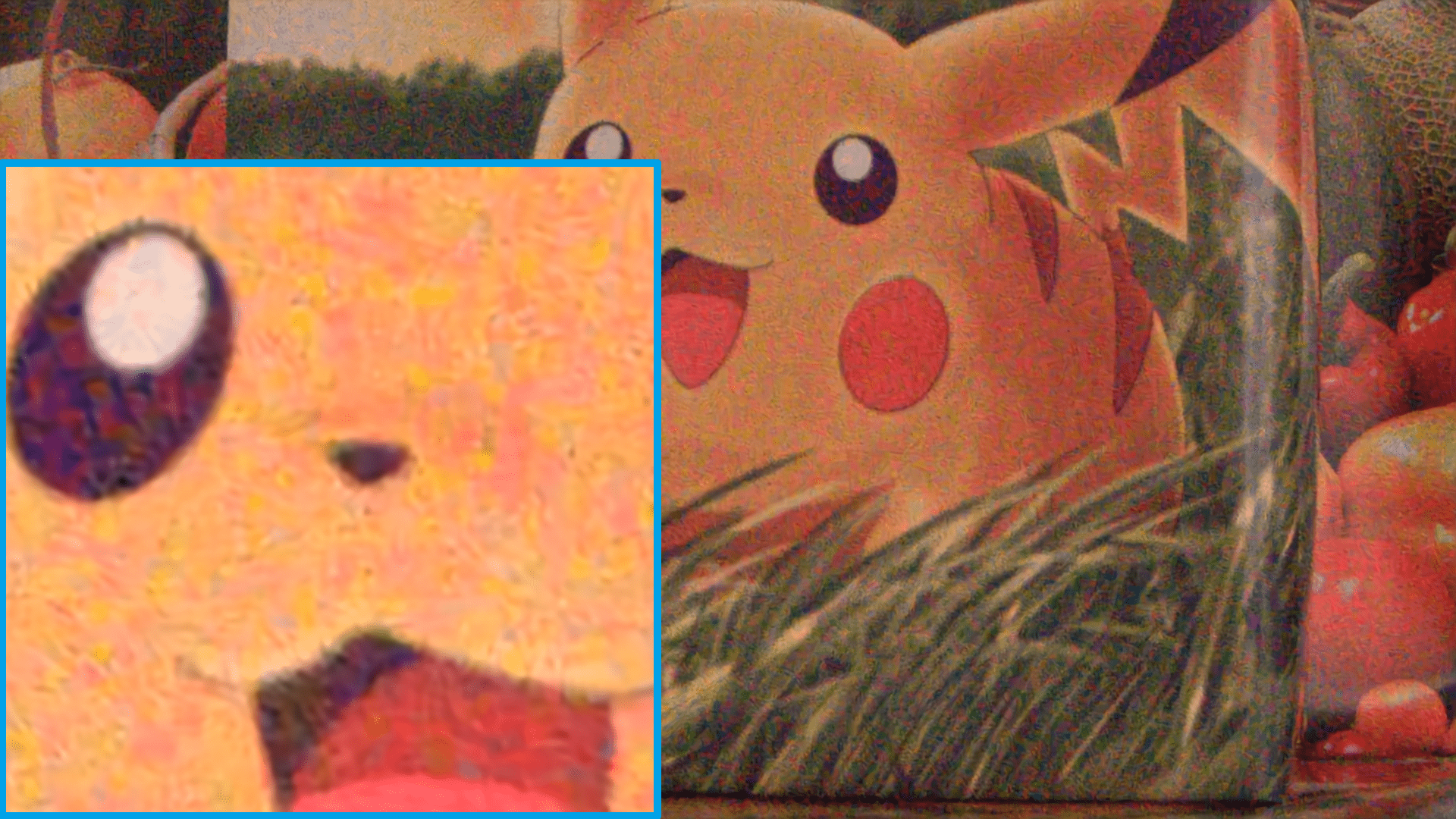}} \,
  \subfloat[RVRT]{\includegraphics[width=1.1299in, height = 0.6356in]{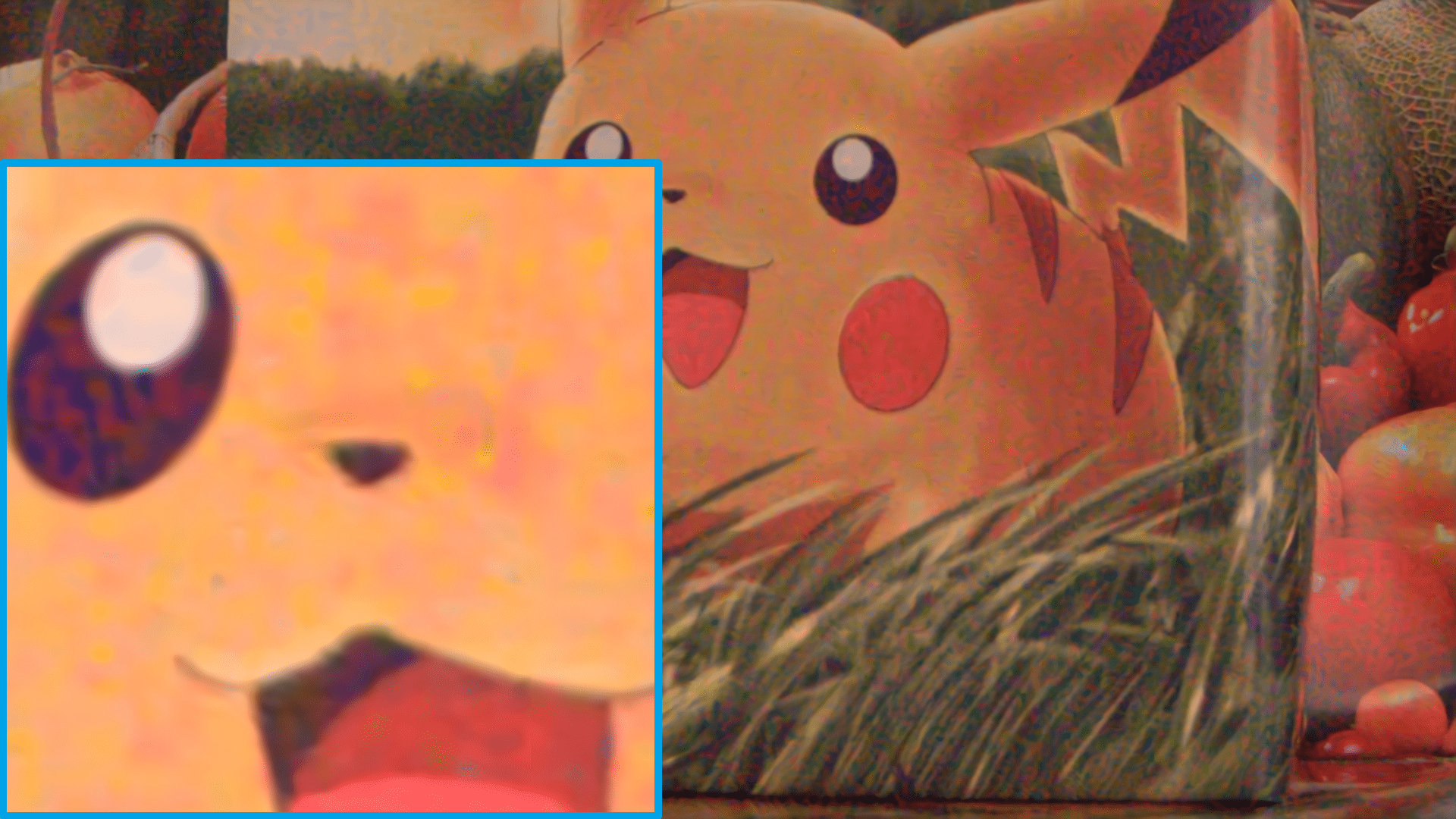}} \,
  \subfloat[VNLNet]{\includegraphics[width=1.1299in, height = 0.6356in]{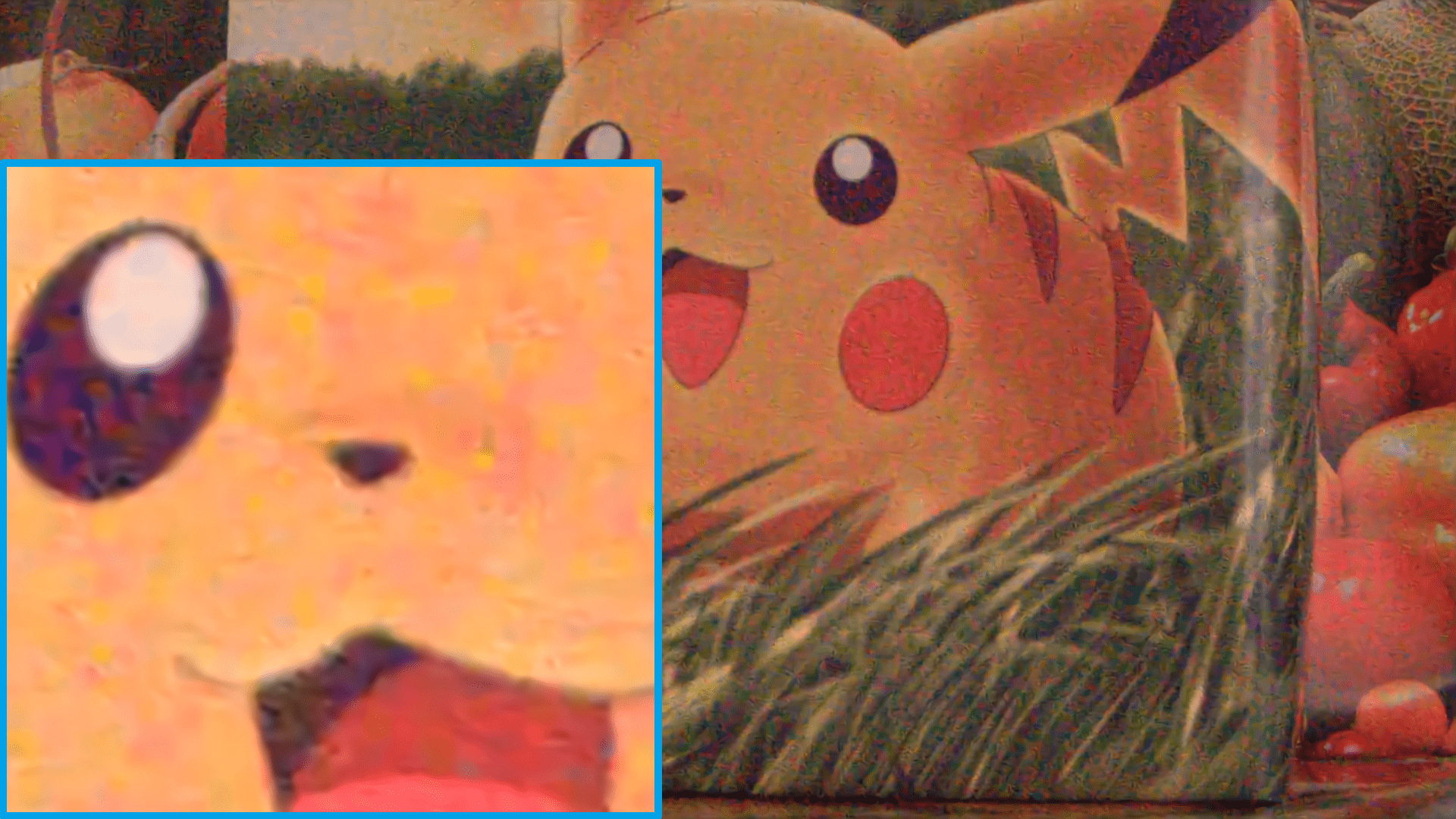}} \,
  \subfloat[GCP-ID + CNN]{\includegraphics[width=1.1299in, height = 0.6356in]{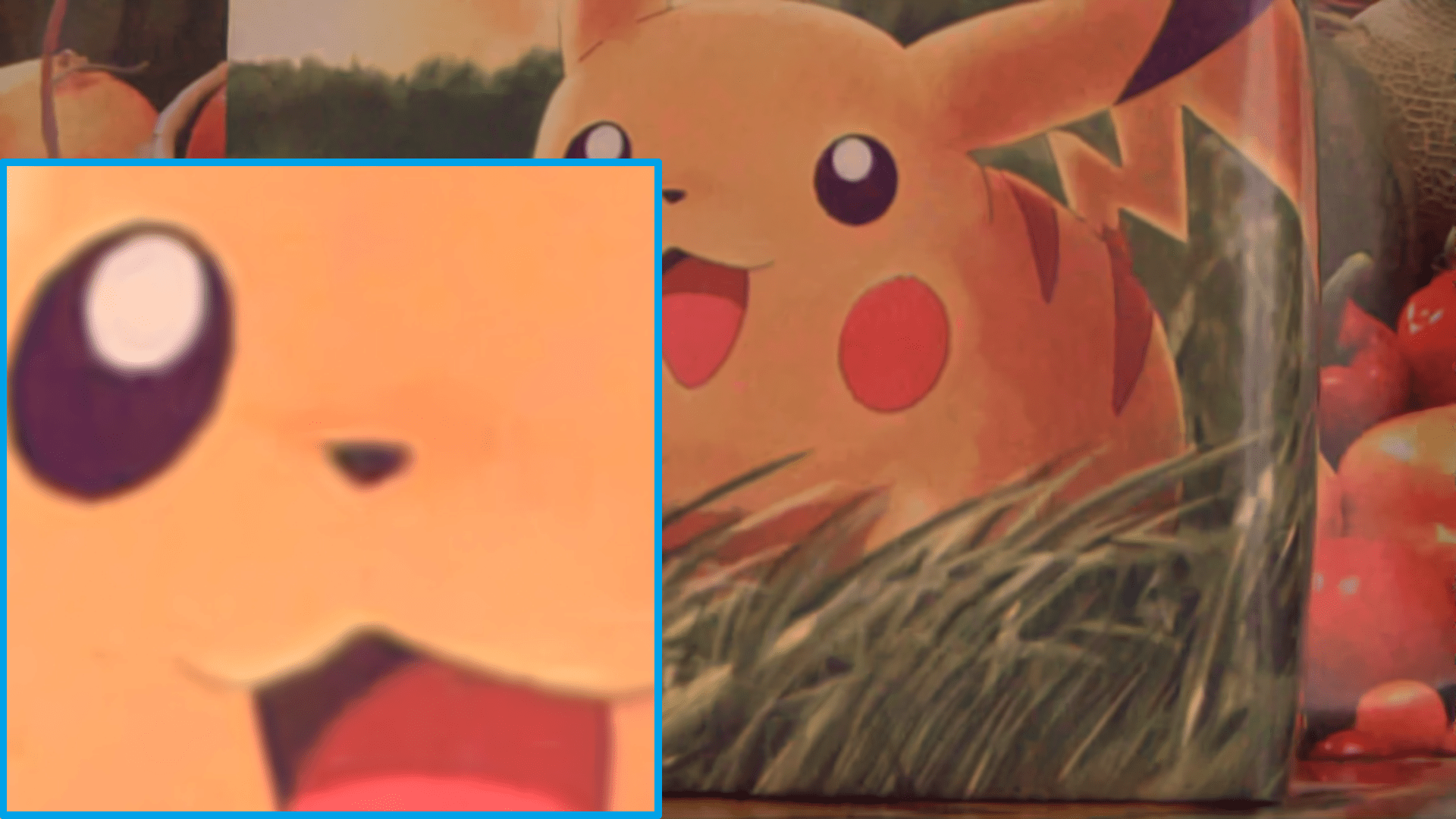}}  
  \vspace{-3.8pt}
  \caption{Denoising results of compared methods on the CRVD (sRGB) dataset. Top to bottom: ISO3200, ISO12800, ISO25600.}
  \label{Fig_CRVDsRGB_indoor_new}
  \vspace{-15.8pt}
\end{figure*}

%% file: Fig_CVRD_outdoor.tex
\begin{figure}[htbp]
\vspace{-8.8pt}
\centering
  \graphicspath{{Figs/CRVD/CRVD_indoor_raw2sRGB_comparison/Sample1/combined/}}
  \subfloat{\includegraphics[width=1.098in, height = 0.8506in]{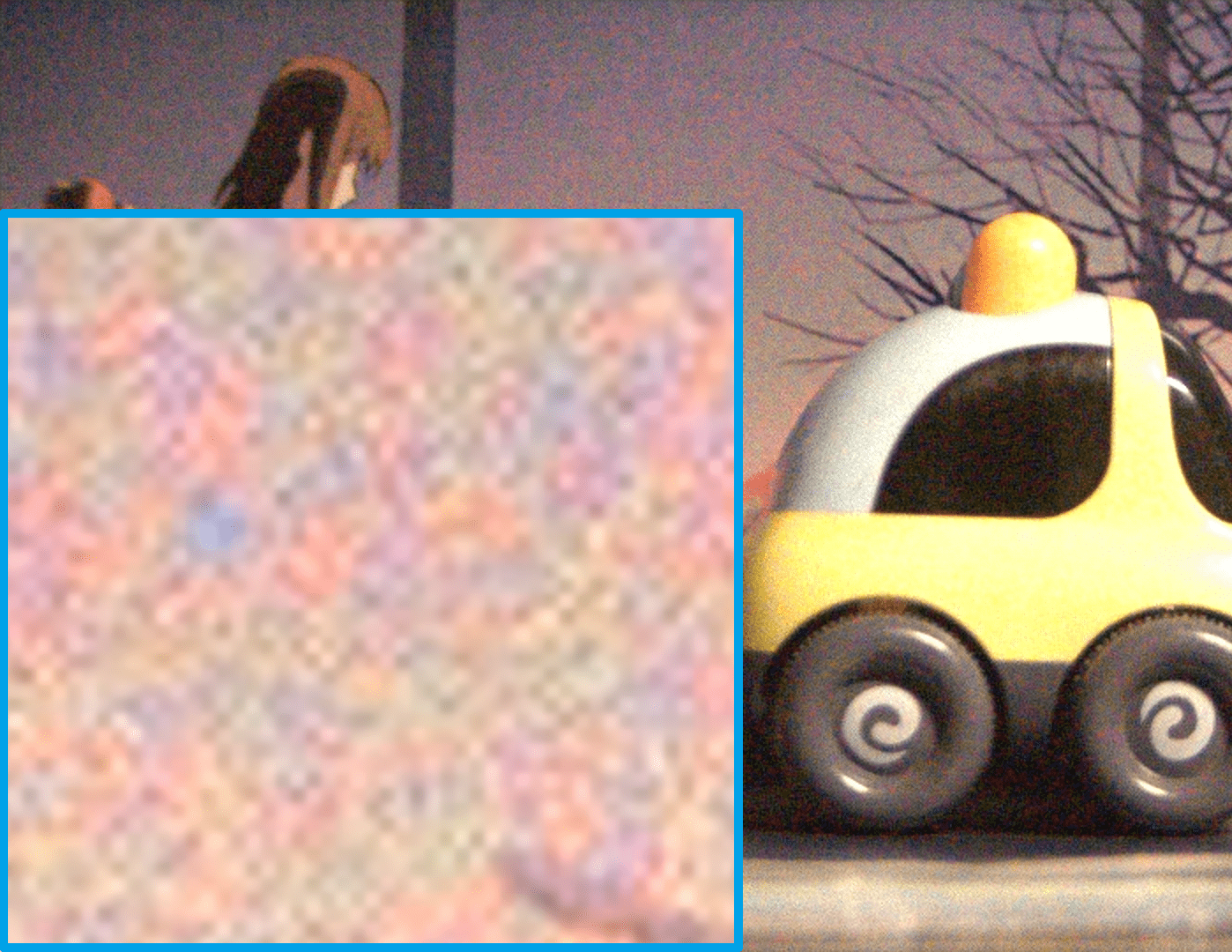}}\,
  \subfloat{\includegraphics[width=1.098in, height = 0.8506in]{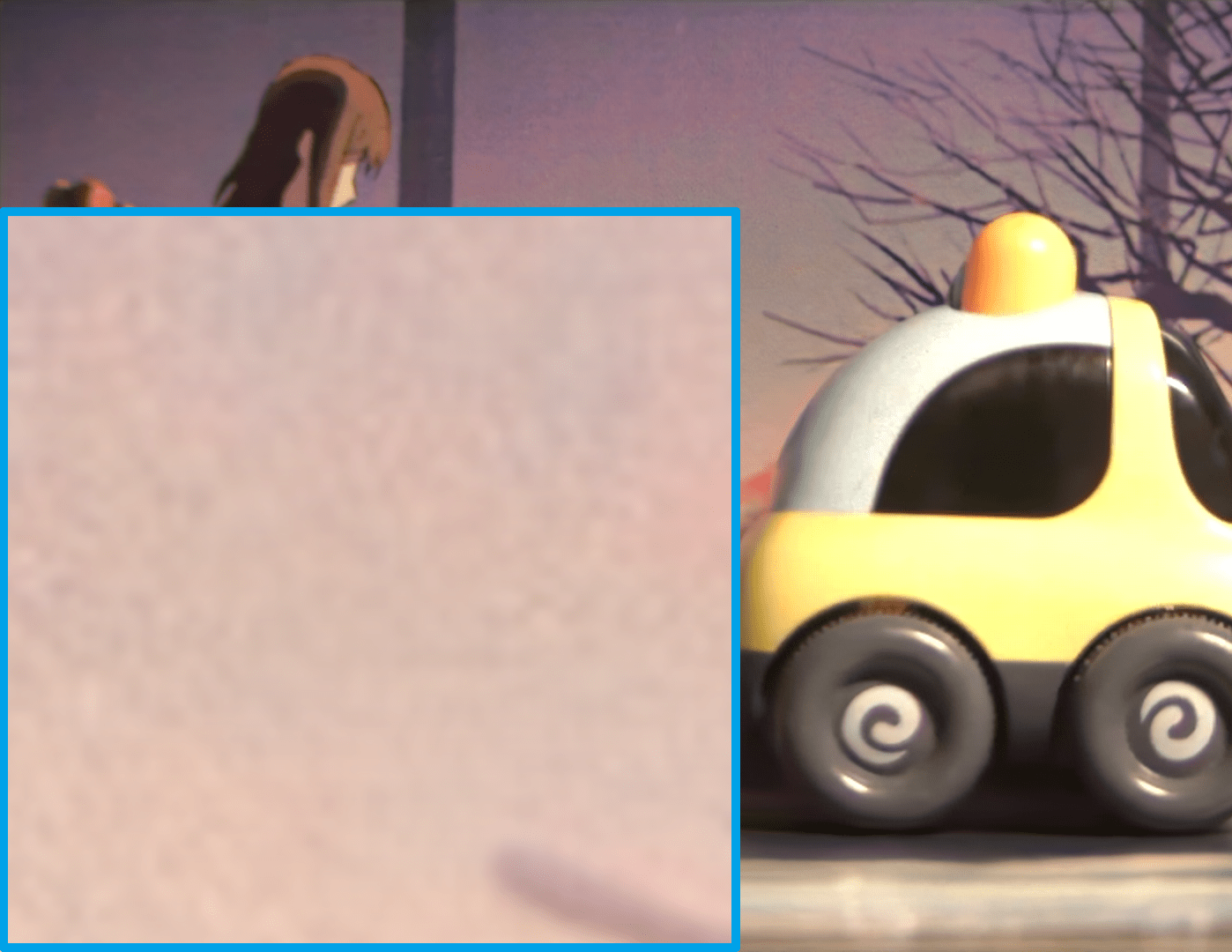}} \, 
  \subfloat{\includegraphics[width=1.098in, height = 0.8506in]{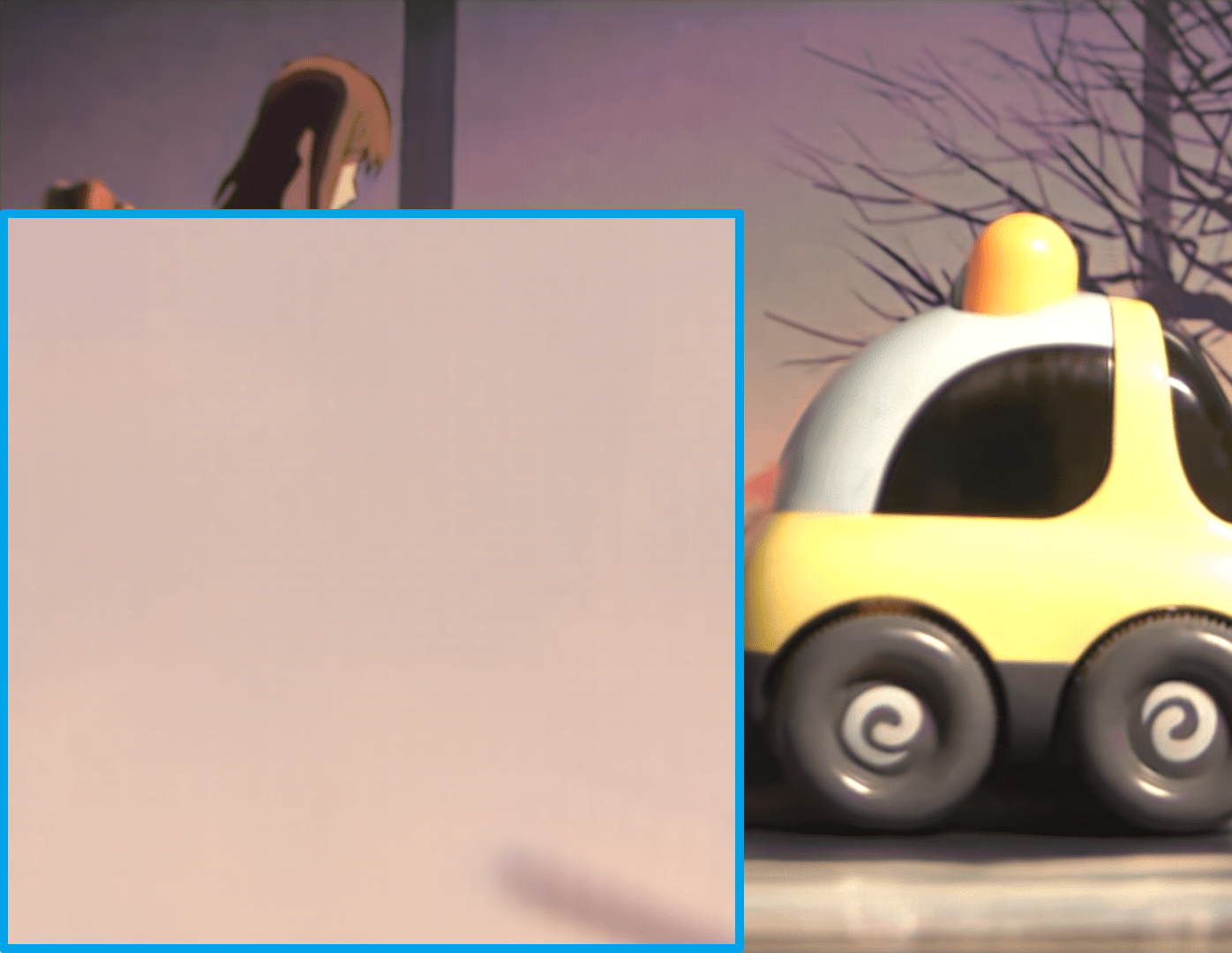}} \\
  \vspace{-8pt}

  \graphicspath{{Figs/CRVD/CRVD_outdoor_raw2sRGB_comparison/combined_new/}}
  \subfloat{\includegraphics[width=1.098in, height=0.7816in]{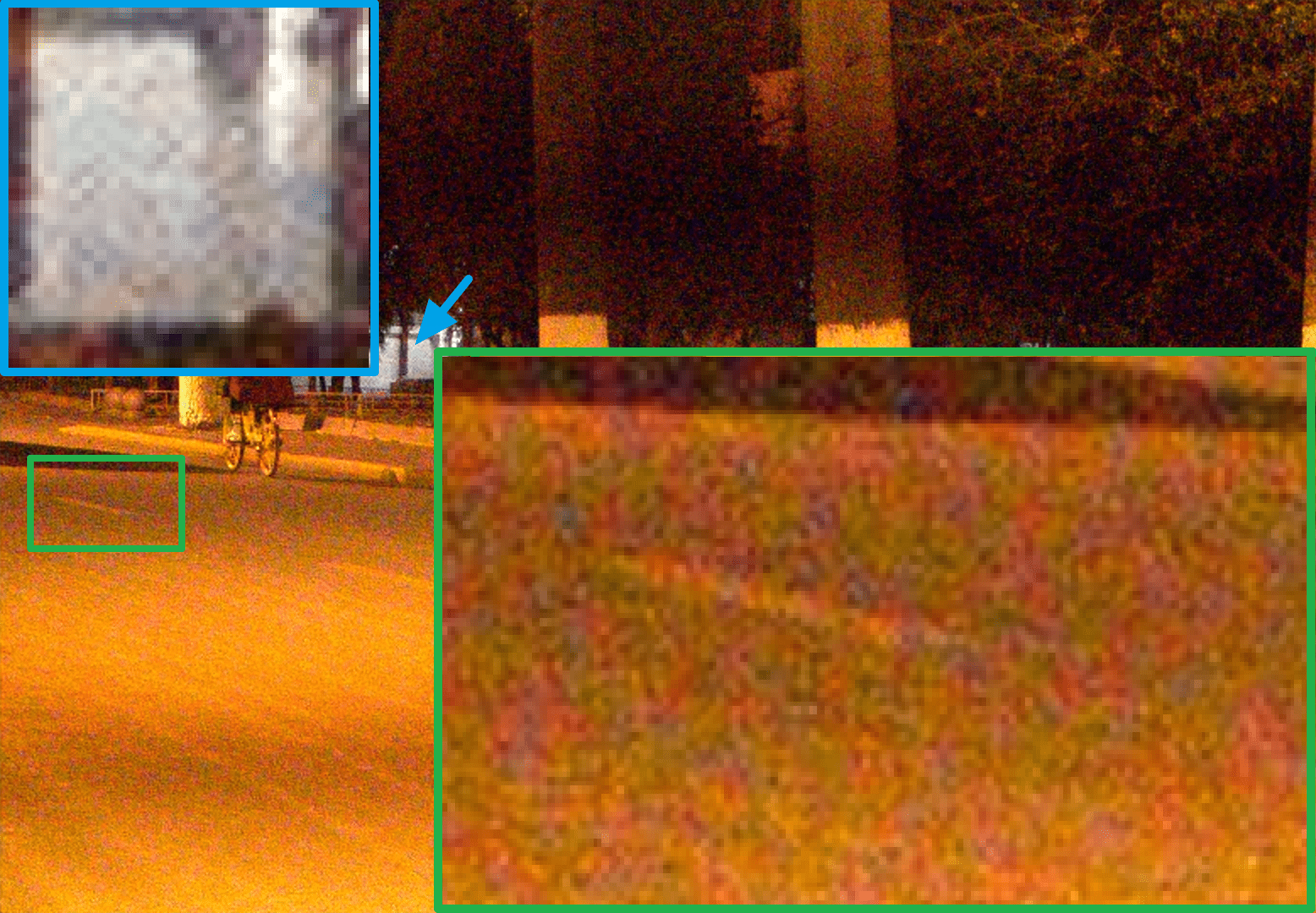}}\,
  \subfloat{\includegraphics[width=1.098in, height=0.7816in]{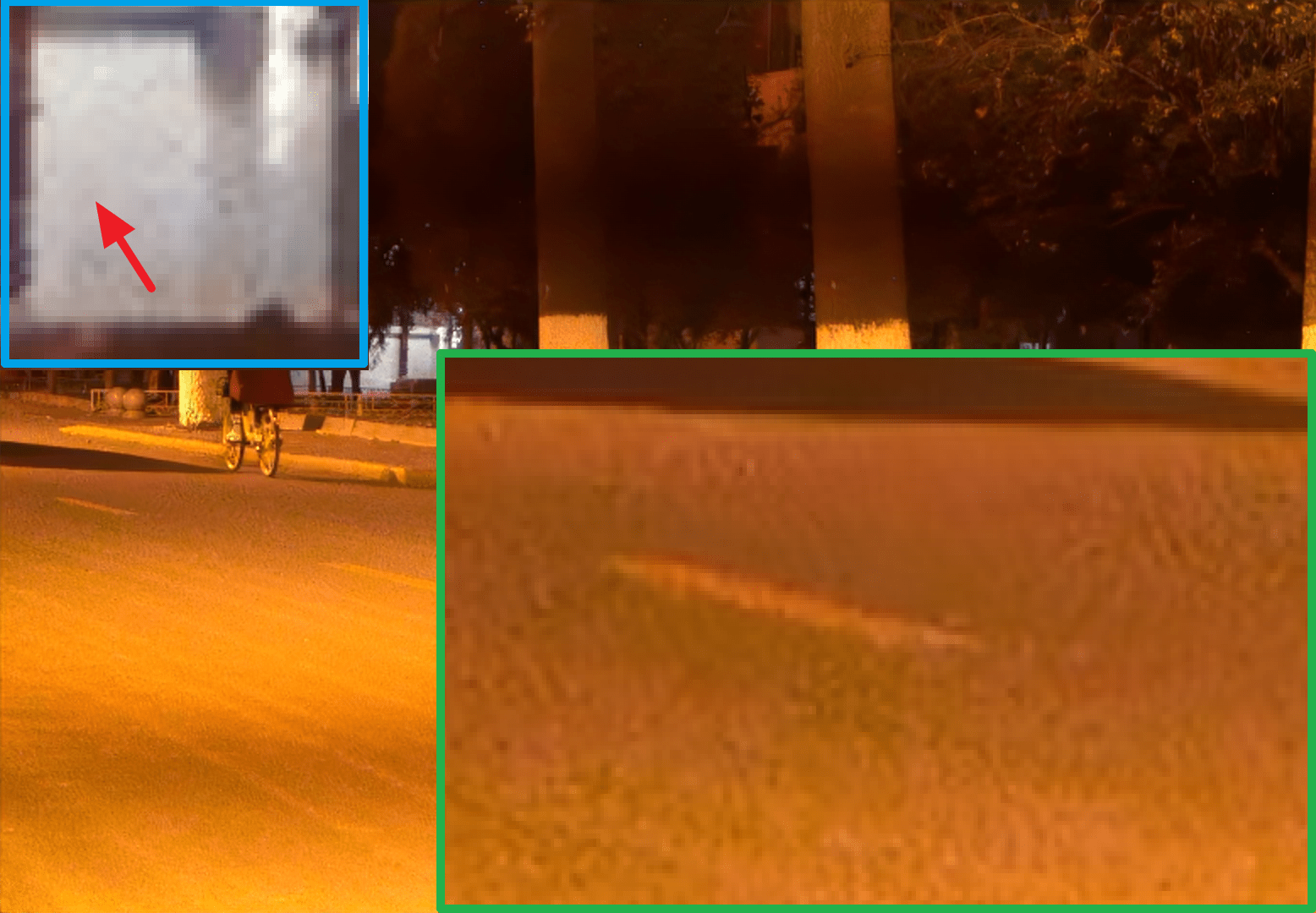}} \, 
  \subfloat{\includegraphics[width=1.098in, height=0.7816in]{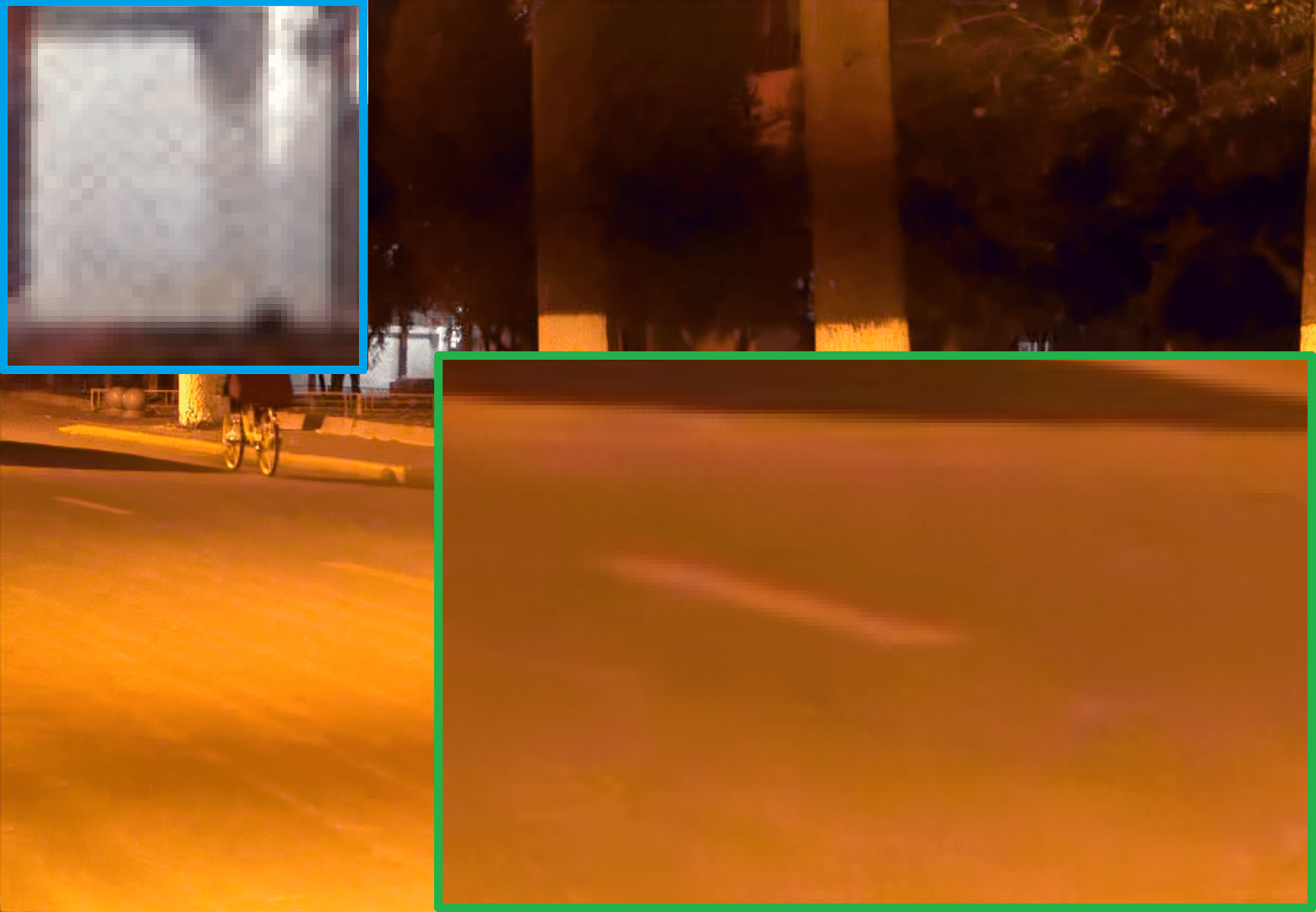}} \\
  \vspace{-8pt}
  
  \setcounter{subfigure}{0}
  \subfloat[Noisy]{\includegraphics[width=1.098in, height=0.868in]{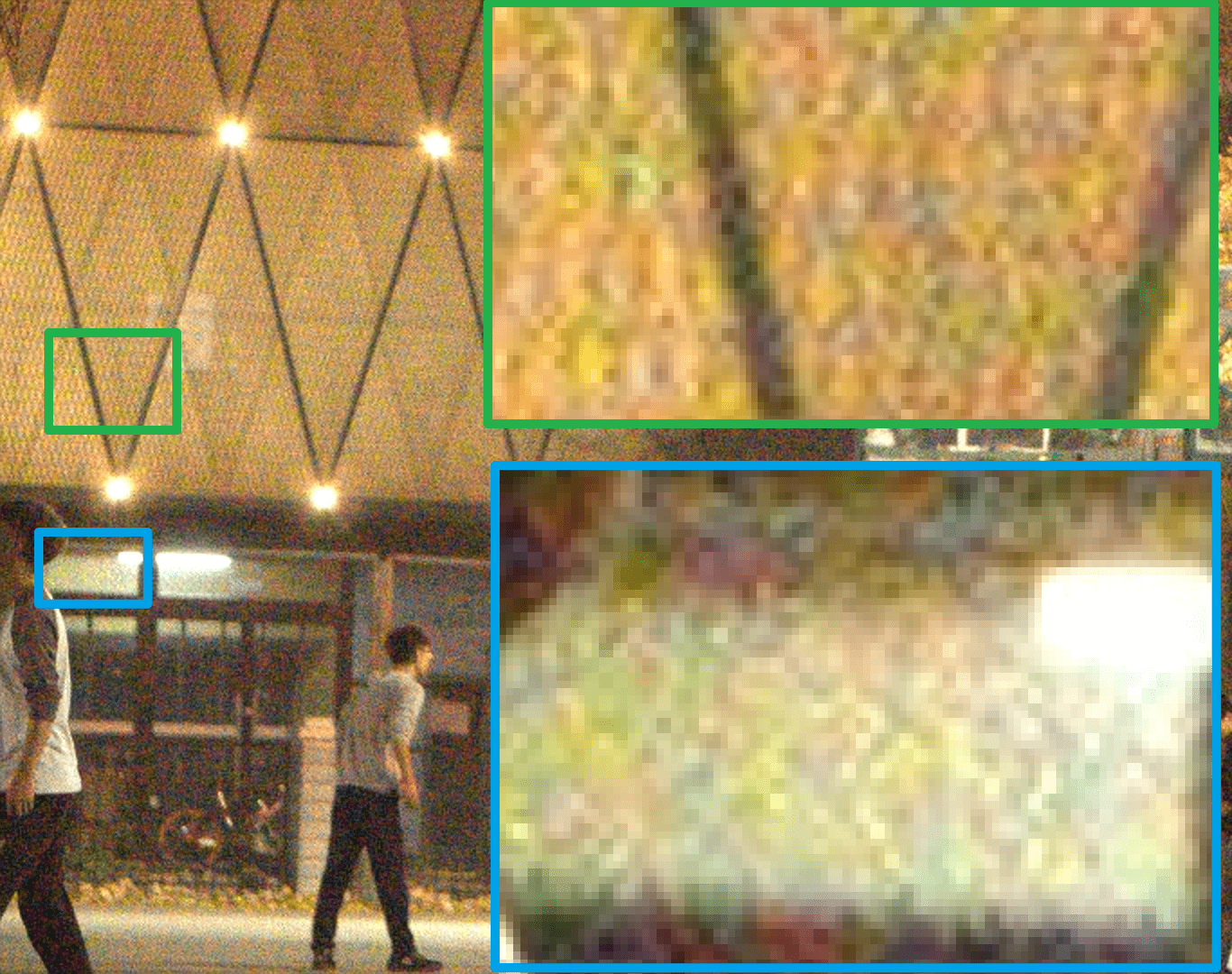}}\,
  \subfloat[MaskDnGAN]{\includegraphics[width=1.098in, height=0.868in]{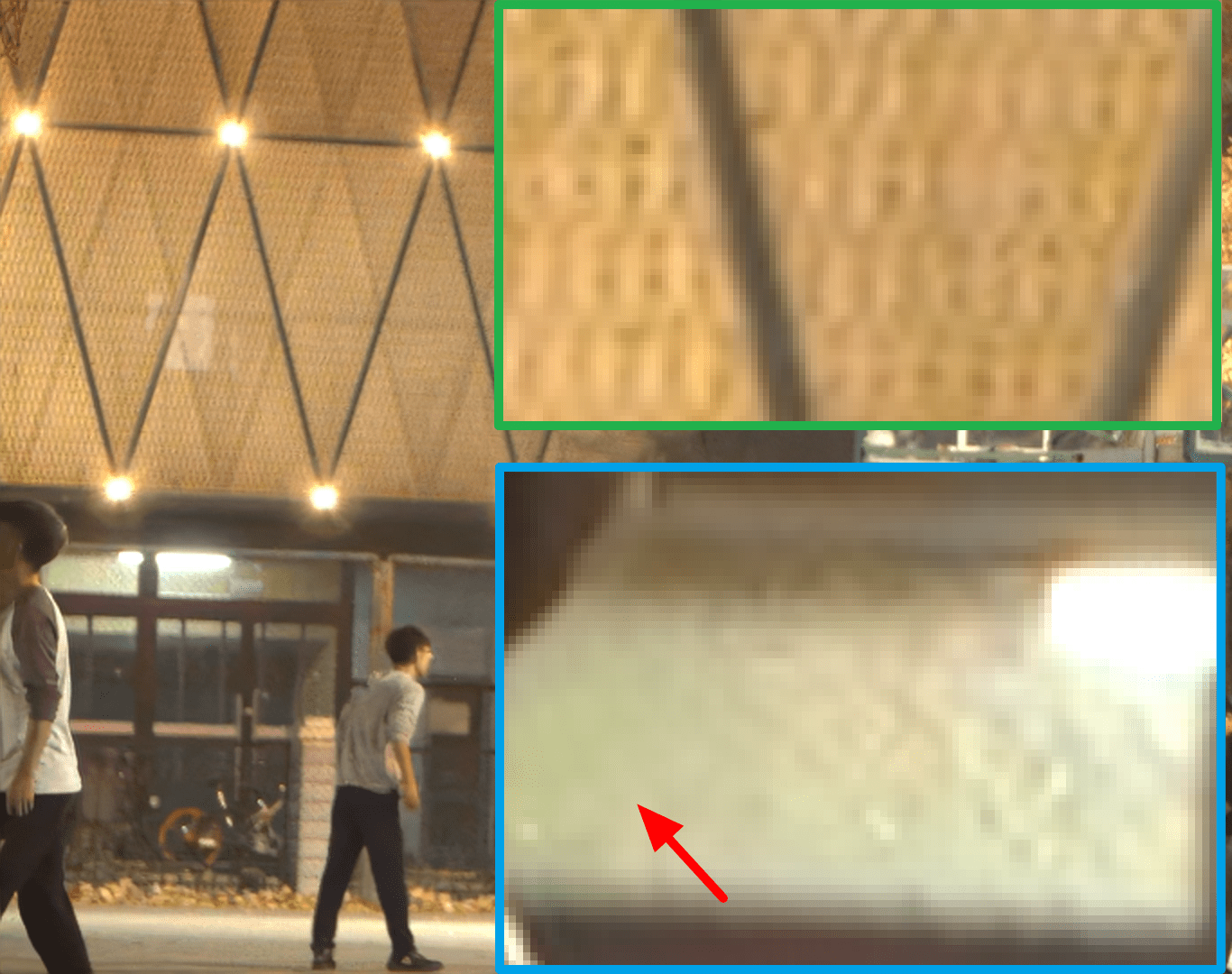}} \, 
  \subfloat[GCP-ID + CNN]{\includegraphics[width=1.098in, height=0.868in]{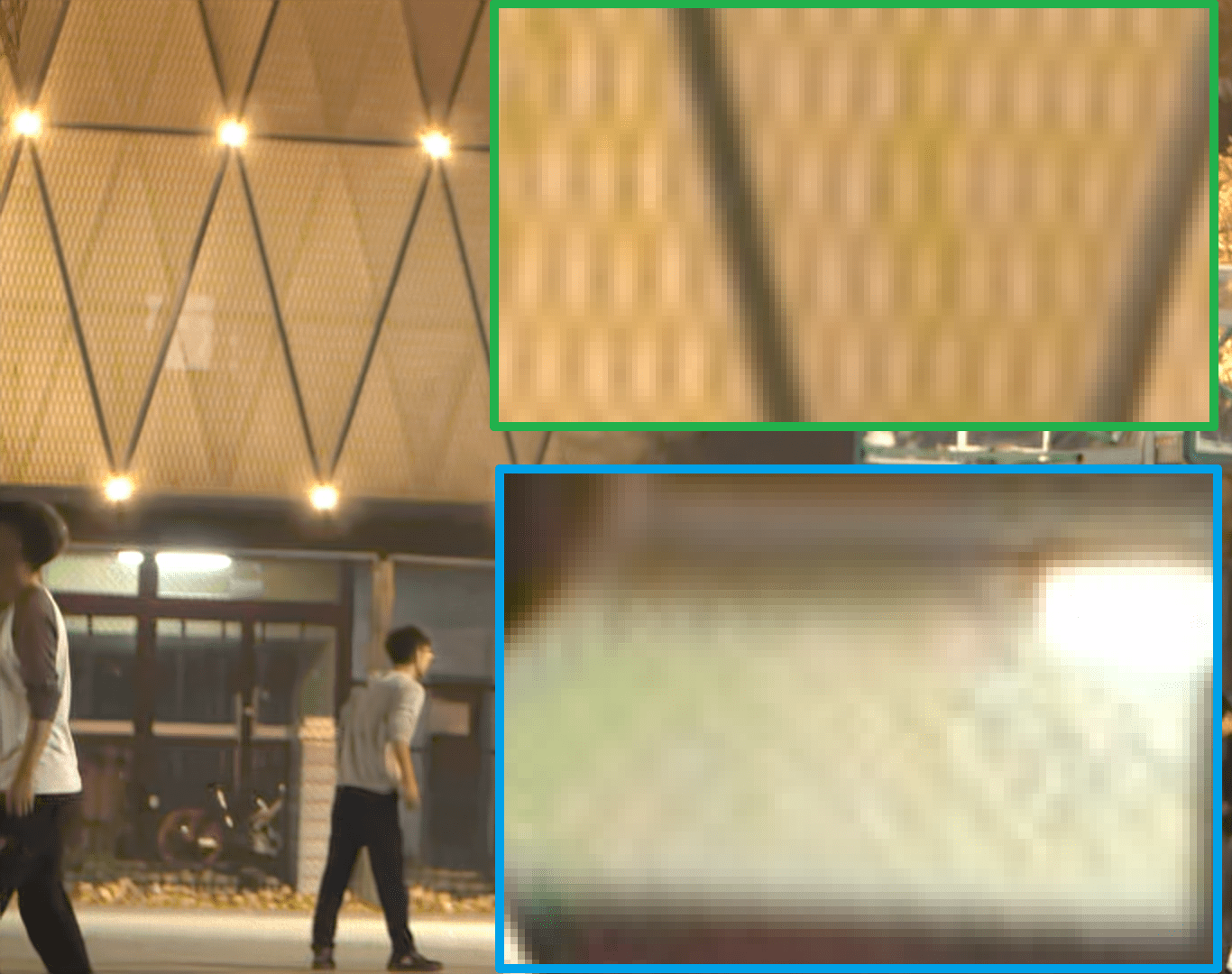}} \\
  \vspace{-3pt}
  \caption{Illustration of denoising results on the CRVD (Raw) dataset with high ISO values. Please zoom in for a better view.}
  \label{Fig_CRVD_outdoor}
  \vspace{-18pt}
\end{figure}

%% file: Fig_parameter_analysis.tex
\begin{figure}[htbp]
\vspace{-6.8pt}
\centering
  \graphicspath{{Figs/Discussion/parameter_sensitivity/}}
  \subfloat[GCP-guided search weight \label{Fig_guided_search_weight_analysis}]{\includegraphics[width=1.668in, height = 1.418in]{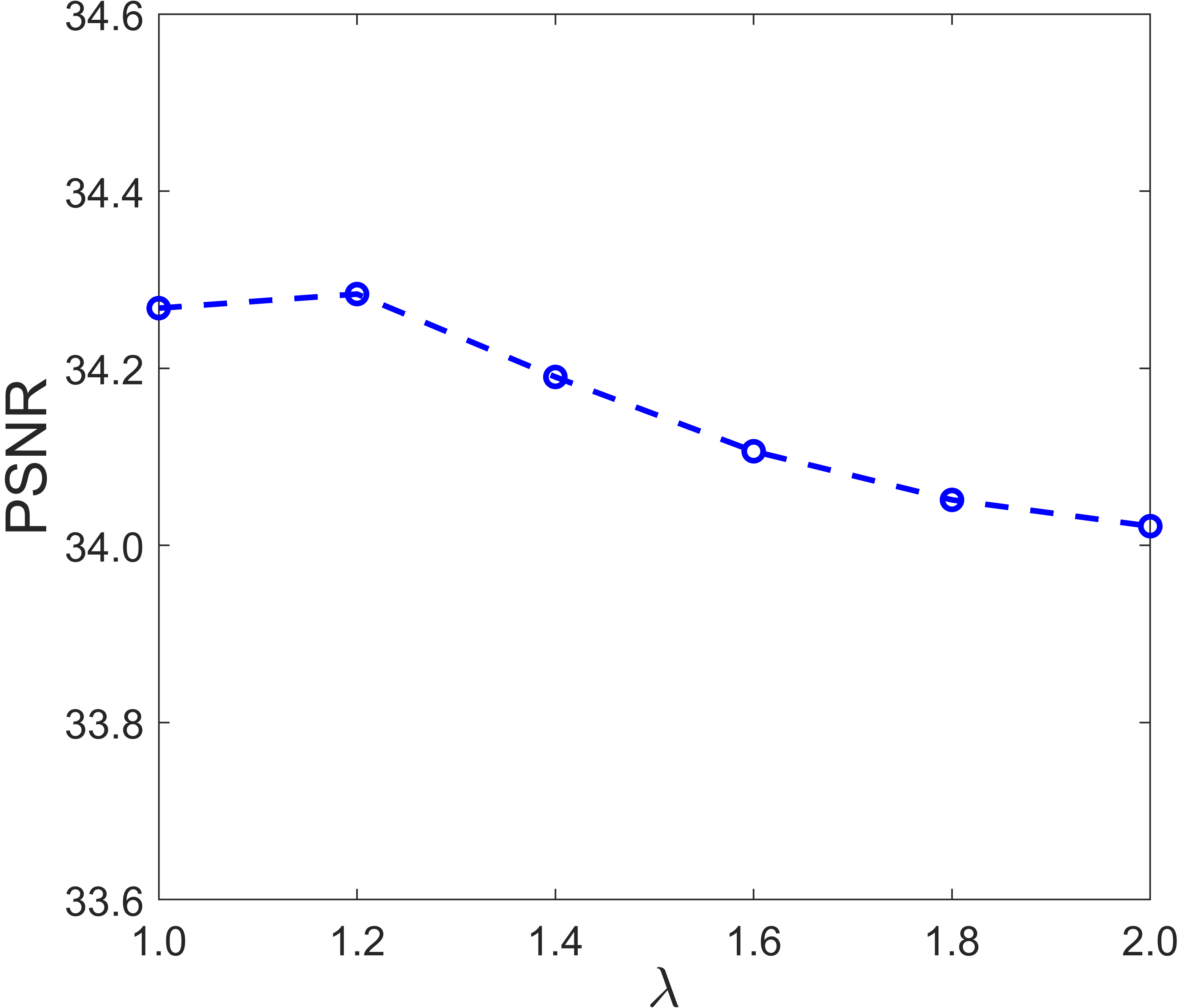}} \, 
  \subfloat[Number of patches \label{Fig_number_of_patches_analysis}]{\includegraphics[width=1.668in, height = 1.418in]{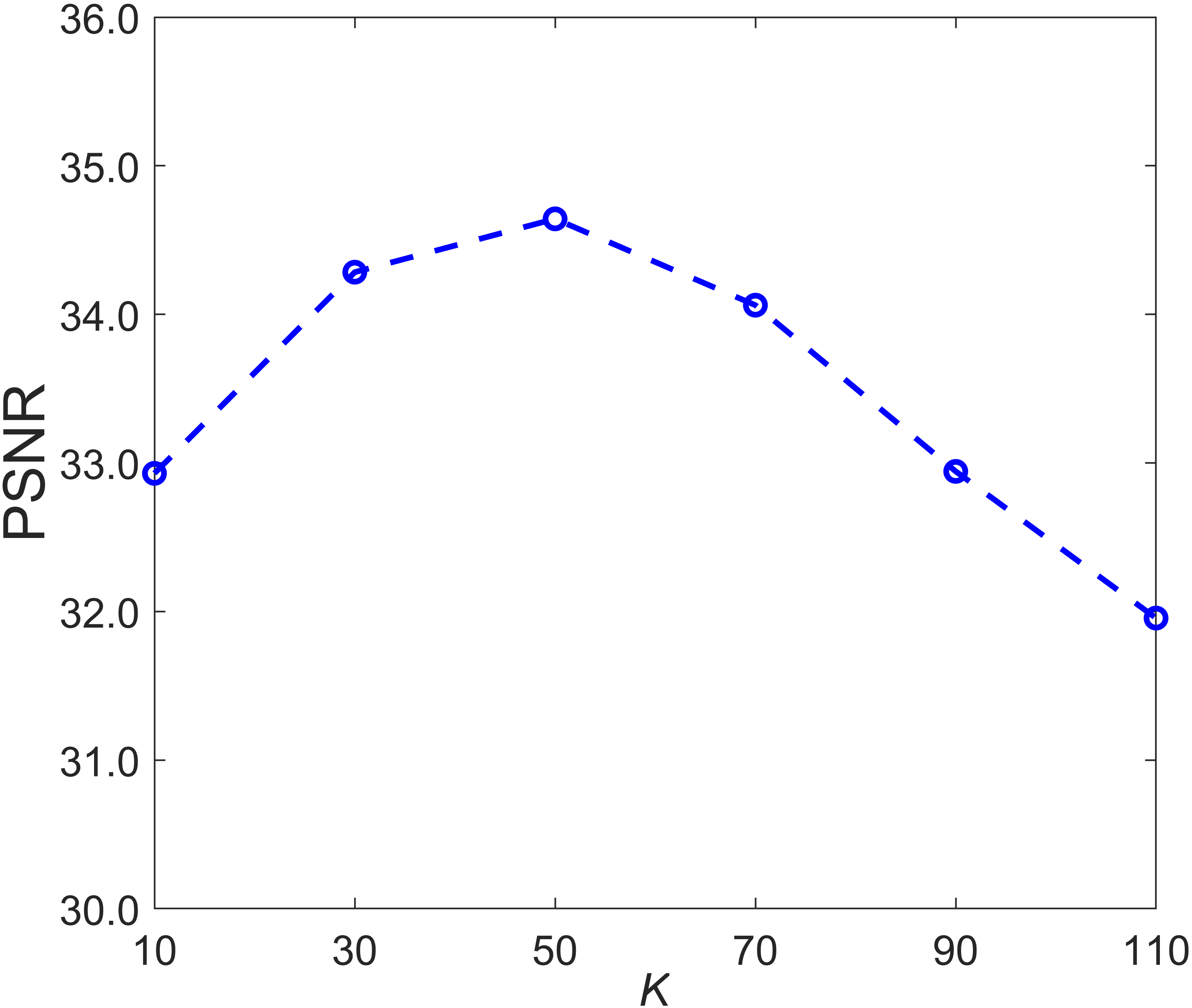}}\\
  \vspace{-2.6pt}
  \caption{Impact of different parameters on GCP-ID.}
  \label{Fig_parameter_analysis}
  \vspace{-6.8pt}
\end{figure}

%% file: Table_ablation_study.tex
\begin{table}[htbp]
\vspace{-2pt}
\scriptsize
  \centering
  \caption{The effectiveness of each step of GCP-ID}
  \scalebox{0.939}{
    \begin{tabular}{cccc}
    \toprule
    Dataset & GCP-guided patch search  & RGGB representation & GCP-ID \\
    \midrule
    SIDD validation & 34.18/0.873 & 34.41/0.865 & 34.66/0.881 \\
    \bottomrule
    \end{tabular}}%
    \vspace{-2pt}
  \label{Table_ablation_study}%
  \vspace{-0pt}
\end{table}%

%% file: Fig_Ablation_study_GCP_RGGB.tex

\begin{figure}[!htb]
\vspace{-6.6pt}
\centering
\graphicspath{{Figs/Discussion/Inpact_two_steps/Selected/img1/combined/}}
  \subfloat{\includegraphics[width=0.8in, height = 0.8in]{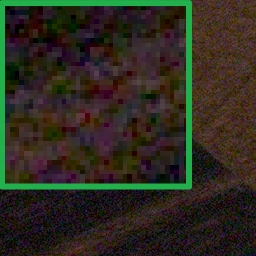}} \, 
  \subfloat{\includegraphics[width=0.8in, height = 0.8in]{GCP_guided_combined_marked.png}} \, 
  \subfloat{\includegraphics[width=0.8in, height = 0.8in]{RGGB_combined_marked.png}} \,
  \subfloat{\includegraphics[width=0.8in, height = 0.8in]{GCP_ID_combined_marked.png}}\\
  \vspace{-6.8pt}
  \setcounter{subfigure}{0}
  \graphicspath{{Figs/Discussion/Inpact_two_steps/Selected/img3/combined/}}
  \subfloat[Noisy]{\includegraphics[width=0.8in, height = 0.8in]{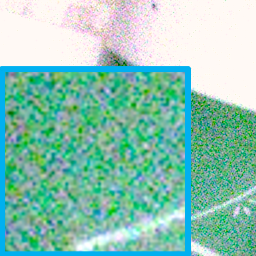}} \, 
  \subfloat[GCP-guided]{\includegraphics[width=0.8in, height = 0.8in]{GCP_guided_combined_marked.png}} \, 
  \subfloat[RGGB]{\includegraphics[width=0.8in, height = 0.8in]{RGGB_combined_marked.png}} \,
  \subfloat[GCP-ID]{\includegraphics[width=0.8in, height = 0.8in]{GCP_ID_combined_marked.png}}\\
  \vspace{-3.8pt}
  \caption{Denoising results of different steps of GCP-ID.}
  \label{Fig_Ablation_study_GCP_RGGB}
  \vspace{-2.6pt}
\end{figure}

%% file: Table_CNN_approximate.tex
\begin{table}[htbp]
\vspace{-1.8pt}
\scriptsize
  \centering
  \caption{Average prediction accuracy ($\%$) of the proposed CNN estimator on SIDD and CC datasets.}
    \begin{tabular}{cccc}
    \toprule
    Estimator & SIDD-validation (raw) & SIDD-validation (sRGB) & CC \\
    \midrule
    CNN   & 85.6  & 72.3  & 70.5 \\
    \midrule
    CNN (approximate) & 98.3  & 85.8  & 83.9 \\
    \bottomrule
    \end{tabular}%
  \vspace{-0.8pt}
  \label{Table_CNN_approximate}%
  \vspace{0pt}
\end{table}%

%% file: Table_compare_CNN_and_Chen.tex
\begin{table}[htbp]
\vspace{-3.8pt}
  \centering
  \caption{Performance of different noise estimators.}
    \begin{tabular}{cccc}
    \toprule
    Method/Estimator & SIDD (sRGB) & DND (sRGB) & CC \\
    \midrule
    GCP-ID + CNN & \textbf{35.56/0.926} & \textbf{38.36/0.943} & \textbf{38.25/0.962} \\
    \midrule
    GCP-ID + Chen & 33.63/0.886 & 36.85/0.893 & 37.32/0.946 \\
    \bottomrule
    \end{tabular}%
  \label{Table_compare_CNN_and_Chen}%
  \vspace{-12.8pt}
\end{table}%

%% file: Fig_CNN_compare_Chen.tex
\begin{figure}[htbp]
\vspace{-3.8pt}
\centering
\graphicspath{{Figs/Discussion/Effectiveness_of_CNN/Sample1/combined/}}
  \subfloat{\includegraphics[width=1.019in, height = 1.019in]{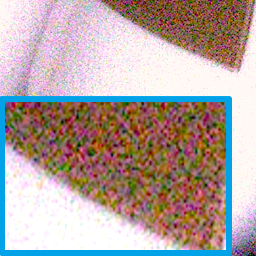}} \, 
  \subfloat{\includegraphics[width=1.019in, height = 1.019in]{GCP_ID_combined_marked}} \, 
  \subfloat{\includegraphics[width=1.019in, height = 1.019in]{GCP_ID_CNN_combined_marked}} \\
  \vspace{-6.6pt}
  \setcounter{subfigure}{0}
  \graphicspath{{Figs/Discussion/Effectiveness_of_CNN/Sample3/combined/}}
  \subfloat[Noisy]{\includegraphics[width=1.019in, height = 1.019in]{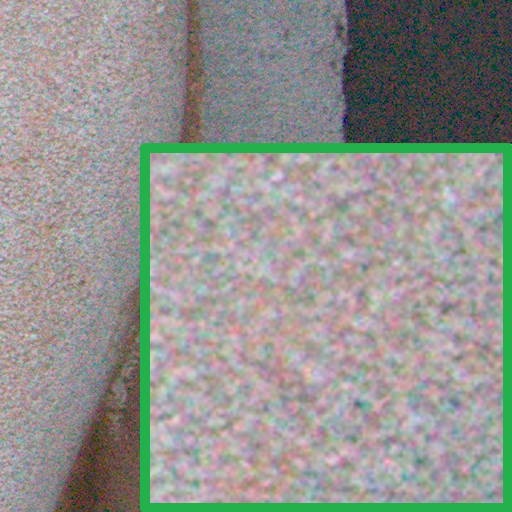}} \, 
  \subfloat[GCP-ID + Chen]{\includegraphics[width=1.019in, height = 1.019in]{GCP_ID_combined_marked}} \, 
  \subfloat[GCP-ID + CNN]{\includegraphics[width=1.019in, height = 1.019in]{GCP_ID_CNN_combined_marked}}
  \vspace{-2.6pt}
  \caption{Visual results of different noise estimators.}
  \label{Fig_CNN_compare_Chen}
  \vspace{-0.8pt}
\end{figure}

%% file: Table_customize_GCP.tex

\begin{table}[htbp]
 \vspace{-3.6pt}
\scriptsize
  \centering
  \caption{Results of different GCP-ID implementations.}
   \vspace{-2.6pt}
    \begin{tabular}{cccc}
    \toprule
    Dataset  & GCP-ID & GCP-ID + CNN  & GCP-ID + CNN (modified) \\
    \midrule
    DND (sRGB) & 38.24/0.941 & 38.36/0.944 & \textbf{38.55/0.948} \\
    \bottomrule
    \end{tabular}%
   \label{Table_customize_GCP}%
   \vspace{-0pt}
\end{table}%

%% file: Fig_GCP_ID_customize.tex
\begin{figure}[htbp]
\vspace{-6.8pt}
\centering
\graphicspath{{Figs/Discussion/GCP_ID_customization/Sample2/combined/}}
  \subfloat{\includegraphics[width=0.8in, height=0.8in]{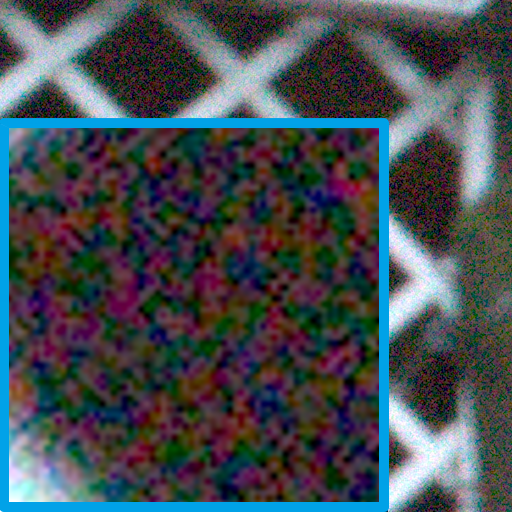}} \, 
  \subfloat{\includegraphics[width=0.8in, height=0.8in]{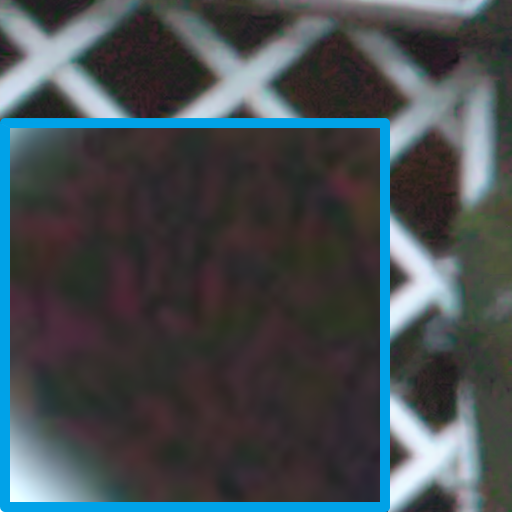}} \, 
  \subfloat{\includegraphics[width=0.8in, height=0.8in]{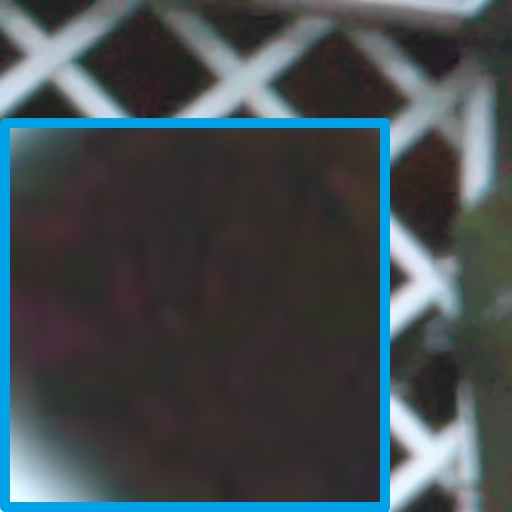}} \,
  \subfloat{\includegraphics[width=0.8in, height=0.8in]{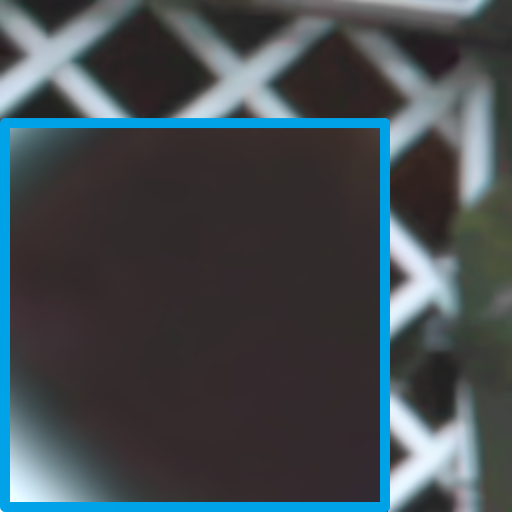}}\\
  
  \vspace{-6.6pt}
  \setcounter{subfigure}{0}
  \graphicspath{{Figs/Discussion/GCP_ID_customization/Sample5/combined/}}
  \subfloat[Noisy]{\includegraphics[width=0.8in, height=0.8in]{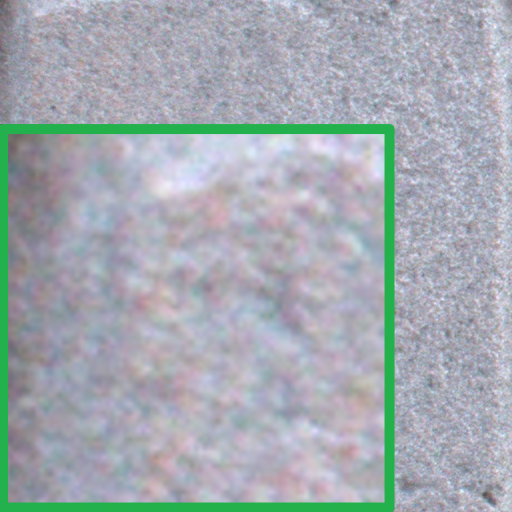}} \, 
  \subfloat[GCP]{\includegraphics[width=0.8in, height=0.8in]{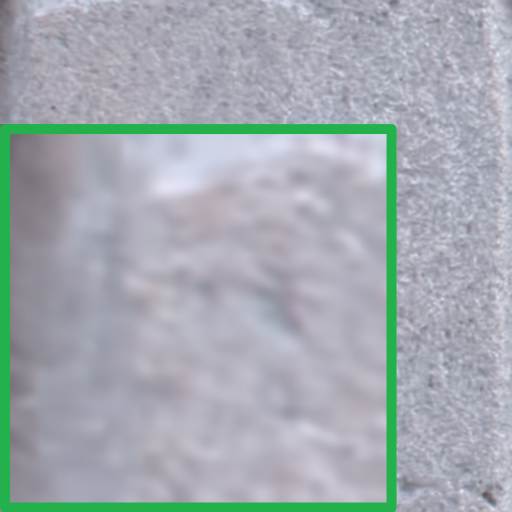}} \, 
  \subfloat[GCP + CNN]{\includegraphics[width=0.8in, height=0.8in]{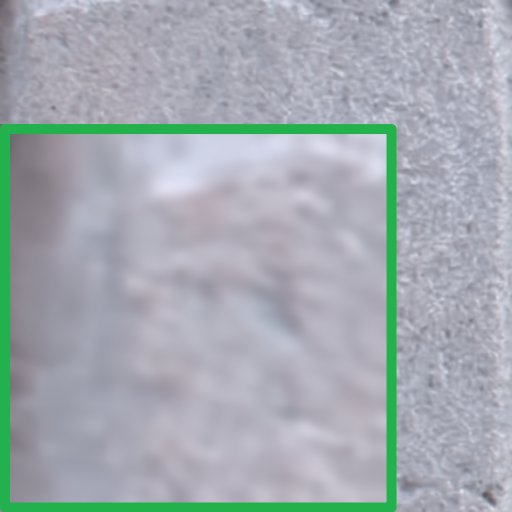}} \,
  \subfloat[Modified]{\includegraphics[width=0.8in, height=0.8in]{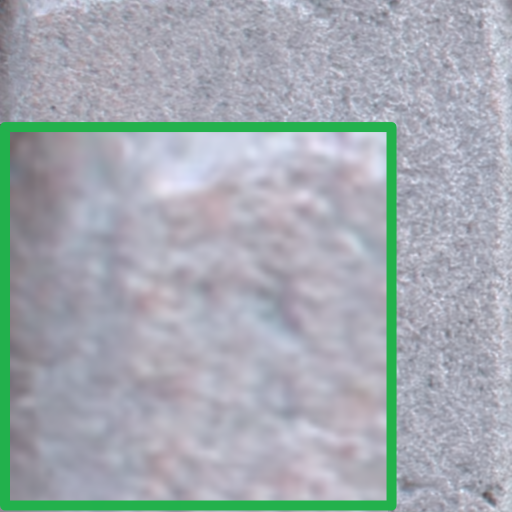}}
  \vspace{-2.6pt}
  \caption{Comparison of different GCP-ID implementations.}
  \label{Fig_GCP_ID_customize}
  \vspace{-0pt}
\end{figure}

%% file: Table_Real-HSI_denoising_results.tex
\begin{table}[htbp]
  \vspace{-3.6pt}
  \centering
  \caption{Denoising results on the Real-HSI dataset.}
  \vspace{-2.6pt}
  \scalebox{0.8516}{
    \begin{tabular}{cccccccc}
    \toprule
    \multirow{3}[3]{*}{Metrics} & \multicolumn{4}{c}{Traditional denoisers} & \multicolumn{3}{c}{DNN models} \\
\cmidrule{2-8}          & BM4D  & LTDL  & OLRT & GCP-ID  & H-DeNet & MAN   & sDeCNN \\
          & \cite{maggioni2012nonlocal} & \cite{gong2020low} & \cite{chang2020hyperspectral} & (ours) & \cite{chang2018hsi} & \cite{lai2023mixed} & \cite{maffei2019single} \\
    \midrule      
    PSNR  & 25.88  & 25.80  & \textbf{25.91} & \textbf{25.91} & 25.63  & 25.82  & 25.70  \\
    \midrule
    SSIM  & 0.865  & 0.851  & \textbf{0.870} & \textbf{0.870} & 0.853  & 0.869  & 0.860  \\
    \midrule
    Time (m) & 4.1   & 35.0  & 24.3   & 1.9  & 0.8   & 0.2   & 0.7  \\
    \bottomrule
    \end{tabular}}%
    \vspace{-3pt}
  \label{Table_Real-HSI_denoising_results}%
  \vspace{-6.8pt}
\end{table}%

%% file: Fig_Real-HSI_visual_evaluation.tex
\begin{figure}[htbp]
\graphicspath{{Figs/Real-HSI/combined/}}
\centering
  \subfloat[Noisy]{\includegraphics[width=1.0516in, height=1.0516in]{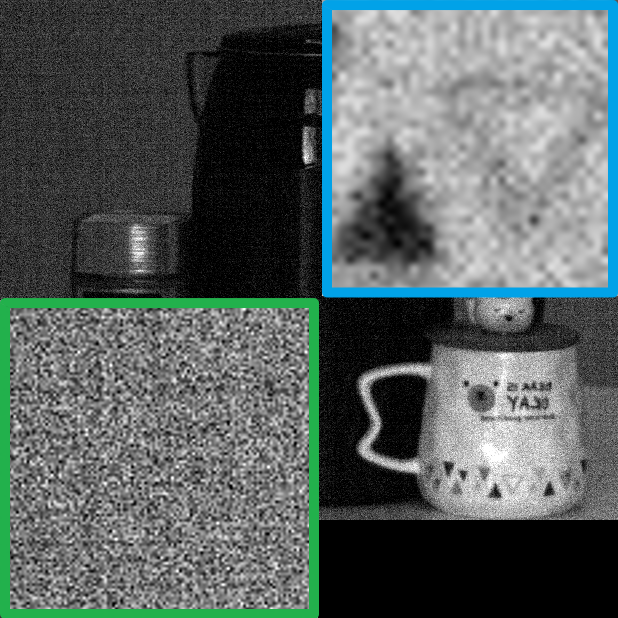}} \,
  \subfloat[BM4D]{\includegraphics[width=1.0516in, height=1.0516in]{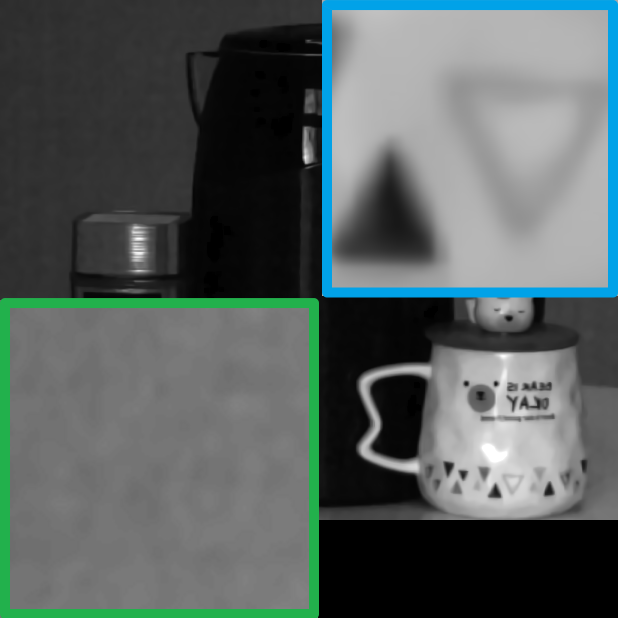}}\,
  \subfloat[LTDL]{\includegraphics[width=1.0516in, height=1.0516in]{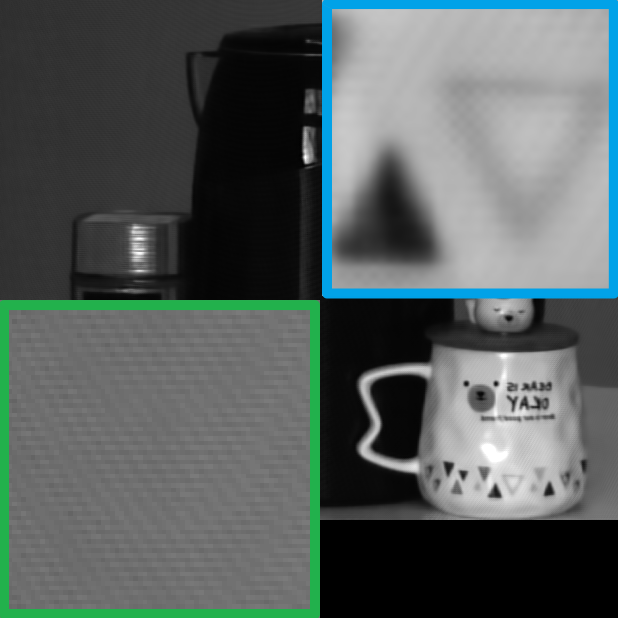}}\\
  \vspace{-10.18pt}
  \subfloat[OLRT]{\includegraphics[width=1.0516in, height=1.0516in]{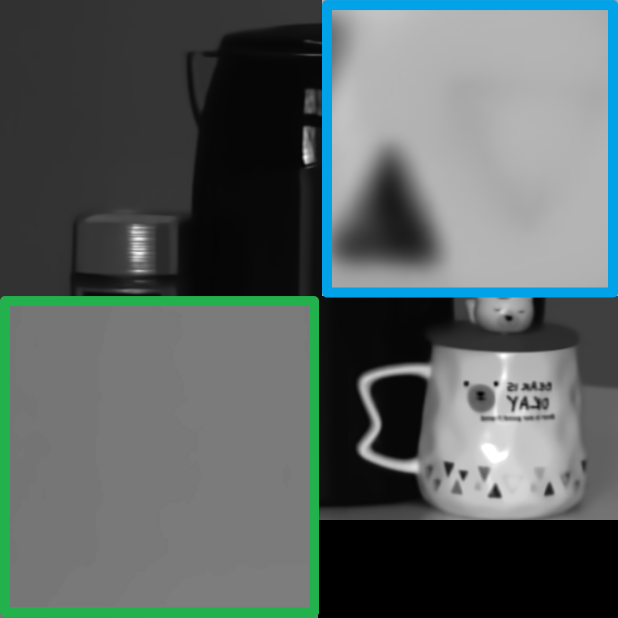}} \,
  \subfloat[MAN]{\includegraphics[width=1.0516in, height=1.0516in]{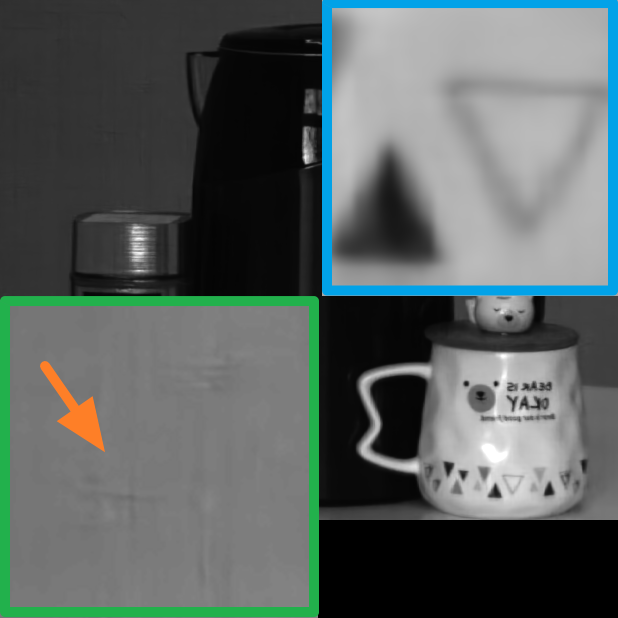}} \,
  \subfloat[GCP-ID]{\includegraphics[width=1.0516in, height=1.0516in]{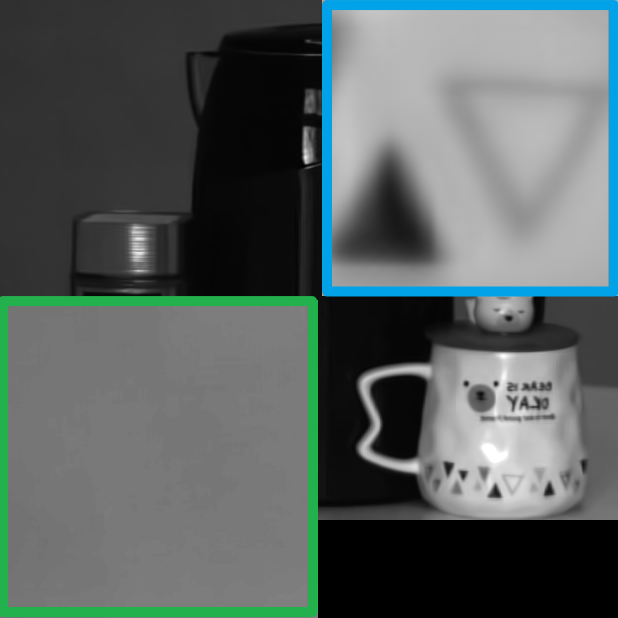}}
  \vspace{-2.8pt}
\caption{Denoising results on the Real-HSI dataset.}
\label{Fig_Real-HSI_visual_evaluation}
\end{figure}